\documentclass[prd,aps,floats,floatfix,eqsecnum,nofootinbib]{revtex4}
\usepackage{graphics}
\usepackage{graphicx}
\usepackage{rotate,psfrag}
\usepackage{color}
\usepackage{rotating}
\usepackage{psfrag}
\usepackage{amsmath,amsthm,amssymb}
\usepackage{epsfig}
\usepackage{slashed}
\usepackage{sidecap}
\newcommand{\be}{\begin{equation}}
\newcommand{\ee}{\end{equation}}
\newcommand{\bea}{\begin{eqnarray}}
\newcommand{\eea}{\end{eqnarray}}
\begin{document}

\title{\Large WARM DARK MATTER IN THE GALAXIES:

THEORETICAL AND OBSERVATIONAL PROGRESSES:

\medskip

HIGHLIGHTS and CONCLUSIONS 

\medskip

of the Chalonge CIAS Meudon Workshop 2011

\medskip

Ecole Internationale d'Astrophysique Daniel Chalonge

Meudon campus of Observatoire de Paris

in the historic Ch\^ateau, 8-10 June 2011.}

\author{\Large \bf   H.J. de Vega$^{(a,b)}$,    N.G. Sanchez$^{(b)}$}

\date{\today}

\affiliation{$^{(a)}$ LPTHE, Universit\'e
Pierre et Marie Curie (Paris VI) et Denis Diderot (Paris VII),
Laboratoire Associ\'e au CNRS UMR 7589, Tour 24, 5\`eme. \'etage, 
Boite 126, 4, Place Jussieu, 75252 Paris, Cedex 05, France. \\
$^{(b)}$ Observatoire de Paris,
LERMA. Laboratoire Associ\'e au CNRS UMR 8112.
 \\61, Avenue de l'Observatoire, 75014 Paris, France.}

\begin{abstract}
Warm Dark Matter (WDM) research is progressing fastly, the subject is new and WDM essentially {\it works}, 
naturally reproducing the astronomical observations over all the scales, small (galactic) as well as large and cosmological scales 
($\Lambda$WDM). Evidence that Cold Dark Matter ($\Lambda$CDM) and its proposed tailored cures do not work at 
small scales is staggering. The Chalonge Workshop `Warm Dark Matter in the Galaxies: Theoretical and 
Observational Progresses', was held at the Meudon Ch\^ateau of Observatoire de Paris on 8-10 June 2011. 
The Workshop approached DM in a fourfold coherent way: astronomical observations of DM structures (galaxy   
and cluster properties, haloes, rotation curves, density profiles, surface density and scaling laws), 
$\Lambda$WDM N-body simulations in agreement with observations, WDM theoretical astrophysics and cosmology 
(kinetic theory, Boltzmann-Vlasov evolution, halo model, improved perturbative approachs), WDM particle physics 
(sterile neutrinos) and its experimental search. Fedor Bezrukov, Pier-Stefano Corasaniti, Hector J. de Vega, 
Stefano Ettori, Frederic Hessmann,  Ayuki Kamada, Marco Lombardi, Alexander Merle, 
Christian Moni Bidin, Angelo Nucciotti on behalf 
of the MARE collaboration, Sinziana Paduroiu, Henri Plana, Norma G. Sanchez, Patrick Valageas, 
Shun Zhou present here their highlights of the Workshop. 
Cored (non cusped) DM halos and WDM (keV scale mass) are clearly determined from theory and 
astronomical observations, they naturally produce the observed structures at all scales; keV sterile neutrinos 
are the most suitable candidates, they naturally appear in minimal extensions of the standard 
model of particle physics.
$\Lambda$WDM simulations with keV particles remarkably reproduce the observations, the small and large structures,
sizes of local minivoids and velocity functions. The summary and conclusions by H. J. de Vega and N. G. Sanchez stress among other points the impressive evidence that DM particles have a mass in the keV scale and that those keV scale 
particles naturally produce the small scale structures observed in galaxies. Wimps (DM particles heavier than 
1 GeV) are strongly disfavoured combining theory with galaxy astronomical observations. keV scale sterile 
neutrinos are the most serious DM candidates and deserve dedicated 
experimental searchs and simulations. Astrophysical constraints including Lyman alpha bounds put the mass in 
the range $ 1 < m < 13 $ keV. Predictions for EUCLID and PLANCK have been presented. Interestingly enough, 
MARE -and hopefully an adapted KATRIN- experiment could provide a sterile neutrino signal. It will be a a 
fantastic discovery to detect dark matter in a beta decay. There is a formidable WDM work to perform ahead of us, these highlights point some of the directions where it is worthwhile to put the effort. Photos of the Workshop are included. 
\end{abstract}

\maketitle

\tableofcontents

\begin{figure}[htbp]
\epsfig{file=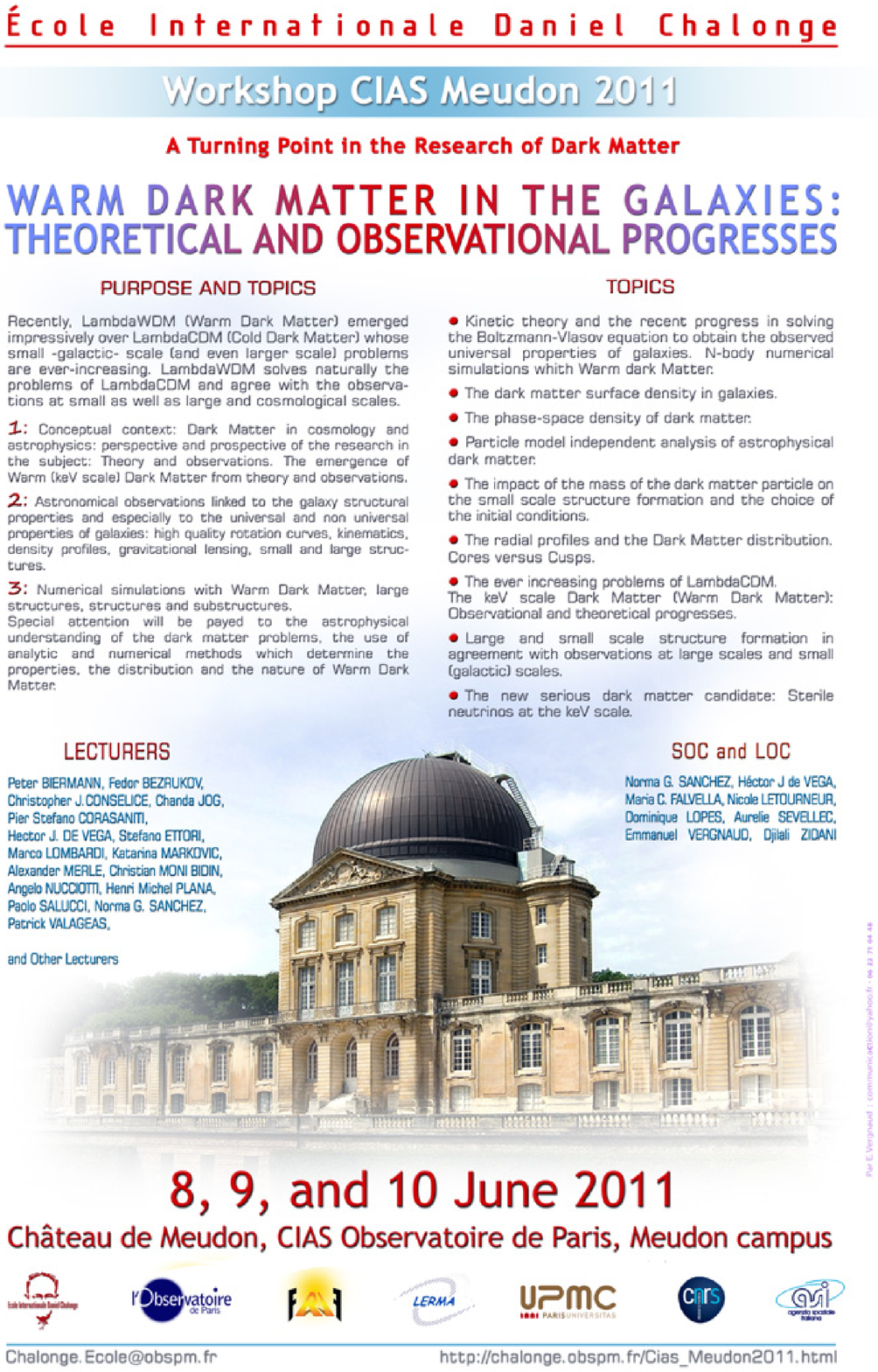,width=14cm,height=18cm}
\caption{Poster of the Workshop}
\end{figure}

\begin{figure}[h]
\begin{turn}{-90}
\includegraphics[width=100mm,height=150mm]{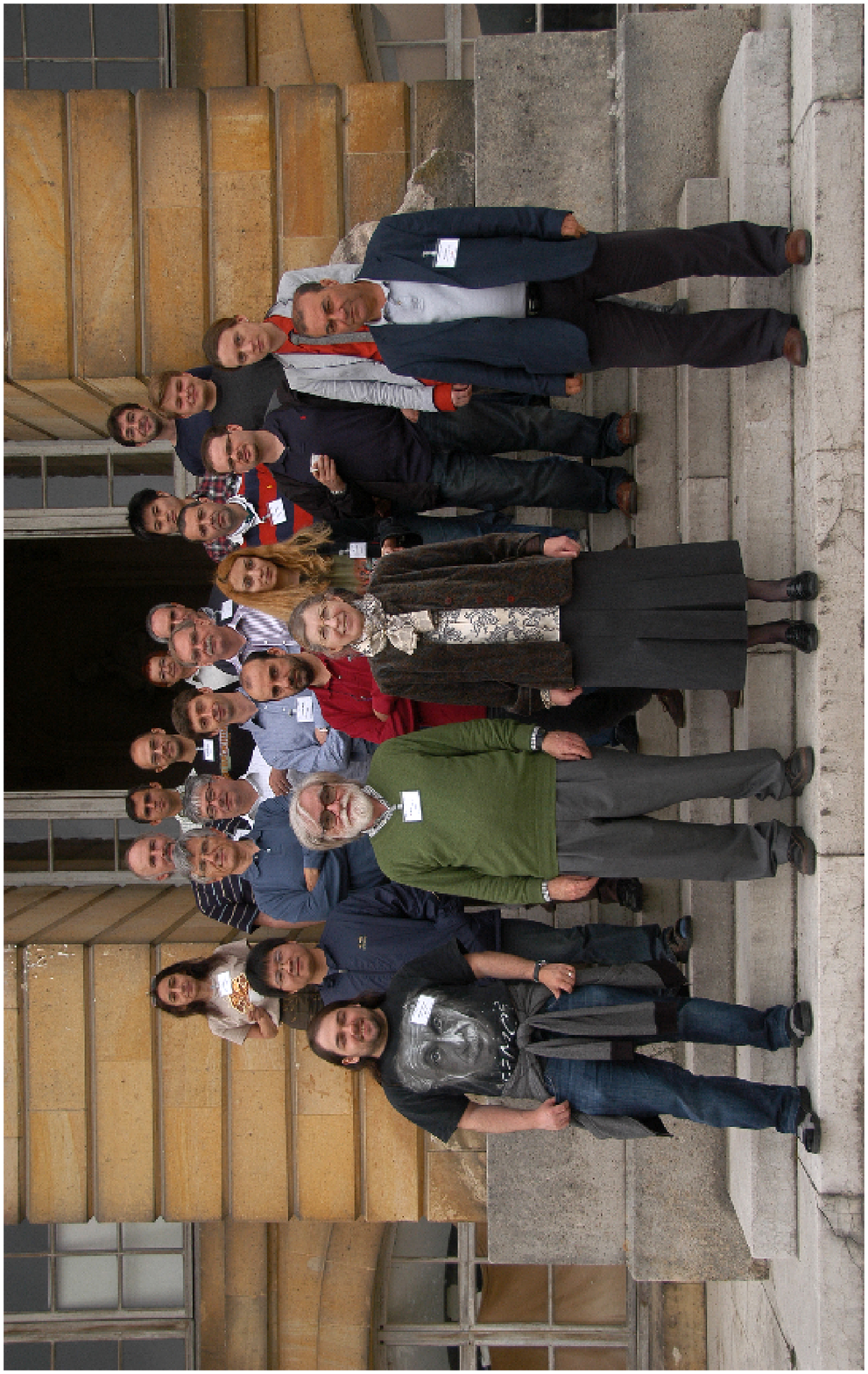}
\end{turn}
\caption{Photo of the Group}
\end{figure}

\section{Purpose of the Workshop, Context and Introduction}

This Workshop addresses for the first time the turning point in the research of Dark Matter represented by 
Warm Dark Matter (WDM) putting together astrophysical, cosmological and particle WDM,
astronomical observations, theory and WDM numerical simulations which naturally reproduce the observations, 
as well as the experimental search for the WDM particle candidates (sterile neutrinos).

\medskip

Recently, $\Lambda$WDM  emerged impressively over $\Lambda$CDM
($\Lambda$-Cold Dark Matter) whose small -galactic- scale problems are staggering. 
$\Lambda$WDM solves naturally the problems of $\Lambda$CDM and {\it agrees} with the observations 
at small as well as large and cosmological scales.

\medskip

This Workshop is the second of a new series dedicated to Dark Matter. 

\medskip

The first Workshop of this series in the Meudon Castle CIAS in June 2010 allowed to identify  and understand 
the issues of the serious problems faced by Cold Dark Matter (CDM) to reproduce the galactic 
(and even cluster of galaxies) observations. \\

The 2010  and 2011 Workshops served as well to verify and better understand the always growing 
amount of confusion in the 
CDM research, namely the increasing number and cyclic change of arguments, counter-arguments and ad-hoc 
mechanisms introduced in the CDM simulations over most of twenty years, in trying to deal with the CDM small 
scale crisis: Cusped profiles and overabundance of substructures. Too many satelites are predicted by CDM 
simulations while cored profiles and no such overabundant substructures are seen by astronomical observations. 
A host of ad-hoc mechanisms are proposed and advocated to cure the CDM problems. `Baryon and supernovae feedbacks',
non circular motions, triaxiality, mergers, `cusps hidden in cores', 
`strippings' are some of such mechanisms tailored or exagerated for the purpose of obtaining the desired result
without having a well established physical basis. For example, the strong "baryon and 
supernovae feedback" introduced to transform the CDM cusps into cores in baryon+CDM simulations
corresponds to a large star formation rate contradicting the observations.

\medskip

On the CDM particle physics side, 
the problems are no less critical: So far, all the {\it dedicated} experimental searches after most of 
twenty years to find the theoretically proposed CDM particle candidate (WIMP) have {\bf failed}. The
CDM indirect searches (invoking CDM annihilation) to explain cosmic ray positron excesses, are in crisis as 
well, as wimp annihilation models are plagued with growing tailoring or fine tuning, and in any case, 
such cosmic rays excesses are well explained and reproduced by natural astrophysical process and sources, 
as non-linear acceleration, shocks and magnetic winds around massive explosion stars, quasars. The so-called 
and repeatedealy invoked `wimp miracle' is nothing but been able to solve one equation with three unknowns  
(mass, decoupling temperature, and annhiliation cross section) within WIMP models
theoretically motivated by SUSY model building twenty years ago
(SUSY was very fashionable at that time and believed a popular motivation for many proposals).

\medskip

After more than twenty years -and as often in big-sized science-, CDM research has by now its own internal 
inertia: growing simulations involve large super-computers and large number of people working with, CDM particle 
wimp search involve large and longtime planned experiments, huge number of people, (and huge budgets); one should not be surprised in 
principle, if a fast strategic change would not yet operate in the CDM and wimp research, although
they would progressively decline. 

\bigskip

In contrast to the CDM situation, the WDM research situation is progressing fast, the subject is new and 
WDM essentially {\it works}, naturally reproducing the observations over all the scales, small as well as 
large and cosmological scales ($\Lambda$WDM).

\bigskip

This 2011 Workshop addressed the last progresses made in WARM DARK MATTER and the Universal and Non Universal properties of Galaxies. In the tradition of the Chalonge School, an effort of clarification and synthesis is  made by combining in a conceptual framework, theory, analytical, observational and numerical simulation results which reproduce observations.

\bigskip

\begin{center} 

{\bf The subject have been approached in a fourfold coherent way:}

\end{center}

\bigskip

(I) Conceptual context 

\bigskip

(II) Astronomical observations linked to the galaxy structural properties and to structure formation at different large and small (galactic) scales.

\bigskip

(III) WDM Numerical simulations which reproduce observations at large and small (galactic) scales.

\bigskip

(IV) WDM particle candidates, keV sterile neutrinos: particle models and astrophysical constraints on them

\bigskip

\begin{center}

{\bf The Topics covered included:}

\end{center}

\medskip

Recent progress in solving the Boltzmann-Vlasov equation to obtain the observed properties of galaxies (and clusters of galaxies).
N-body numerical simulations with Warm Dark Matter; the surface density; scaling laws, universality and Larson laws;  the phase-space density. Particle model independent analysis of astrophysical dark matter. The impact of the mass of the dark matter particle on the small scale structure formation.

\medskip

The radial profiles and the Dark Matter distribution. Observed cored density profiles. The ever increasing problems 
of $\Lambda$CDM at small scales. The keV scale Dark Matter (Warm Dark Matter): Observational and theoretical progresses. 
Perturbative  approachs and the Halo model. Large and small scale structure formation in agreement with 
observations at large and small (galactic) scales. The new serious dark matter candidate: Sterile neutrinos 
at the keV scale:  Particle physics models of sterile neutrinos, astrophysical constraints (Lyman alpha, 
Supernovae, weak lensing surveys),  experimental searches of keV sterile neutrinos

\medskip

Fedor Bezrukov, Pier-Stefano Corasaniti, Hector J. de Vega, Stefano Ettori, Frederic Hessmann,  Ayuki Kamada, 
Marco Lombardi, Alexander Merle, Christian Moni Bidin, Angelo Nucciotti on behalf 
of the MARE collaboration, Sinziana Paduroiu, Henri Plana, Norma G. Sanchez, Patrick Valageas, 
J. Shun Zhou present here their highlights of the Workshop. 

\medskip

\begin{center}

{\bf The Summary and Conclusions:}

\end{center}

Summary and conclusions are presented at the end by H. J. de Vega and N. G. Sanchez in three subsections: 

\medskip

A. General view and clarifying remarks. 

B. Conclusions. 

C. The present context and future in the DM research.

\medskip

The conclusions stress among other points the 
impressive evidence that DM particles have a mass in the keV scale and that those keV scale particles naturally 
produce the small scale structures observed in galaxies. Wimps (DM particles heavier than 1 GeV) are strongly 
disfavoured combining theory with galaxy astronomical observations. keV scale sterile neutrinos are the most 
serious DM candidates and deserve dedicated 
experimental searchs and simulations. Astrophysical constraints including Lyman alpha bounds put the 
sterile neutrinos mass in 
the range $ 1 < m < 13$ keV. Predictions for EUCLID and PLANCK have been presented. MARE -and hopefully 
an adapted KATRIN- experiment could provide a sterile neutrino signal. The experimental search for these 
serious DM candidates appears urgent and important: It will be a a fantastic discovery to 
detect dark matter in a beta decay. 

\medskip

There is an encouraging and formidable WDM work to perform ahead us, these highlights point some of the directions
where to put the effort.

\bigskip

Sessions lasted for three full days in the beautiful Meudon campus of Observatoire de Paris, where CIAS 
`Centre International d'Ateliers Scientifiques' is located. All sessions took place in the historic 
Chateau building, (built in 1706 by the great architect Jules-Hardouin Mansart in orders by King Louis XIV 
for his son the Grand Dauphin).

\bigskip

The meeting was open to all scientists interested in the subject. 
All sessions were plenary followed by discussions. 
The format of the Meeting was intended to allow easy and 
fruitful mutual contact and communication. Large time was devoted  
to discussions. All informations about the meeting, are displayed at

\begin{center}

{\bf http://www.chalonge.obspm.fr/Cias\_Meudon2011.html}

\end{center}

The presentations by the lecturers are available on line 
(in .pdf format) in `Programme and Lecturers' in the above link, 
as well as the Album of photos of the Workshop: 

\begin{center}

{\bf http://www.chalonge.obspm.fr/Programme\_CIAS2011.html}

\bigskip

{\bf http://www.chalonge.obspm.fr/albumCIAS2011/index.html}

\end{center}

We thank all again, both lecturers and participants, 
for having contributed so much to the great success of this 
Workshop and look forward to seeing you again in the next 
Workshop of this series.     

\medskip

We thank the Observatoire de Paris and its Scientific Council, CNRS-GISP2I and UPMC supports, as well as the
CIAS, secretariat and logistics assistance and all those who contributed so efficiently to the
successful organization of this Workshop.

\medskip
      
\begin{center}
                                   
With compliments and kind regards,

\bigskip  
                                            
Hector J de Vega, Norma G Sanchez

\end{center}

\newpage

\section{Programme and Lecturers}

\begin{itemize}

\item{{\bf Fedor BEZRUKOV} (Ludwig-Maximilians-Universitat M\"unchen, Germany)
Light sterile neutrino dark matter in extensions of the Standard Model}

\item{{\bf Christopher J. CONSELICE} (School for Physics and Astronomy, University of Nottingham, UK)
Correlations between baryons and dark matter during galaxy formation}

\item{{\bf Pier Stefano CORASANITI} (CNRS LUTH Observatoire de Paris, Meudon, France)
Theorical modeling of the halo mass function and halo bias: a path-integral approach to the Excursion Theory}

\item{{\bf Hector J. DE VEGA} (CNRS LPTHE Univ de Paris VI, France)
Warm Dark Matter from theory and galaxy observations}

\item{{\bf Stefano ETTORI} (INAF, Osservatorio Astronomico Bologna, Italy)
Mass profiles and concentrations in X-ray galaxy clusters}

\item{{\bf Frederic V. HESSMAN} \& Monika ZIEBART (Inst. F. Astrophysik, 
Georges-August-Univ. G\"ottingen, G\"ottingen Germany)
The Bosma effect revisited - Correlations between the Interstellar Medium and Dark Matter in galaxies}

\item{{\bf Ayuki KAMADA} (IPMU, Institute for the Physics and Mathematics of the Universe, Tokyo, Japan)
keV-mass sterile neutrino dark matter and the structure of galactic haloes}

\item{{\bf Marco LOMBARDI} (University of Milano, Italy)
Larson' laws and the universality of molecular cloud structures}

\item{{\bf Katarina MARKOVIC} (Max-Planck-Ludwigs-Maximilians UniversitÃ¤t, Munich, Garching, Germany )
Constraining Warm Dark Matter with future cosmic shear data}

\item{{\bf Alexander MERLE} (Royal Institute of Technology KTH Stockholm, Sweden)
Neutrino model building and keV sterile neutrino Dark Matter}

\item{{\bf Christian MONI BIDIN} (Universidad de Concepci\'on, Departamento de Astromomia, Chile)
The lack of evidence of a Dark Matter disk in the Milky Way}

\item{{\bf Angelo NUCCIOTTI} (INFN Milano Biccoca, Italy)
The MARE experiment in Milano Bicocca and its capabilities to measure the mass of light (active) and heavy (sterile) neutrinos}

\item{{\bf Sinziana PADUROIU}, Observatoire de Geneve, Switzerland Numerical simulations with Warm Dark Matter: The effects of free streaming on Warm Dark Matter haloes of Galaxies}

\item{{\bf Henri Michel PLANA} (Laboratorio de Astrofisica Teorica e Observacional, 
UESC Santa Cruz, Bahia, Brazil) Mass Distribution in Groups of Galaxies}

\item{{\bf Oleg RUCHAYSKIY} (CERN-Teory, Geneva, Switzerland)
Sterile neutrinos as dark matter candidates and the constraints on them from Lyman-alpha forest data}

\item{{\bf Paolo SALUCCI} (SISSA-Astrophysics, Trieste, Italy)
Structural properties of Galaxies and Cored Density Profiles from Galaxy observations}

\item{{\bf Norma G. SANCHEZ} (CNRS LERMA Observatoire de Paris, Paris, France)
Warm Dark Matter and Galaxy properties from primordial fluctuations and observations.}

\item{{\bf Joop SCHAYE} (Leiden University, The Netherlands)
Dark matter haloes, central AGN black holes, and the effect of baryons on the clustering of matter}

\item{{\bf Patrick VALAGEAS} (Inst. de Physique Th\'eorique, Orme de Merisiers, 
CEA-Saclay, Gif-sur-Yvette, France) Perturbation theories for large-scale structures}

\item{{\bf Shun ZHOU} (Max-Planck-Heinsenberg Inst., Munich, Germany)
Supernova bound on keV-mass sterile neutrinos}

\end{itemize}

\newpage

\section{Highlights by the Lecturers}

\subsection{Fedor Bezrukov}

\vskip -0.3cm

\begin{center}

Department f\"ur Physik, Ludwig-Maximilians-Universit\"{a}t M\"{u}nchen, Germany\\
Institute for Nuclear Research of RAS, Moscow, Russia

\bigskip

{\bf Light sterile neutrino dark matter in extensions of the Standard Model}

\end{center}

\medskip

In this talk I would like to summarize the possibilities of having a
light sterile neutrino DM in the extensions of the SM.  Such a
candidate is naturally very weakly interacting.  At the same time, if
sterile neutrino is light (of the order of several keV), its lifetime
can exceed significantly the Universe lifetime without exceptional
fine tuning.  However, obtaining the proper observed abundance of the
neutrino DM is quite constraining.  The usual DM production mechanism
is the tehrmal one, which corresponds to a particle that enters the
thermal equilibrium and freezes out later.  Then its abundance (number
to entropy density ratio) is constant after freeze out (if no
additional entropy is generated later), and depends on how
relativistic the particle was at decoupling.  If the particle freezes
out while relativistic (what is the case for a keV scale particle),
then its abundance is $n/s=135\zeta(3)/4\pi^4g_{*f}$, with $g_{*f}$
being the effective number of degrees of freedom at freeze out.  Then
the mass $M$ of the particle is fixed from the requirement of its
energy density being equal to the observed DM value
\[
  \frac{\Omega_N}{\Omega_\mathrm{DM}}
  \simeq
  \frac{1}{S}\left(\frac{10.75}{g_{*\mathrm{f}}}\right)
  \left(\frac{M}{1~\mathrm{keV}}\right)\times100
  \;,
\]
where $S$ stands for the possible entropy generation factor.

The possible situations are the following.

\begin{figure}[h]
  \centering
  \includegraphics[width=0.39\textwidth]{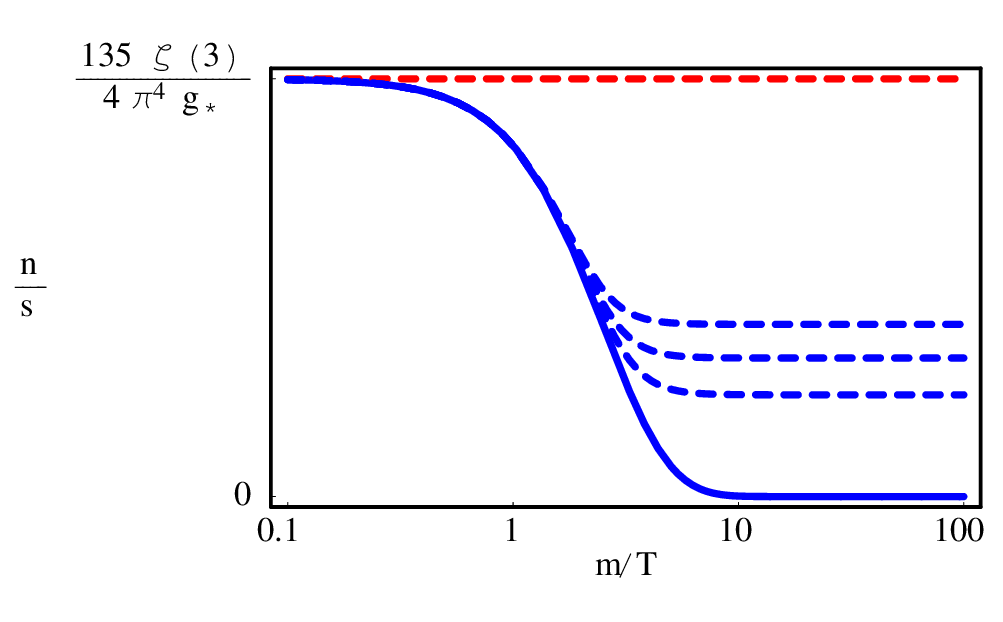}
  \includegraphics[width=0.51\textwidth]{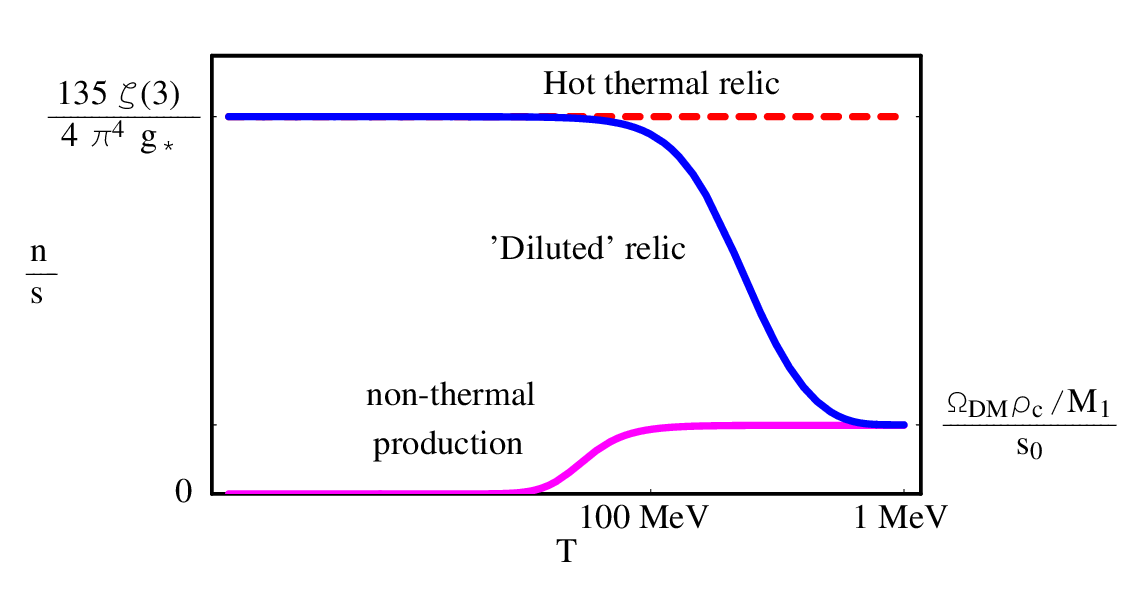}
  \caption{Schematic evolution of the relic abundance in the Universe.
    Left plot depicts the standard CDM mechanism, where the abundance
    is controlled by the decoupling moment (several dashed curves).
    Right plot illustrates various non-thermal mechanisms.  The dashed
    line is a thermal relic decoupled while being relativistic (hot
    thermal relic), leading to the over closure of the Universe.  The
    blue decreasing line is the same hot thermal relic, but with the
    abundance diluted by rapid expansion of the Universe (entropy
    production), leading to correct DM abundance.  The lowest
    (magenta) line depicts the evolution of the non-thermally produced
    particle with zero primordial abundance.}
\end{figure}

\begin{figure}[t]
  \centering
  \includegraphics[width=0.33\textwidth]{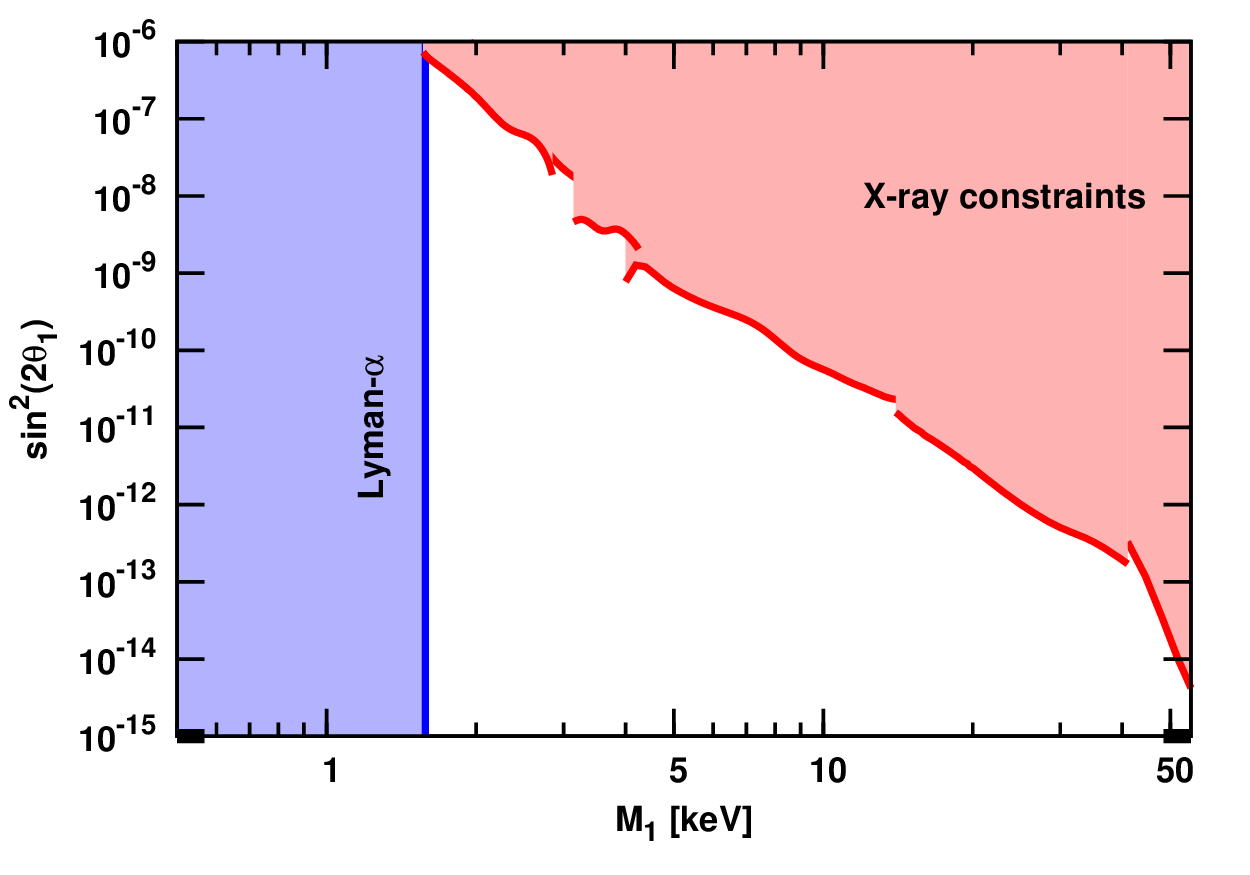}
  \includegraphics[width=0.33\textwidth]{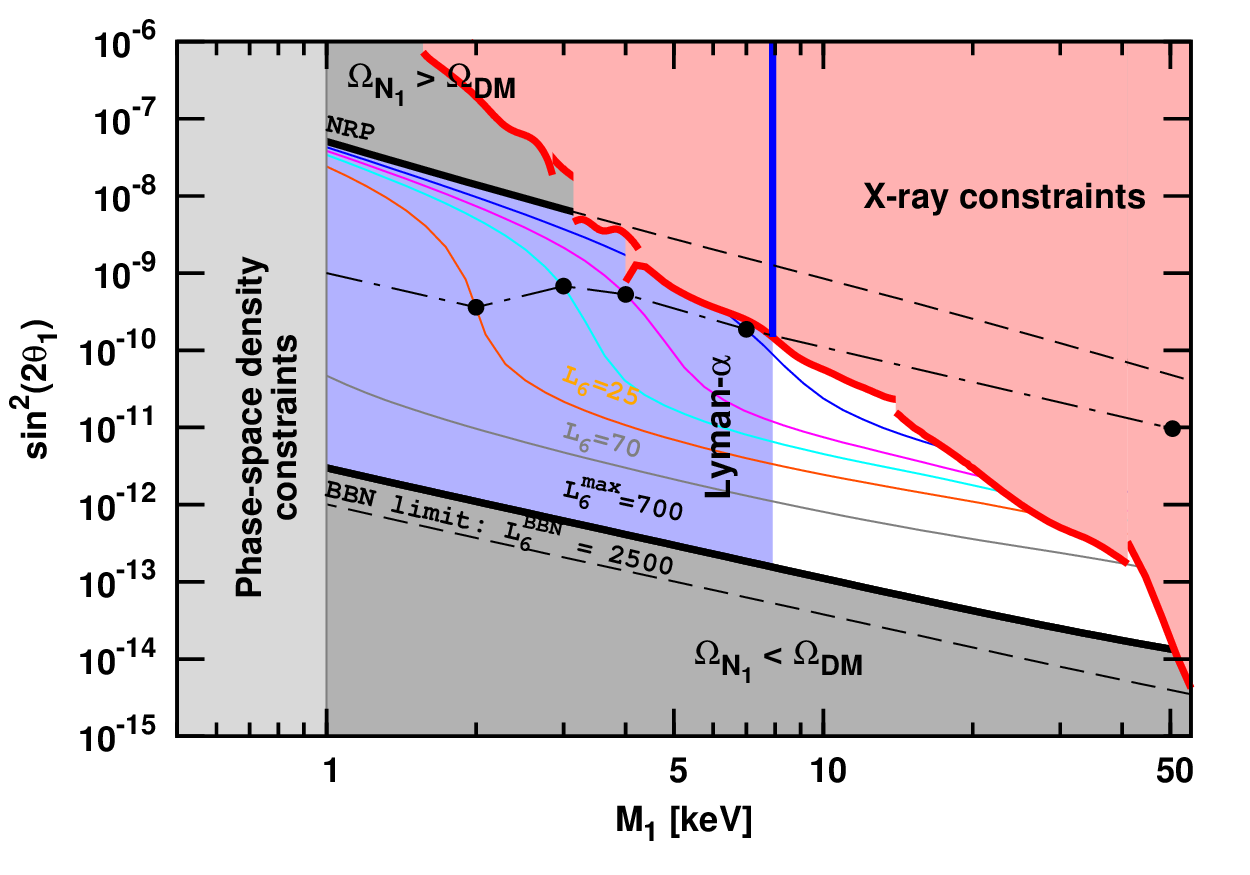}%
  \includegraphics[width=0.33\textwidth]{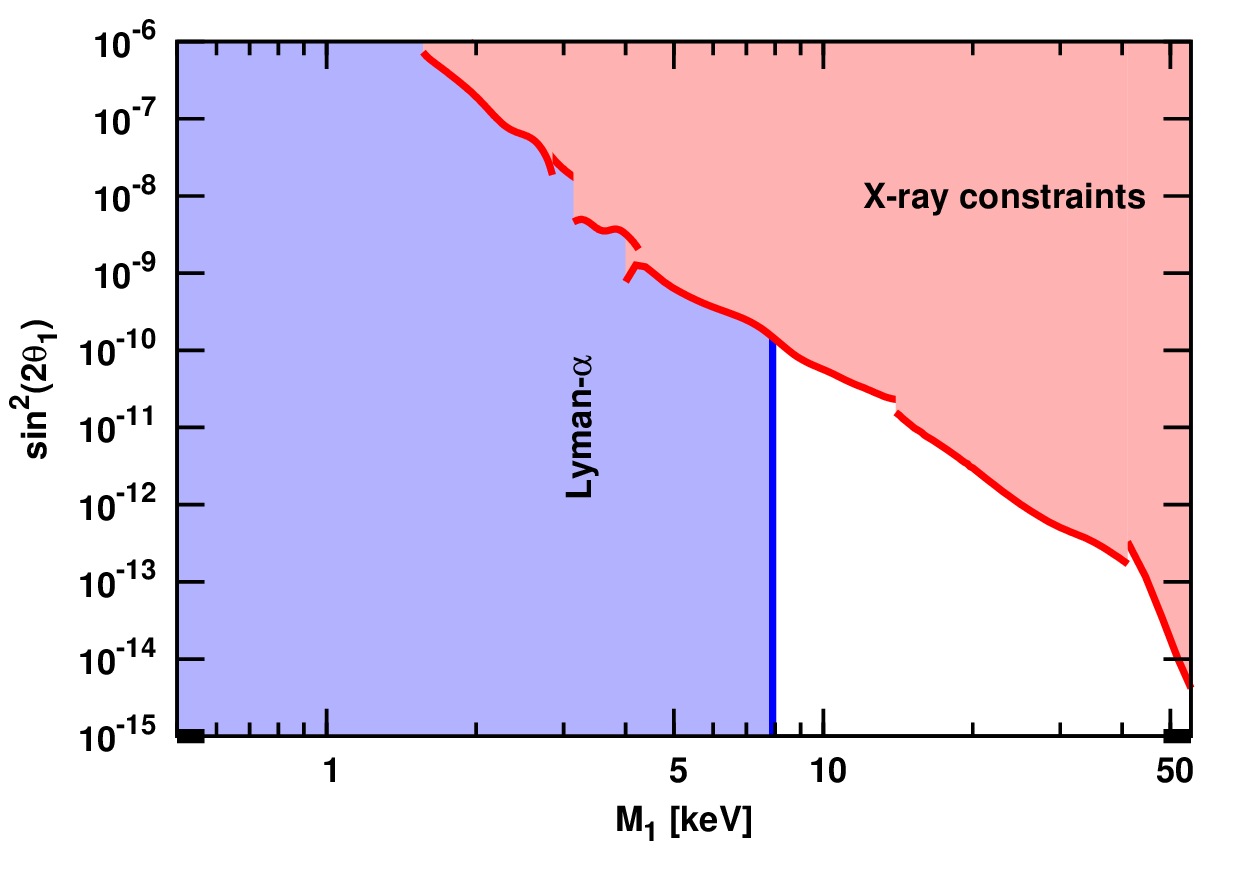}%
  \caption{Bounds on the mass $M_1$ and the mixing angle $\theta_1$ of
    the DM sterile neutrino for (left to right) the model with entropy
    dilution, $nu$MSM, and production in the inflaton decays.}
\label{bezrukov}
\end{figure}

\textbf{Hot Dark Matter.}
If neutrino enters thermal equilibrium and there is no entropy
dilution $S=1$, then the DM mass is $M\sim10~\mathrm{eV}$.  This gives
us HDM, which is excluded now from the structure formation
observations.

\textbf{Entropy dilution.}
However, the constraint on the mass may be weakened if there is large
subsequent entropy dilution $S$.  This entropy production happens if
there is some other particle with long lifetime, which first decouples
while still relativistic and then decays when already non-relativistic
[1].  Then the proper DM abundance is controlled by the properties of
this long-lived particle through the entropy dilution factor $S\simeq
0.76\frac{\bar{g}_{*}^{1/4}M_2}{g_{*f} \sqrt{\Gamma M_\mathrm{pl}}}$,
where $g_*$ is an averaged number of d.o.f.\ during entropy
generation.

This situation can be realized if the sterile neutrinos are charged
under some beyond the SM gauge group [2].  Note, that teh gauge scale
should be below GUT, otehrwise the neutrinos do not enter the thermal
equilibrium after reheating and the situation is similar to those of
the $\nu$MSM (described in the next paragraph). In this setup one of
the sterile neutrinos $N_1$ is light and is the DM particle, while
other sterile neutrinos $N_{2,3}$ should dilute its abundance up to
the correct amount by out-of-equilibrium decay.

The parameters of the DM sterile neutrino are constrained by two
considerations.  First, its lifetime in the radiative decays
$N_1\to\nu\gamma$ should be large enough not to contradict the X-ray
observations.  This constraints the mixing angle $\theta_1$ of $N_1$
and active neutrinos from above (see figure).  At the same time the
structure formation from the Lyman-$\alpha$ analysis constraints its
mass from below $M_1>1.6$~keV (its velocity distribution is that of
the cooled thermal relic, so the bound is weaker, than in the other
cases, analyzed in this note).

All other constraints in this scenario correspond to the heavier
sterile neutrinos and to the gauge sector.  The right abundance of the
sterile neutrino requires entropy dilution.  To provide proper the
entropy dilution, $N_2$ should decouple while relativistic and has
decay width
\(
  \Gamma \simeq
  0.50\times10^{-6}
  \frac{g_N^2}{4} \frac{g_{*\mathrm{f}}^2}{g_{*}^2}
  \bar{g}_{*}^{1/2}\frac{M_2^2}{M_\mathrm{pl}}
  \left(\frac{1~\mathrm{keV}}{M_1}\right)^2
\).
At the same time, the heavy neutrino $N_2$ should decay before BBN,
which bounds its lifetime to be shorter than approximately
$0.1\div2$~s.  Then, the proper entropy can be generated only if its
mass is larger than
\begin{equation}
  M_2 > \left(\frac{M_1}{1~\mathrm{keV}}\right)
  (1.7\div10)~\mathrm{GeV}
  \;.
\end{equation}
The entropy is effectively generated by out-of-equilibrium decay, if
the particle decoupled while still relativistic.  The bound on the
decoupling temperature leads to the bound on the right-handed gauge
boson mass \( M >
\frac{1}{g_{*\mathrm{f}}^{1/8}}\left(
  \frac{M_2}{1~\mathrm{GeV}t}\right)^{3/4}(10\div16)~\mathrm{TeV} \).

\textbf{Non-thermal production within $\nu$MSM.}
Another option to escape the overproduction of the light sterile
neutrinos is to make them not enter the thermal equilibrium at all.
This is the case for the simple extension of the SM by three right
handed sterile neutrinos---$\nu$MSM [3--5].  In this case the
production of the sterile neutrinos proceed via oscillations from the
active neutrinos.  These oscillations are \emph{non-resonant} if there
are no lepton asymmetries during the DM generation, or
\emph{resonantly enhanced} if the lepton asymmetries are present.
Then, the amount of produced DM is governed by the mixing angle
$\theta_1$.  However for nonzero lepton asymmetries, there is an
allowed region [5,6] in $\theta_1$ (see figure \ref{bezrukov}), 
with the upper dashed
line corresponding to the non-resonant production, and the lower
shaded area corresponds to lepton asymmetries which are too large and
contradict BBN.  The usual X-ray bound also bounds $\theta_1$.
Together with the Lyman-$\alpha$ bound this forbids the
\emph{non-resonant} production mechanism, however there is an open
region for the \emph{resonant} mechanism.  Note, that the
Lyman-$\alpha$ bound depends strongly on the properties of the
resonance, and can be as small as $M_1>2$~keV for some values of
lepton asymmetry, see [6] for details.

\textbf{Production in inflaton decays.}
Finally, the neutrinos may be produced by some other mechanism, like
inflaton decay [7--9].  In this case the production is again not
connected with $\theta_1$, so the only bounds present for the DM
neutrino properties are the X-ray bound and the Lyman-$\alpha$ bound.
It is interesting to note, that the mass of the DM sterile neutrino
can be related to the mass of the inflaton, which is light in the
model and can be searched for in B-meson decays [9].

\bigskip

{\bf References}

\begin{description}

\item[1] R.~J.~Scherrer, M.~S.~Turner,
  Phys.\ Rev.\  {\bf D31 } (1985)  681.

\vspace{-0.3cm}

\item[2] F.~Bezrukov, H.~Hettmansperger, M.~Lindner,
  Phys.\ Rev.\  {\bf D81 } (2010)  085032.
  [arXiv:0912.4415].

\vspace{-0.3cm}

\item[3] T.~Asaka, S.~Blanchet, M.~Shaposhnikov,
  Phys.\ Lett.\  {\bf B631 } (2005)  151-156.
  [hep-ph/0503065].

\vspace{-0.3cm}

\item[4] T.~Asaka, M.~Shaposhnikov,
  Phys.\ Lett.\  {\bf B620 } (2005)  17-26.
  [hep-ph/0505013].

\vspace{-0.3cm}

\item[5] A.~Boyarsky, O.~Ruchayskiy, M.~Shaposhnikov,
  Ann.\ Rev.\ Nucl.\ Part.\ Sci.\  {\bf 59 } (2009)  191-214.
  [arXiv:0901.0011].

\vspace{-0.3cm}

\item[6] A.~Boyarsky, J.~Lesgourgues, O.~Ruchayskiy, M.~Viel,
  JCAP {\bf 0905 } (2009)  012.
  [arXiv:0812.0010].

\vspace{-0.3cm}

\item[7] M.~Shaposhnikov, I.~Tkachev,
  Phys.\ Lett.\  {\bf B639 } (2006)  414-417.
  [arXiv:hep-ph/0604236].

\vspace{-0.3cm}

\item[8] A.~Anisimov, Y.~Bartocci, F.~L.~Bezrukov,
  Phys.\ Lett.\  {\bf B671 } (2009)  211-215.
  [arXiv:0809.1097].

\vspace{-0.3cm}

\item[9]   F.~Bezrukov, D.~Gorbunov,
  JHEP {\bf 1005 } (2010)  010.
  [arXiv:0912.0390].

\end{description}

\newpage

\subsection{Pier Stefano Corasaniti \& Ixandra Achitouv}

\vskip -0.3cm

\begin{center}
Laboratoire Univers et Th\'eories (LUTh),\\ UMR 8102
CNRS, Observatoire de Paris, Universit\'e Paris Diderot, \\ 5 Place
  Jules Janssen, 92190 Meudon, France

\bigskip

{\bf Theoretical Modeling of the Halo Mass Function: A Path-Integral Approach to the Excursion Set Theory} 

\end{center}

\medskip

The computation of the halo mass function is of primary importance in cosmology.Halos are the building blocks of the cosmic structure formation and predicting their distribution over space and time is key to understanding the formation of galxies and the other visible structures in the universe. 

Because of the non-linear gravitational interactions which lead to the formation of halos, their properties have been mainly investigated using large volume numerical N-body simulations. 

The excursion set theory [1] provides a powerful mathematical framework to model the characteristics of the halo mass distribution. In the original formulation an analytical evaluation has been limited to few ideal cases, while requiring the use of Monte Carlo simulations to solve the stochastic model equations for more realistic scenarios. The recent formulation of the excursion set in terms of path-integrals [2] extends the analytical computation of the halo mass function to halo mass definitions which are consistent with those from observations, thus allowing for a direct comparison with numerical simulations as well as observational results. Furthermore, a stochastic modeling of the halo collapse conditions can be easily implemented in this formalism [3]. 

In this talk we review the derivation of the halo mass function in the case of a model of ellipsoidal collapse of halos [4,5]. We show that the inferred mass function can be written in terms of physically motivated quantities. Morevoer, the result of the theoretical computation is in excellent agreement with results from N-body simulation data.

\bigskip

{\bf References}

\begin{description}

\item[1] J. R. Bond, S. Cole, G. Efstathiou and G. Kaiser,
  {\it Excursion set mass functions for hierarchical Gaussian
  fluctuations}, Astrophys. J. {\bf 379} (1991) 440.

\vspace{-0.3cm}

\item[2] M. Maggiore \& A. Riotto, {\it The Halo
  Mass Function from Excursion Set Theory. I. Gaussian Fluctuations
  with Non-Markovian Dependence on the Smoothing Scale},
  Astrophys. J. {\bf 711} (2010) 907 (2010).

\vspace{-0.3cm}

\item[3] M. Maggiore and A. Riotto, 
{\it The Halo mass function from Excursion Set Theory. II. The Diffusing Barrier}, Astrophys. J. {\bf 717}, 515 (2010).

\vspace{-0.3cm}

\item[4] P. S. Corasaniti and I. Achitouv, {\it Toward a
  Universal Formulation of the Halo Mass Function},
  Phys. Rev. Lett. 106, 241302 (2011), arXiv:1012.3468
\vspace{-0.3cm}

\item[5] P. S. Corasaniti and I. Achitouv, {\it Excursion Set Halo Mass Function and Bias in a Stochastic
  Barrier Model of Ellipsoidal Collapse}, PRD in press, arXiv:1107.1251

\end{description}

\newpage

\subsection{Hector J. de Vega and Norma G. Sanchez}

\vskip -0.3cm

\begin{center}

HJdV: LPTHE, CNRS/Universit\'e Paris VI-P. \& M. Curie \& Observatoire de Paris.\\
NGS: Observatoire de Paris, LERMA \& CNRS

\bigskip

{\bf keV scale dark matter from theory and observations and 
galaxy properties from linear primordial fluctuations} 

\end{center}

In the context of the standard Cosmological model the nature of dark matter (DM)
is unknown. Only the DM gravitational effects are noticed and they are necessary
to explain the present structure of the Universe. DM particle candidates are not present in the standard model (SM) of particle physics. Particle model independent theoretical analysis combined with astrophysical data from
galaxy observations points towards a DM particle mass in
the {\bf keV scale} (keV = 1/511 electron mass) [1-4]. Many extensions of the SM can be envisaged to include a DM particle with mass in the keV scale and
weakly enough coupled to the Standard Model particles to fulfill
all particle physics experimental constraints.

DM particles can decouple being ultrarelativistic (UR) at 
$ \; T_d \gg m $ or non-relativistic $ \; T_d \ll m $.
They may  decouple at or out of local thermal equilibrium (LTE).
The DM distribution function: $ F_d[p_c] $ freezes out at decoupling
becoming a function of the  comoving momentum $ p_c = $.
$ P_f(t) = p_c/a(t) = $ is the physical momentum. 
Basic physical quantities can be expressed in terms of the
distribution function as the velocity fluctuations,
$ \langle \vec{V}^2(t) \rangle = \langle \vec{P}^2_f(t) \rangle/m^2 $ 
and the DM energy density $ \rho_{DM}(t) $
where $ y = P_f(t)/T_d(t) = p_c/T_d $ is the integration variable and
$ g $  is the number of internal degrees of freedom of the DM 
particle; typically $ 1 \leq g \leq 4 $.

\medskip

{\bf Two} basic quantities characterize DM: its particle mass $ m $
and the temperature $ T_d $ at which DM decouples.  $ T_d $
is related by entropy conservation to the number of
ultrarelativistic degrees of freedom $ g_d $ at decoupling by
$ \quad T_d = \left(2/g_d \right)^\frac13 \; T_{cmb} \; ,
\; T_{cmb} = 0.2348 \; 10^{-3} \; $ eV.
One therefore needs {\bf two} constraints to determine the values of
$ m $ and $ T_d $ (or $ g_d $).

\medskip

One constraint is to reproduce the known cosmological DM density today.
$\rho_{DM}({\rm today})= 1.107 \; {\rm keV/cm}^3 $.

Two independent further constraints are considered in refs. [1-4].
First, the phase-space density $ Q=\rho/\sigma^3 $ [1-2] and second the
surface acceleration of gravity (surface density) in DM dominated galaxies [3-4].
We therefore provide {\bf two} quantitative ways to derive the value $ m $ 
and $ g_d $ in refs. [1-4].

\medskip

The phase-space density $ Q $ is invariant under the
cosmological expansion and can {\bf only decrease} 
under self-gravity interactions 
(gravitational clustering). The value of $ Q $ today follows
observing dwarf spheroidal satellite galaxies of the Milky Way (dSphs):
$ Q_{today} = (0.18 \;  \mathrm{keV})^4 $ (Gilmore et al. 07 and 08).
We compute explicitly $ Q_{prim} $ (in the primordial universe) and it turns
to be proportional to $ m^4 $ [1-4].

\medskip

During structure formation $ Q $
{\bf decreases} by a factor that we call $ Z $. Namely, 
$ Q_{today} = Q_{prim}/Z $. The value of $ Z $ is galaxy-dependent.
The spherical model gives $ Z \simeq 41000 $
and $N$-body simulations indicate: $ 10000 >  Z > 1 $ (see [1]).
Combining the value of $ Q_{today} $ and $\rho_{DM}({\rm today}) $ with 
the theoretical analysis yields that $ m $ must be in the keV scale and 
$ T_d $ can be larger than 100 GeV. More explicitly, we get 
general formulas for $ m $ and $ g_d $ [1]:
$$ 
m = \frac{2^\frac14 \; \sqrt{\pi}}{3^\frac38 \; g^\frac14 } \; 
Q_{prim}^\frac14
\; I_4^{\frac38} \; I_2^{-\frac58} \; , \quad
g_d = \frac{2^\frac14 \; g^\frac34}{3^\frac38 \; 
\pi^\frac32 \; \Omega_{DM}} \; 
 \; \frac{T_{\gamma}^3}{\rho_c} \; Q_{prim}^\frac14 \; 
\left[I_2 \; I_4\right]^{\frac38}
$$
where $ I_{2 \, n} = \int_0^\infty y^{2 \, n} \; F_d(y) \; dy 
\quad , \quad n=1, 2 $ 
and $ Q_{prim}^\frac14 = Z^\frac14 \; \; 0.18 $ keV using the dSphs data,
$T_{\gamma} = 0.2348 \; {\rm meV } 
\; , \; \Omega_{DM} = 0.228 $ and $ \rho_c = (2.518 \; {\rm meV})^4$.
These formulas yield for relics decoupling UR at LTE:
$$ 
m = \left(\frac{Z}{g}\right)^\frac14 \; \mathrm{keV} \; 
\left\{\begin{array}{l}
         0.568 \\
              0.484      \end{array} \right. \; , \;
 g_d = g^\frac34 \; Z^\frac14 \; \left\{\begin{array}{l}
         155~~~\mathrm{Fermions} \\
              180~~~\mathrm{Bosons}      \end{array} \right. \; . 
$$
Since $ g = 1-4 $, we see that 
$ g_d \gtrsim 100 \Rightarrow  T_d \gtrsim 100 $ GeV.
Moreover, $ 1 < Z^\frac14 < 10 $ for $ 1 < Z < 10000 $.
For example for DM Majorana fermions $ (g=2) \; m \simeq 0.85 $ keV.

\medskip

We get results for $ m $ and $ g_d $ on the same scales for DM particles 
decoupling UR out of thermal equilibrium [1]. For a specific model of 
sterile neutrinos where decoupling is out of thermal equilibrium:
$$
0.56 \; \mathrm{keV} \lesssim m_{\nu} \;  
Z^{-\frac14} \lesssim 1.0 \; \mathrm{keV}
\quad ,  \quad 15 \lesssim g_d  \;  Z^{-\frac14}\lesssim 84
$$
For relics decoupling non-relativistic
we obtain similar results for the DM particle mass: keV 
$ \lesssim m \lesssim $ MeV [1].

\medskip

Notice that the dark matter particle mass $ m $ and decoupling temperature 
$ T_d $ are {\bf mildly} affected by the uncertainty in the factor $ Z $ through a 
power factor 
$ 1/4 $ of this uncertainty, namely, by a factor $ 10^{\frac14} \simeq 1.8 $

\medskip

The comoving free-streaming) wavelength, and the Jeans' mass are obtained in the range
$$
\frac{0.76}{\sqrt{1+z}} \; {\rm kpc} <\lambda_{fs}(z) <
\frac{16.3}{\sqrt{1+z}} \; {\rm kpc} \; , \; 0.45 \; 10^3 \; M_{\odot} 
< \frac{M_J(z)}{(1+z)^{+\frac32}} < 0.45 \; 10^7  \; 
\; M_{\odot} \; .
$$

These values at $ z = 0 $ are consistent with the $N$-body simulations 
and are of the order of the small dark matter structures observed today .
By the beginning of the matter dominated era $ z \sim 3200 $, the masses are of the 
order of galactic masses $ \sim 10^{12} \; M_{\odot} $ and the comoving free-streaming 
wavelength scale turns to be of the order of the galaxy sizes today 
$ \sim 100 \; {\rm kpc}$.

\medskip

 Lower and upper bounds for the dark matter annihilation cross-section $ \sigma_0 $ 
are derived: $ \sigma_0 > (0.239-0.956) \; 10^{-9} \; \mathrm{GeV}^{-2} $ and 
$ \sigma_0 < 3200 \; m \; \mathrm{GeV}^{-3} \; . $ There is at least five orders of 
magnitude between them, the dark matter non-gravitational self-interaction is 
therefore negligible (consistent with structure formation and observations, as 
well as by comparing X-ray, optical and lensing observations of the merging of 
galaxy clusters with $N$-body simulations).

\medskip

Typical `wimps' (weakly interacting massive particles) with mass $ m = 100 $ GeV 
and $ T_d = 5 $ GeV  would require a huge $ Z \sim 10^{23} $, well above
the upper bounds obtained and cannot reproduce the observed galaxy properties. 
They produce an extremely short free-streaming or Jeans length $ \lambda_{fs} $ today $ 
\lambda_{fs}(0) \sim 3.51 \; 10^{-4} \; {\rm pc} = 72.4  \; {\rm AU} \; $ that would
correspond to unobserved structures much smaller than the galaxy structure.
Wimps result are strongly disfavoured. 

\medskip

Galaxies are described by a variety of physical quantities:

\medskip

(a) {\bf Non-universal} quantities: mass, size, luminosity, fraction of DM,
DM core radius $ r_0 $, central DM density $ \rho_0 $.

(b) {\bf Universal} quantities: surface density $ \mu_0 \equiv r_0 \; \rho_0 $
and DM density profiles. $M_{BH}/M_{halo}$ (or halo binding energy).
The galaxy variables are related by
{\bf universal} empirical relations. Only one variable remains free. 
That is, the galaxies are a one parameter family of objects.
The existence of such universal quantities may be explained by the
presence of attractors in the dynamical evolution. 
The quantities linked to the attractor always reach the same value
for a large variety of initial conditions. This is analogous
to the universal quantities linked to fixed points in critical
phenomena of phase transitions.The universal DM density profile in Galaxies has the scaling property: 
$$
\rho(r) = \rho_0 \; F\left(\displaystyle\frac{r}{r_0}\right) \quad , \quad  
F(0) = 1 \quad , \quad x \equiv  \displaystyle\frac{r}{r_0} \; ,
$$
where $ r_0 $ is the DM core radius.
As empirical form of cored profiles one can take Burkert's form for $ F(x) $.
Cored profiles {\bf do reproduce} the astronomical observations.

\medskip

The surface density for dark matter (DM) halos 
and for luminous matter galaxies is defined as: 
$ \mu_{0 D} \equiv r_0 \; \rho_0, $ 
$ r_0 = $ halo core radius, $ \rho_0 = $ central density for DM galaxies.
For luminous galaxies  $ \rho_0 = \rho(r_0) $
(Donato et al. 09, Gentile et al. 09).
Observations show an Universal value for $ \mu_{0  D} $: independent of 
the galaxy luminosity for a large number of galactic systems 
(spirals, dwarf irregular and spheroidals, elliptics) 
spanning over $14$ magnitudes in luminosity and of different Hubble types.
Observed values:
$$
\mu_{0  D} \simeq 120 \; \frac{M_{\odot}}{{\rm pc}^2} = 
5500 \; ({\rm MeV})^3 = (17.6 \; {\rm Mev})^3 \quad , \quad
5 {\rm kpc}  < r_0 <  100 {\rm kpc} \; .
$$
Similar values $ \mu_{0  D} \simeq 80  \; \frac{M_{\odot}}{{\rm pc}^2} $ are observed in 
interstellar molecular clouds of size $ r_0 $ of different type and composition over 
scales $ 0.001 \, {\rm pc} < r_0 < 100 $ pc (Larson laws, 1981).
Notice that the surface gravity acceleration is 
given by $\mu_{0  D}$ times Newton's constant.

\bigskip

We combine in refs.[3-4] the theoretical evolution of density fluctuations 
computed from first principles since decoupling till today to the observed properties of
galaxies as the surface density and core radius. We obtain that 
(i) the dark matter particle mass must be in the keV scale both for in and out
of thermal equilibrium decoupling (ii) the density profiles are cored for 
keV scale DM particles and cusped for GeV scale  DM particles (wimps). 

\bigskip

{\bf References}

\begin{description}

\item[1]  H. J. de Vega, N. G. Sanchez,  arXiv:0901.0922, 
Mon. Not. R. Astron. Soc. 404, 885 (2010).

\item[2] D. Boyanovsky, H. J. de Vega, N. G. Sanchez, 	
arXiv:0710.5180, Phys. Rev. {\bf D 77}, 043518 (2008).

\item[3] H. J. de Vega, N. G. Sanchez, Int. J. Mod. Phys. A26: 1057 (2011),
arXiv:0907.0006.

\item[4] H. J. de Vega, P. Salucci, N. G. Sanchez, arXiv:1004.1908.

\end{description}

\newpage

\subsection{Stefano Ettori}

\newcommand{\chandra}{{\it Chandra}}
\newcommand{\xmm}{{\it XMM-Newton}}
\newcommand{\rhog}{${\rho_g}$}

\vskip -0.3cm

\begin{center}

INAF, Osservatorio Astronomico di Bologna, via Ranzani 1, I-40127 Bologna, Italy
\\ INFN, Sezione di Bologna, viale Berti Pichat 6/2, I-40127 Bologna, Italy

\bigskip

{\bf
Mass profiles and $c-M_{\rm DM}$ relation
in X-ray luminous galaxy clusters 
}

\end{center}

\medskip

The distribution of the total and baryonic mass in galaxy clusters
is a fundamental ingredient to validate the scenario of structure
formation in a Cold Dark Matter (CDM) Universe. Within this scenario,
the massive virialized objects are powerful cosmological tools
able to constrain the fundamental parameters of a given CDM model.
The $N-$body simulations of structure formation in CDM models indicate
that dark matter halos aggregate with a typical mass density profile
characterized by only 2 parameters,
the concentration $c$ and the scale radius $r_{\rm s}$ (e.g. [8],
hereafter NFW).
The product of these two quantities fixes the radius within which the mean cluster
density is 200 times the critical value at the cluster's redshift
[i.e. $R_{200} = c_{200} \times r_{\rm s}$ and the cluster's volume
$V = 4/3 \pi R_{200}^3$ is equal to $M_{200} / (200 \rho_{c,z})$,
where $M_{200}$ is the cluster gravitating mass within $R_{200}$].
With this prescription, the structural properties of DM halos from galaxies to
galaxy clusters are dependent on the halo mass, with systems at higher masses
less concentrated. Moreover, the concentration depends upon the
assembly redshift (e.g. [1]),
which happens to be later in cosmologies with lower matter density,
$\Omega_{\rm m}$, and lower normalization of the linear power spectrum
on scale of $8 h^{-1}$ Mpc, $\sigma_8$, implying less concentrated DM halos
of given mass.
The concentration -- mass relation, and its evolution in redshift, is therefore
a strong prediction obtained from CDM simulations of structure formation
and is quite sensitive to the assumed cosmological parameters
(e.g. NFW; [1]; [5]).

{\it Recent X-ray studies (see reference in [2], [6], [9])
have shown good agreement between observational constraints at low
redshift and theoretical expectations from Cold Darm Matter 
over a wide range of halo mass.
}
For instance, it is confirmed at high significance that
the concentration decreases with increasing mass, requiring
a $\sigma_8$, the dispersion of the mass
fluctuation within spheres of comoving radius of 8 $h^{-1}$ Mpc,
in the range $0.76-1.07$ (99\% confidence) definitely in contrast to
the lower constraints obtained, for instance, from the analysis
of the {\it WMAP} 3 years data (see [2]).
However, a possible tension between the observational
constraints and the numerical predictions is also observed, 
in the sense that either the relation
is steeper than previously expected or some redshift evolution has to
be considered ([9]).
Moreover, there is evidence that strong lensing measurements of the concentration
are systematically larger than the ones
estimated in the X-ray band, and 55 per cent higher, on average, than
the rest of the cluster population ([3]).

In [6], we extend the spectral analysis presented in
[7] for a sample of 44 X-ray luminous galaxy clusters
located in the redshift range $0.1 - 0.3$ with a detailed spatial analysis
of the surface brightness profiles with the aim to (1) recover
their total and gas mass profiles, (2) constraining the cosmological parameters
$\sigma_8$ and $\Omega_{\rm m}$ through the analysis of the measured distribution
of $c_{200}$, $M_{200}$ and baryonic mass fraction in the mass
range above $10^{14} M_{\odot}$.
We note that this is the statistically largest sample for which this study
has been carried on up to now between $z=0.1$ and $z=0.3$.

We use the profiles of the spectroscopically determined ICM temperature and
of the PSF--corrected surface brightness estimated to recover the X-ray gas,
the dark and the total mass profiles,
under the assumptions of the spherical geometry distribution
of the Intracluster Medium (ICM) and that the hydrostatic equilibrium
holds between ICM and the underlying gravitational potential.

Our dataset is able to resolve the temperature profiles up to about $0.6-0.8 R_{500}$
and the gas density profile, obtained from the geometrical deprojection
of the PSF--deconvolved surface brightness, up to a median radius of $0.9 R_{500}$.
Beyond this radial end, our estimates are the results of an
extrapolation obtained by imposing a NFW profile for the total mass
and different functional forms for $M_{\rm gas}$.

We estimate, with a relative statistical uncertainty of $15-25\%$,
the concentration $c_{200}$ and the mass $M_{200}$ of the dark matter
(i.e. total$-$gas mass) halo.
We constrain the $c_{200}-M_{200}$ relation to have a normalization
$c_{15} = c_{200} \times (1+z) \times \left( M_{200}/10^{15} M_{\odot} \right)^{-B}$
of about $2.9-4.2$ and a slope $B$ between $-0.3$ and $-0.7$ (depending on the
methods used to recover the cluster parameters and to fit the linear
correlation in the logarithmic space),
with a relative error of about 5\% and 15\%, respectively.

We put constraints on the cosmological parameters ($\sigma_8, \Omega_{\rm m}$)
by using the measurements of $c_{200}$ and $M_{200}$ and by comparing
the estimated values with the predictions tuned from numerical simulations of CDM universes.
In doing that, we propagate the statistical errors (with a relative value
of about $15-25$\% at $1 \sigma$ level) and consider the systematic uncertainties
present both in the simulated datasets ($\sim 20$\%) and in our measurements.

When the subsample of 11 LEC clusters, that are expected to be more relaxed
and with a well-formed central cooling core, is considered, we measure
$\gamma = 0.56 \pm 0.04$, $\Gamma=0.39\pm0.02$
$\sigma_8 = 0.83 \pm 0.1$ and $\Omega_{\rm m}=0.26 \pm 0.02$
(at $2\sigma$ level).

We show that the study
of the distribution of the measurements in the $c-M_{\rm DM}-f_{\rm gas}$
plane provides a valid technique already mature and competitive
in the present era of precision cosmology.

However, we highlight the net dependence of our results on the models
adopted to relate the properties of a DM halo to the background cosmology.
In this context, we urge the $N-$body community to generate
cosmological simulations over a large box to properly predict the expected concentration
associated to the massive ($>10^{14} M_{\odot}$) DM halos as function of
$\sigma_8$, $\Omega_{\rm m}$ and redshift.
The detailed analysis of the outputs of these datasets will provide
the needed calibration to make this technique more reliable and robust.

\begin{figure*}
\hbox{
 \epsfig{figure=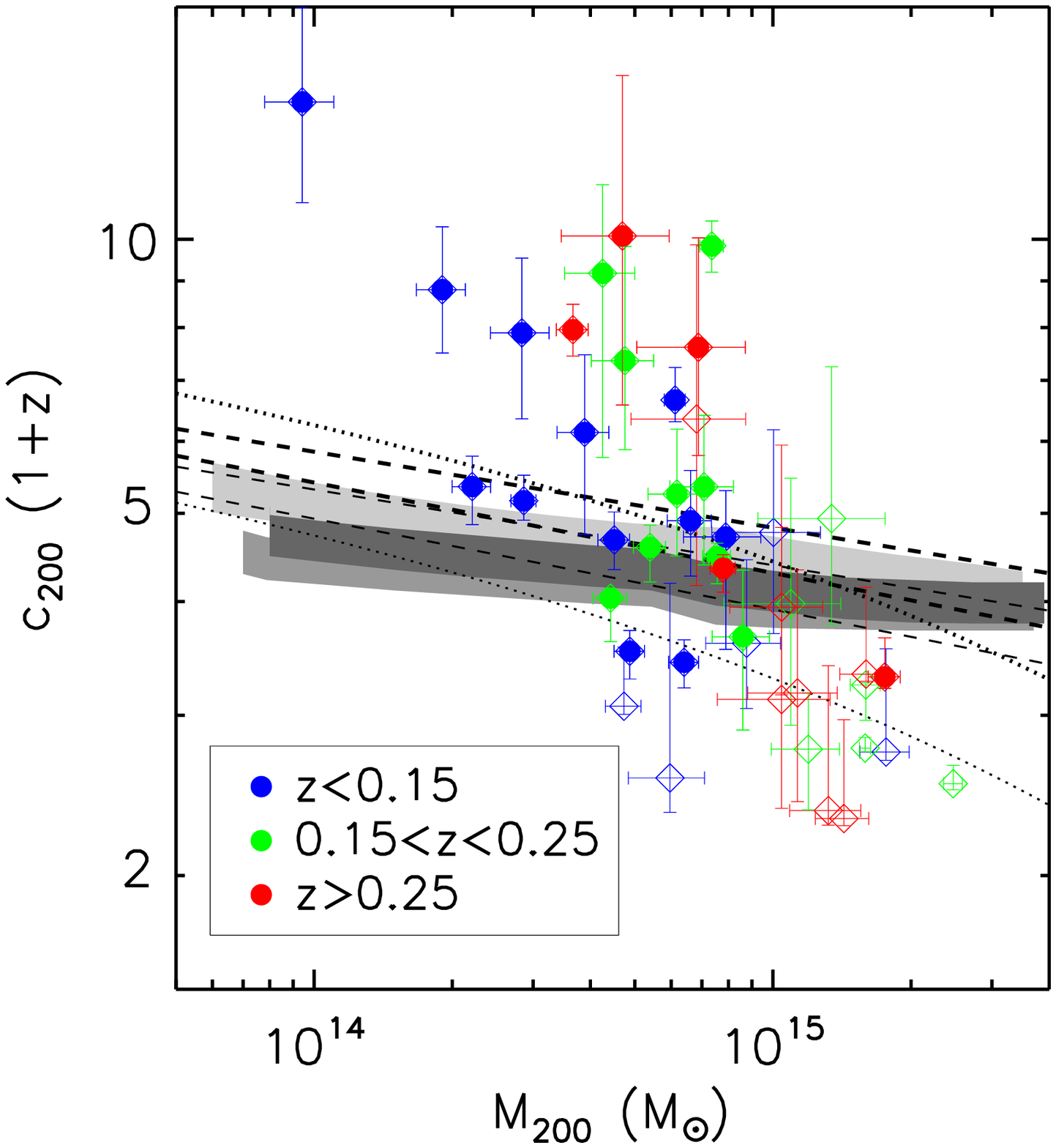,width=0.5\textwidth}
 \epsfig{figure=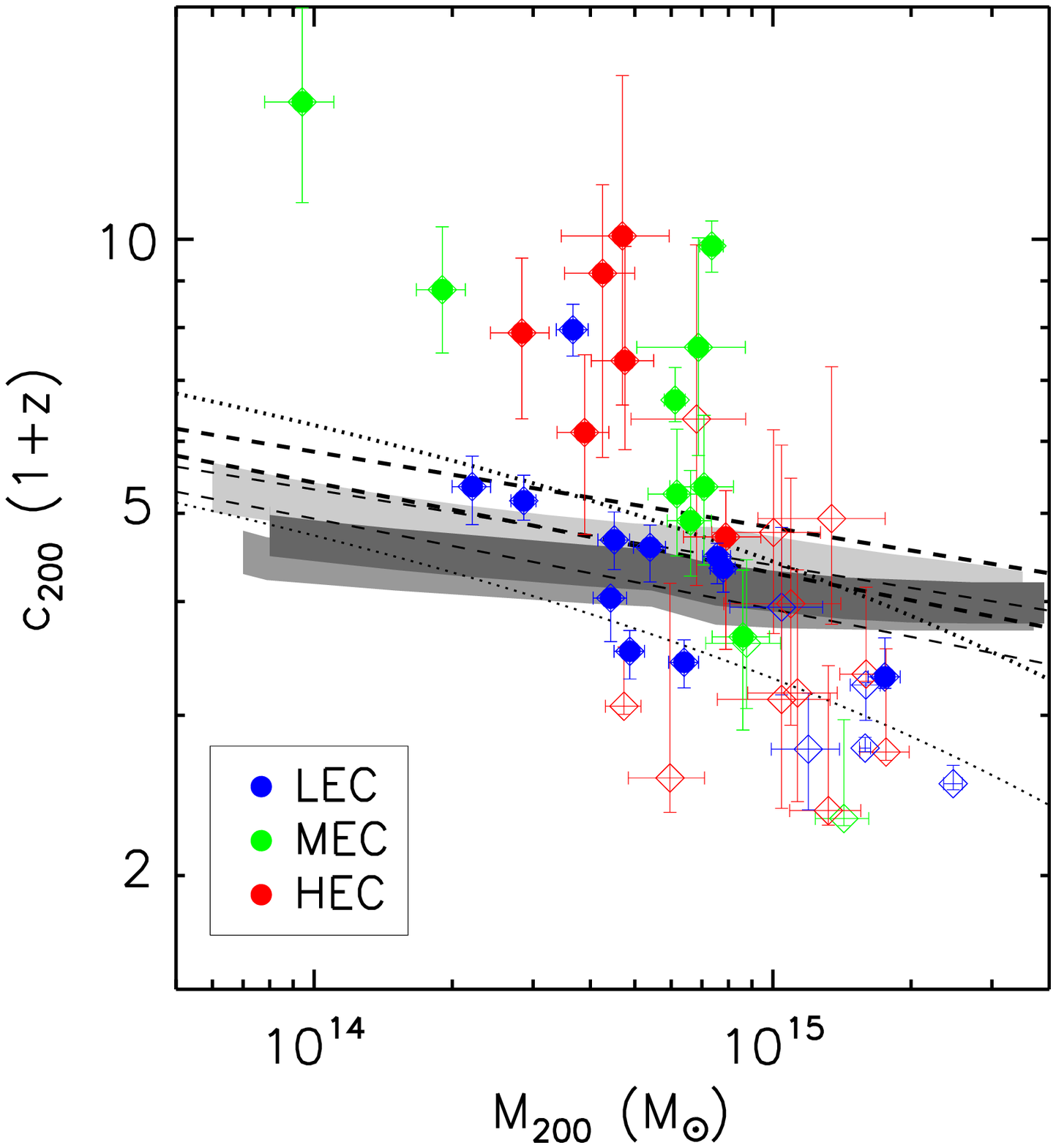,width=0.5\textwidth}
}
\caption{Data in the plane $(c_{200}, M_{200})$ used to constrain the cosmological
parameters $(\Omega_{\rm m}, \sigma_8)$.
The dotted lines show the predicted relations from [1]
for a given $\Lambda$CDM cosmological model at $z=0$ (from top to bottom:
$\sigma_8=0.9$ and $\sigma_8=0.7$).
The shaded regions show the predictions in the redshift range $0.1-0.3$
for an assumed cosmological model in agreement with
WMAP-1, 5 and 3 years (from the top to the bottom, respectively) from [5].
The dashed lines indicate the best-fit range at $1\sigma$ obtained for relaxed halos in a
WMAP-5 years cosmology from [4; thin lines: $z=0.1$, thick lines: $z=0.3$].
({\it Left})
The color code indicates the objects at $z<0.15$ (blue), in the range
$0.15<z<0.25$ (green) and at $z>0.25$ (red).
({\it Right})
Distribution of Low (LEC), Medium (MEC), High (HEC) Entropy Core
systems.
Filled diamonds indicate the data where a more robust 
(i.e. with well defined and constrained free parameters) 
mass reconstruction is achievable.
}
\label{fig:cm}
\end{figure*}

\bigskip

{\bf References}

\begin{description}

\item[1] Bullock J.~S. et al., 2001, MNRAS, 321, 559; 
Macci\`o A.V. et al., 2008, MNRAS, 391, 1940
\vspace{-0.3cm}
\item[2] Buote D.A. et al., 2007, ApJ, 664, 123
\vspace{-0.3cm}
\item[3] Comerford J.M., Natarajan P., 2007, MNRAS, 379, 190
\vspace{-0.3cm}
\item[4] Duffy A.R. et al., 2008, MNRAS, 390, L64
\vspace{-0.3cm}
\item[5] Eke V.~R., Navarro J.~F., Steinmetz M., 2001, ApJ, 554, 114
\vspace{-0.3cm}
\item[6] Ettori S. et al., 2010, A\&A, 524, 68
\vspace{-0.3cm}
\item[7] Leccardi A., Molendi S., 2008, A\&A, 486, 359
\vspace{-0.3cm}
\item[8] Navarro J.F., Frenk C.S., White S.D.M., 1997, ApJ, 490, 493
\vspace{-0.3cm}
\item[9] Schmidt R.W., Allen S.W., 2007, MNRAS, 379, 209

\end{description}

\newpage

\subsection{Frederic V. Hessman, Monika Ziebart}
\vskip -0.5cm
\begin{center}
Institut f\"ur Astrophysik, Georg-August-Universit\"at, G\"ottingen, Germany

\bigskip

{\bf The ``Bosma Effect'' Revisited - Correlations between the ISM and DM in Galaxies}

\end{center}

\medskip

The problem of explaining motions within galaxies -- the internal stellar dynamics of the Sun's neighborhood or within an elliptical galaxy and the rotation curves of spiral galaxies -- was one of the original reasons for the creation of the cold Dark Matter paradigm, since it is on the scales of galaxies that the masses of the putative DM particles are most relevant.
With the creation of the $\Lambda CDM$ paradigm, much of the emphasis has shifted onto the larger scales of galaxy clusters and cosmological large-scale structure, but DM in galaxies is one of the critical links in the Borromean chain of the Concordance Cosmological Model (CCM) of the Universe: astronomical DM; Dark Energy and inflationary cosmology; and the standard model of particle physics.

Despite the astounding performance of the CCM, there are many signs that the simple picture of CDM in galaxies is not working.
The most troubling signs of the failure of the CDM paradigm have to do with the tight coupling between baryonic matter and the dynamical signatures of DM in galaxies, e.g. the Tully-Fisher relation [1], the stellar disc-halo conspiracy [2], the maximaum disc phenomenon [3], the MOdified Newtonian Dynamics (MOND) phenomenon [4], the baryonic Tully-Fisher relation [5], the baryonic mass discrepancy-acceleration relation [6], the 1-parameter dimensionality of galaxies [7], and the presence of both a DM and a baryonic mean surface density [8][9].
The strangest of these relations is the ``Bosma effect'' [10][11]: the centripetal contribution of the dynamically insigificant interstellar medium (ISM) in spiral galaxies is directly proportional to that of the dominant DM.
The constant of proportionality has been determined for about 100 galaxies, with dwarf galaxies showing a smaller and late-type spirals showing a larger factor.

Hoekstra, van Albada \& Sancisi [12] set out to test the Bosma effect, showed that it indeed allowed a very detailed fit to the rotation curves of many well-studied galaxies, but concluded that it was not real.
Reviewing their arguments, it is clear that their negative judgement was very conservative - by the normal standards of rotation curves, the results were, in fact, very convincing.
Since their fits were performed by hand and not compared with the corresponding CDM model fits, it was not possible to make any formal conclusions and certainly not possible to reject the effect as non-physical.

Using the much better data made available by the {\it Spitzer Infrared Nearby Galaxy Survey} [13], {\it The HI Nearby Galaxy Survey} [14], and the analyses by de Block et al. [15], we have tested the Bosma effect and compared the results against standard CDM models.
The use of infrared photometry and colours permits formal fits to the stellar components nearly independent of extinction corrections and with reasonably reliable mass-to-light estimates.
In addition to the standard bulge, stellar disk, and visible HI components, we fitted the rotation curves with the addition of either one or two Bosma components, using either the HI disc (so-called ``simple Bosma'' models) or both the stellar and the HI discs (``classic Bosma'' models) as proxies, where the stellar disc is used as a proxy for the molecular gas obviously present in regions of previous and current star formation.  
For comparison, we also fit the data with self-consitent NFW models, where the compactness of the halo is a function of the halo mass [16] or with the Burkert halo mass profiles used in the ``Universal Rotation Curve'' model [17].

The ``simple'' Bosma models using only the HI as a proxy are remarkably good in the outer discs, as shown by Bosma, independent of the shape of the rotation curve.
However, the inner disks are not well-fit by pure ``HI-scaling'': the saturation of the HI surface densities above levels around $10\,M_\odot/pc^2$ results in centripetal contributions which are clearly too small.
This problem is not present in ``classic'' Bosma models, since the saturation of the HI profiles occurs exactly where the stellar component starts to dominate, permitting a perfect compensation.

The self-consistent NFW models are clearly much worse than the Bosma models.
Not unexpectedly, the fully unconstrained URC models perform best.
The median ratio of the fitted $\chi^2$ values between the ``classical'' Bosma and URC models is only about 1.5 compared with a value of 3 between the NFW and URC fits.
Thus, the ``classical'' Bosma models are only marginally worse than the best-fitting DM models.
Since the Bosma model works so well in galaxies covering a very wide range of masses, sizes, stellar contributions, and rotation curve shapes, one must conclude that the Bosma effect is, in fact, real.

At the time Bosma initially reported this effect, it was reasonable to assume that the DM might be in the discs, but already at the time of Hoektra, van Albada \& Sancisi's attempted test, the paradigm had shifted to a quasi-spherical DM halo.
However, the centripetal pattern of a spherical halo is dramatically different from that of a disc: in a disc, the matter at all radii contribute to the local centripetal effect, whereas in a spherical halo only the inner mass is felt.
This means that a physical interpretation of the Bosma effect withing the classic DM paradigm is highly unlikely: how can a dominant spherical halo always teach a minor disc mass component to behave as if it were a spherical halo, fully independently of the properties of both.  This is either a bizzare conspiracy or a potent argument against the DM halo paradigm.

If the Bosma effect is interpreted as truly being due to disc DM (presumedly baryonic, since scalable from the ISM), the amount of additional baryonic mass is quite significant: this implies that only 10-70\% of the baryons are visible as stars and gas.
Within the CCM, this is not a fundamental problem, since the standard paradigm has a ``missing baryon'' problem anyway [18].
The reason that a relatively modest amount of disc DM can replace a large amount of CDM is very simple: a disc provides much more centripetal effect per unit mass than a highly extended spherical halo.

The disc DM solution has the power to explain a wide range of otherwise inexplicable phenomena, e.g. the utility of ``maximal discs'' (using the stellar disc as a proxy for disc DM is nearly paramount to fitting ``maximal discs'', particularly since the flat HI profiles often produce a negative centripetal rotation curve contribution at small radii), the extended baryonic Tully-Fisher relation, the relative value of the mean URC/DM and baryonic surface densities, 
the MOND phenomenon, the mass discrepancy-acceleration relation.
Since cold gas in the ISM is very hard to detect -- a situation which is being confirmed by new far infrared data taken by the Herschel satellite -- the $H_2$ ``clumpuscule'' model of Pfenniger, Combes \& Martinet [19] provides a natural explanation for the Bosma effect.
The main argument against disc DM is that such a massive disc must be unstable: this argument must be taken very seriously, but the standard arguments are not nearly as tight as is normally assumed: this will be one of the topics in a future paper [20].

The Bosma effect does not show that there is no DM, since it only probes the gravitational potentials and sources within galaxies.
For instance, large total masses are derived for the Milky Way and M31 using the kinematics of satellite galaxies [21], although it is difficult to obtain tight constraints on the true total masses due to a variety of effects [20].
Within the Bosma effect paradigm, these large masses are most easily explained using a DM component whose core radii are as large or larger than the visible galaxies: DM halos with these extents have practically no effect on the dynamics on scales of galaxies and only show up on the scales of groups (e.g. the Local Group), clusters of galaxies, and cosmologically.
The presentation of Moni Bidin at this conference (see [21]) fits very well with this model: there is no sign of DM at the radius of the Sun.

The topic of this workshop -- Warm to rather Hot DM (depending upon where one chooses to set the mass scale) -- provides the most natural explanation for phenomena outside of the scale of visible galaxies while providing a potential reason for the pure dominance of baryons on the scales of visible galaxies.
While the streaming scale of bare $10\,keV$ neutrinos (ca. $10\,kpc$) is still too small, it is conceivable that processes during the formation of galaxies resulted in a transfer of angular momentum from the dissipative baryons to the W/HDM.

\bigskip

{\it We would like to thank the THINGS consortium, particularly E. de Blok, for making their data and rotation curve analyses available.}

\bigskip

{\bf References}

\begin{description}
\item[1] Tully. R. B., Fisher, J. R., 1977, A\&A 54, 66
\vspace{-0.3cm}
\item[2] Bahcall, J. N., Caseterno, S., 1985, ApJ 293, 7
\vspace{-0.3cm}
\item[3] van Albada, T. S., Sancisi, R., 1986, RSPTA 320, 447
\vspace{-0.3cm}
\item[4] Sanders, R. H., McGaugh, S. S., ARA\&A 40, 263
\vspace{-0.3cm}
\item[5] McGaugh, 2000,
\vspace{-0.3cm}
\item[6] McGaugh, 2004,
\vspace{-0.3cm}
\item[7] Disney, M. J., et al., 2008, Nature 455, 1082
\vspace{-0.3cm}
\item[8] Donato, M. J., et al., 2009, MNRAS 397, 1169
\vspace{-0.3cm}
\item[9] Gentile, G., Famaey, B., Zhao, H., Salucci, P., 2009, Nature 461, 627
\vspace{-0.3cm}
\item[10] Bosma, A., 1978, disseration, Rijksuniversiteit Groningen
\vspace{-0.3cm}
\item[11] Bosma, A., 1981, AJ 86, 1825
\vspace{-0.3cm}
\item[12] Hoektra H., van Albadaa, T. S., Sancisi, R., 2001, MRAS 323, 453
\vspace{-0.3cm}
\item[13] Kennecutt Jr., R. C., et al., 2003, PASP 115, 928
\vspace{-0.3cm}
\item[14] Walter, F. et al., 2008, AJ 136, 2563
\vspace{-0.3cm}
\item[15] de Blok, W. J. G., et al., 2008, AJ 136, 2648 
\vspace{-0.3cm}
\item[16] Maccio, A. V., Dutton, A. A., van den Bosch, F. C., 2008, MNRAS 391, 1940
\vspace{-0.3cm}
\item[17] Salucci, P., et al., 2007, MNRAS 378, 415
\vspace{-0.3cm}
\item[18] Bregman, J., N., 2007, ARA\&A 45, 221
\vspace{-0.3cm}
\item[19] Pfenniger, D., Combes, F., Martinet, L., 1994, A\&A 285, 79 
\vspace{-0.3cm}
\item[20] Hessman, F. V., 2011, in preparation (Paper III)
\vspace{-0.3cm}
\item[21] Watkins, L. L., Evans, N. W., An, J. H., 2010, MNRAS 406, 264
\vspace{-0.3cm}
\item[22] Moni Bidin, C., Carraro, G., M\'endez, R. A., van Altena, W. F., 2010, ApJL 724, 122
\end{description}

\newpage

\newcommand{\apjl}[0]{The Astrophysical Journal}
\newcommand{\aj}[0]{The Astronomical Journal}
\newcommand{\aap}[0]{Astronomy {\&} Astrophysics}
\newcommand{\aaps}[0]{Astronomy {\&} Astrophysics Supplement Series}
\newcommand{\apss}[0]{Astrophysics and Space Science}
\newcommand{\apjs}[0]{The Astrophysical Journal Supplement Series}
\newcommand{\mnras}[0]{Monthly Notices of the Royal Astronomical Society}
\newcommand{\pasp}[0]{Publications of the Astronomical Society of the Pacific}
\newcommand{\araa}[0]{Annual Review of Astronomy and Astrophysics}

\subsection{Ayuki Kamada}

\vskip -0.3cm

\begin{center}

%
Institute for the Physics and Mathematics of the Universe, TODIAS,\\
The University of Tokyo, 5-1-5 Kashiwanoha, Kashiwa, Chiba 277-8583, Japan

\bigskip


{\bf Light sterile neutrino as warm dark matter and the structure of galactic dark halos} 

\end{center}

\medskip

%
%
%
%
%
%
{\bf Abstract:} We study the formation of nonlinear structure in $\Lambda$ Warm Dark Matter (WDM) 
cosmology using large cosmological N-body simulations.
We assume that dark matter consists of sterile neutrinos that are generated 
through nonthermal decay of singlet Higgs bosons 
near the Electro-Weak energy scale.
Unlike conventional thermal relics, the nonthermal WDM has a peculiar velocity 
distribution, which results in a characteristic shape of the matter
power spectrum.
We perform large cosmological N-body simulations for the nonthermal 
WDM model. We compare the radial distribution of subhalos 
in a Milky Way size halo 
with those in a conventional thermal WDM model.
The nonthermal WDM with mass of 1 keV predicts
the radial distribution of the subhalos that is
remarkably similar to the observed distribution
of Milky Way satellites.

\medskip

{\bf Introduction}


Alternative models to the standard $\Lambda$ Cold Dark Matter model have been suggested as a solution 
of the so-called 'Small Scale Crisis'.
One of them is $\Lambda$ Warm Dark Matter cosmology, in which Dark Matter particles had non-zero velocity dispersions. 
The non-zero velocities smooth out primordial density perturbations below its free-streaming length of sub-galactic sizes. 
The formation of subgalactic structure is then suppressed. Moreover, The large phase space density may prevent dark matter 
from concentrating into galactic center. Particle physics models provide promising WDM candidates such as gravitinos, 
sterile neutrinos and so on. 
Gravitinos with mass of $\sim$keV in generic Supergravity theory are produced in thermal bath immediately 
after reheating and are then decoupled kinematically
from thermal bath similarly to the Standard Model (SM) neutrinos 
because gravitinos interact with SM particles only through gravity. 
The thermal relics have a Fermi-Dirac (FD) momentum distribution. 
Sterile neutrinos are initially proposed in the see-saw mechanism to explain the masses of the SM neutrinos. 
If the mass of sterile neutrinos is in a range of $\sim$keV 
and their Yukawa coupling is of order $\sim 10^{-10}$, then 
it cannot be in equilibrium with SM particles 
throughout the thermal history of the universe. 
There are several peculiar production mechanisms for sterile neutrinos,
such as Dodelson-Widrow (DW) mechanism~[1] and EW scale Boson Decay (BD)~[2]. 
In DW mechanism, sterile neutrinos 
are produced through oscillations of active neutrinos, and its velocity distribution has a Fermi-Dirac form 
just like gravitinos.
In the BD case, sterile neutrinos are produced via decay 
of singlet Higgs bosons and they have generally a nonthermal velocity distribution. 
The nonthermal velocity distribution imprints particular features in the transfer function of the density fluctuation 
power spectrum~[3]. The transfer function has a cut-off at the corresponding free-streaming length,
but it decreases somewhat slowly than thermal WDM models.
In this article, we study the formation of nonlinear structure for a cosmological model with nonthermal sterile neutrino WDM.
We perform large cosmological N-body simulations. 
There are several models of nonthermal WDM. For example, gravitinos can be produced via decay of inflaton~[4] 
or long-lived Next Lightest Supersymmetric Particle (NLSP)~[5]. 
Our result can be generally applied to the formation of nonlinear structure in these models as well. 



\medskip

{\bf Nonthermal sterile neutrino}

The clustering properties of the above BD sterile neutrino model is investigated by Boyanovsky~[3],
who solved the linearized Boltzmann-Vlasov equation in the matter dominant era when WDM has already 
become non-relativistic. 
Firstly, solving the Boltzmann equation for sterile neutrinos produced by the singlet boson decay process, 
we find the most of contribution to the present number of sterile neutrinos comes when the temperature 
decreases to the EW scale. We then get the velocity distribution of sterile neutrinos. 
The BD distribution has a distinguishable feature from usual Fermi-Dirac distributions at small velocities, $y=P/T(t)$:
\begin{align}
f_{\rm BD}(y)\propto \frac{1}{y^{\frac{1}{2}}}, \ \ \ \ \ f_{\rm FD}(y)\propto 1. \notag
\end{align}
The velocity distribution is imprinted in the power spectrum, which decreases slowly across the free-streaming length scale. 
Following Boyanovsky~[3], we solve the linearlized Boltzmann-Vlasov equation. We define the comoving free-streaming wavenumber as
$$
 k_{\rm fs}={\Big[}\frac{3H_{0}^2\Omega_{M}}{2\langle\vec{V}^2\rangle(t_{\rm eq})}{\Big]}^{\frac{1}{2}} \notag
$$
akin to the Jeans scale at the matter-radiation equality.
In the BD case, the present energy density of dark matter is determined
by the Yukawa coupling.
Assuming the relativistic degree of freedom $\bar{g}$ is the usual SM value 
$\bar{g}\simeq 100$ for the $m=1$ keV sterile neutrino dark matter, we find
\begin{align}
k_{\rm fs}^{\rm BD}=18 \ h \ \mathrm{Mpc^{-1}}, \notag
\end{align}
while for the $m=1$ keV gravitino dark matter, which has the FD distribution, 
we should take $\bar{g}\simeq1000$ to get the present energy density of dark matter.
Then, for $m=1$ keV gravitinos, 
\begin{align}
k_{\rm fs}^{\rm FD}=32 \ h \ \mathrm{Mpc^{-1}}. \notag
\end{align}
Although these two models have almost the same free-streaming length, 
their linear power spectra show appreciable differences, as seen in 
Fig. \ref{ak1}. There, we adopt the numerically fit transfer function 
in Bode et al.~[6] for the linear power spectrum of the gravitino 
dark matter.
 
The enhancement of the velocity distribution in the low velocity region leads 
to the slower decrease of the linear power spectrum. This implies that we should care not only free-streaming 
scale or velocity dispersion, but also the shape of the velocity distribution in studying the formation 
of the nonlinear object below the cut-off (free-streaming) scale.

Using these linear power spectrum, we have performed direct numerical simulations to study observational 
signatures of the imprinted velocity distribution in the subgalactic structures. 
We start our simulations from a redshift of $z=9$.
We use $N = 256^{3}$ particles in a comoving volume of $10 \ h^{-1}$ Mpc on a side. The mass of a dark 
matter particle is $4.53\times10^{6} \ h^{-1} \ \mathrm{M_{\odot}}$ and the gravitational softening
length is $2 \ h^{-1}$ kpc. 

\medskip

{\bf Simulation Results}

Fig. \ref{ak2} shows the projected distribution of dark matter in and
 round a Milky Way size halo at $z=0$. 
The plotted region has a side of $2 \ h^{-1}$ Mpc. 
Dense regions appear bright. 
Clearly, there are less subgalactic structures for the WDM models. 
The 'colder' property of the nonthermal WDM shown in the linear 
power spectrum (see Fig. \ref{ak1}) can also be seen in the abundance 
of subhalos.


In Fig. \ref{ak3}, we compare the cumulative radial distribution of the subhalos in our 'Milky Way' halo at $z=0$ 
with the distribution of the observed Milky Way satellites~[7]. 
Interestingly, the nonthermal WDM model reproduces the radial distribution of the observed Milky Way satellites 
in the range above $\sim 40$ kpc. 
Contrastingly, the CDM model overpredicts the number of subhalos by a factor of $5$ than the 
observed Milky Way satellites. This is another manifestation of the so-called 
'Missing satellites problem'. The thermal WDM model 
surpresses subgalactic structures perhaps too much, by a factor of $2-4$ than the observation.

\medskip

{\bf Summary}

We have studied the formation of the nonlinear structures in a $\Lambda$WDM cosmology using 
large cosmological N-body simulations. We adopt the sterile neutrino dark matter produced via the decay of 
singlet Higgs bosons with a mass of EW scale. The sterile neutrinos have a nonthermal velocity distribution, 
unlike the usual Fermi-Dirac distribution. The distribution is a little skewed to low velocities. 
The corresponding linear matter power spectrum decreases
slowly across the cut-off scale compared to the thermal WDM, such as gravitino dark matter.
Both of the two models have the same mass of $1$ keV and an approximately the same cut-off (free-streaming) scale 
of a few 10 $h \ \mathrm{Mpc^{-1}}$.  We have shown that this 'colder' property of the nonthermal WDM can be seen 
in richer subgalactic structures. The nonthermal WDM model with mass of $1$ keV appears to reproduce the radial 
distribution of the observed Milky Way satellites. 

\medskip

{\bf References}

\begin{description}

\item[1]
  S.~Dodelson and L.~M.~Widrow,
  Phys.\ Rev.\ Lett.\  {\bf 72}, 17 (1994)
  [arXiv:hep-ph/9303287].

\vspace{-0.3cm}

\item[2]
  K.~Petraki and A.~Kusenko,
  Phys.\ Rev.\  D {\bf 77}, 065014 (2008)
  [arXiv:0711.4646 [hep-ph]].

\vspace{-0.3cm}

\item[3]
  D.~Boyanovsky,
  Phys.\ Rev.\  D {\bf 78}, 103505 (2008)
  [arXiv:0807.0646 [astro-ph]].

\vspace{-0.3cm}

\item[4]
  F.~Takahashi,
  Phys.\ Lett.\  B {\bf 660}, 100 (2008)
  [arXiv:0705.0579 [hep-ph]].

\vspace{-0.3cm}

\item[5]
  K.~Sigurdson and M.~Kamionkowski,
  Phys.\ Rev.\ Lett.\  {\bf 92}, 171302 (2004)
  [arXiv:astro-ph/0311486].
  
\vspace{-0.3cm}





\item[6]
  P.~Bode, J.~P.~Ostriker and N.~Turok,
  Astrophys.\ J.\  {\bf 556}, 93 (2001)
  [arXiv:astro-ph/0010389].

\vspace{-0.3cm}



  

  
 
 \item[7]
  E.~Polisensky and M.~Ricotti,
  Phys.\ Rev.\  D {\bf 83}, 043506 (2011)
  [arXiv:1004.1459 [astro-ph.CO]].



\begin{figure}[htbp]
\begin{center}
\includegraphics[height=45mm]{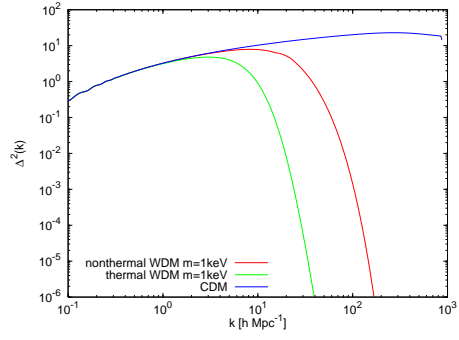}
\end{center}
\caption{The linear power spectra for the nonthermal WDM (red line), thermal WDM (green line) and CDM (blue line). 
The dark matter mass of the both WDM model is $m=1$ keV.}
\label{ak1}
\end{figure}

\begin{figure}[htbp]
\begin{minipage}{0.32\hsize}
\begin{center}
\includegraphics[width=35mm,height=35mm]{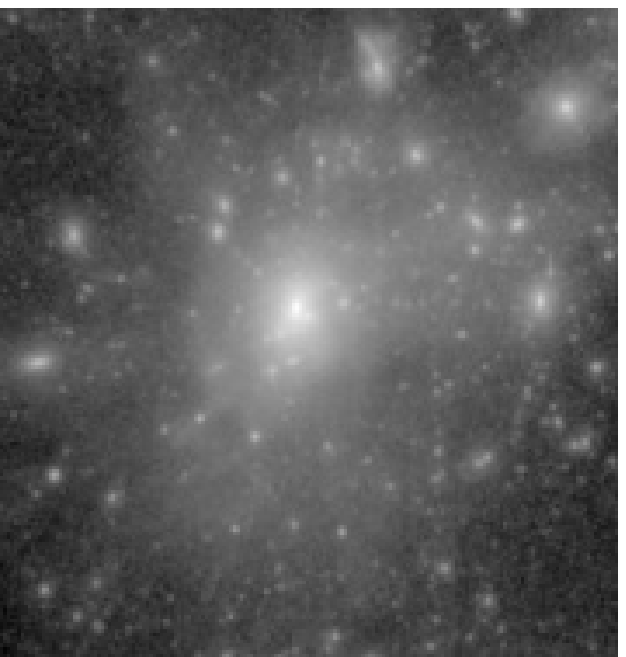}
\end{center}
\end{minipage}
\begin{minipage}{0.32\hsize}
\begin{center}
\includegraphics[width=35mm,height=35mm]{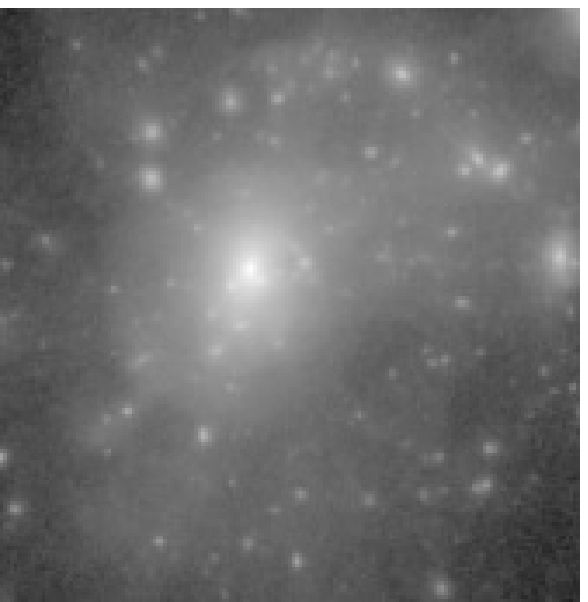}
\end{center}
\end{minipage}
\begin{minipage}{0.32\hsize}
\begin{center}
\includegraphics[width=35mm,height=35mm]{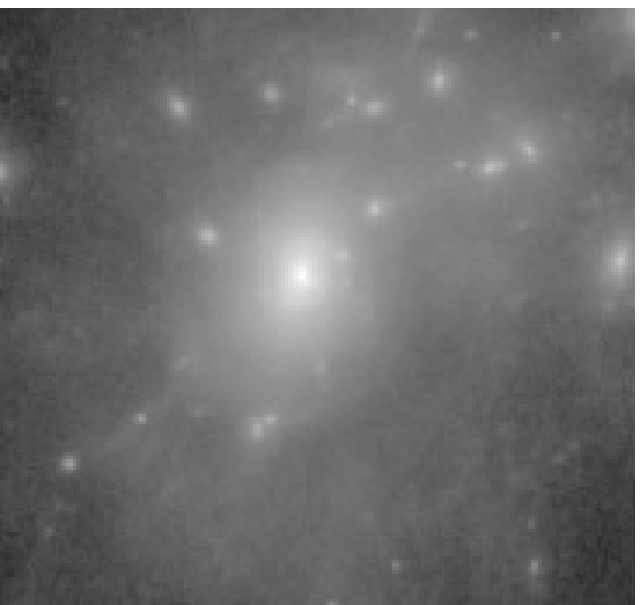}
\end{center}
\end{minipage}
\caption{The projected distributions of the substructures in our 'Milky Way' halo at $z=0$. 
The sidelength of the shown region is $2 \ h^{-1}$ Mpc. 
The results are for three dark matter models, CDM,  nonthermal WDM with 1 keV mass and thermal WDM with 1keV mass, respectively, 
from left to right.}
\label{ak2}
\end{figure}



\begin{figure}[htbp]
\begin{center}
\includegraphics[height=50mm]{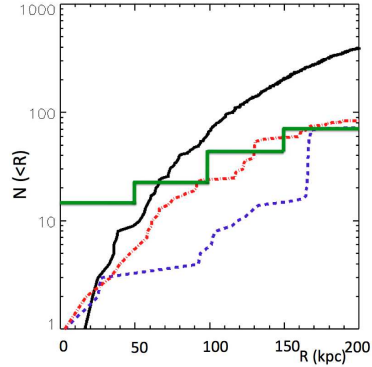}
\end{center}
\caption{The radial distribution of the subhalos in our 'Milky Way' halo at $z=0$. 
For simulation results for CDM (black), 1keV nonthermal WDM (red), and 
1keV thermal WDM (blue). Green bars show the distribution of the observed Milky Way satellites.}
\label{ak3}
\end{figure}

\end{description}

\clearpage

\subsection{Marco Lombardi, Jo\~ao Alves, and Charles Lada}

\vskip -0.3cm

\begin{center}

  M.L.: University of Milan, Department of Physics, via Celoria 16,
  I-20133
  Milan, Italy \\
  J.A.: University of Vienna, T\"urkenschanzstrasse 17, 1180 Vienna,
  Austria \\
  C.L.: Harvard-Smithsonian Center for Astrophysics, Mail Stop 72, 60
  Garden Street, Cambridge, MA 02138
  
  \bigskip

  {\bf Larson' laws and the universality of molecular cloud structures} 

\end{center}

\medskip

It has long been recognized that star formation is inextricably linked
to the molecular clouds where the process is taking place, and
therefore it is important to study the structure of these objects.
One of the first attempts in this direction has been carried out by
Larson 1981.  In his seminal work, Larson used
molecular line data available from earlier studies (mostly millimeter
data of nearly objects) and showed that molecular clouds obey three
scaling relations: (1) a power-law relationship between the length $L$
of the cloud and its velocity dispersion $\sigma_\mathrm{v}$, with
$\sigma_\mathrm{v} \propto L^{0.38}$; (2) approximate virial
equilibrium, with $2 G M / \sigma_\mathrm{v}^2 L \simeq 1$; and (3) a
relationship between the density $n$ of the cloud and its length, with
$n \propto L^{-1.1}$.  Larson's third law, which is the main focus of
this talk, implies that molecular clouds have approximately constant
column densities $\Sigma$, since $\Sigma \sim n L \propto L^{-0.1}$.

Since their formulation, Larson's laws have been the subject of
several observational and theoretical studies.  From the observational
point of view, Solomon et al. 1987 presented ${}^{12}$CO data
for a 273 nearby clouds from the University of Massachusetts-Stony
Brook (UMSB) Galactic Plane Survey Sanders et al. 1986.  They
found a size-line width relationship with a power index (0.5) steeper
than the one derived by Larson 1981.  Additionally, in
agreement with Larson's third law, they found that the molecular gas
surface density is approximately constant for all clouds with
$\Sigma(\mathrm{H}_2) = 170 \mbox{ M}_\odot \mbox{ pc}^{-2}$.
Recently, the same sample of clouds has been reanalysed by
Heyer et al. 2009 using data from the Boston
University-FCRAO Galactic Ring Survey Jackson et al. 2006.
The use of ${}^{13}$CO ($J = 1â0$) emission instead of ${}^{12}$CO
ensures that a large fraction of the data are optically thin;
additionally, the data used have a much higher angular sampling and
spectral resolution. Heyer et al. 2009 confirmed Larson's
third law with a relative scatter (approximately a factor 3) similar
to previous studies.  However, surprisingly they found a median mass
surface density of molecular hydrogen for this sample of $42 \mbox{
  M}_\odot \mbox{ pc}^{â2}$, thus significantly smaller than 
the one derived by Solomon et al. 1987.

On the theoretical side, there have been many attempts to explain
Larson's laws using numerical simulations.  In many cases, the
validity of Larsonâs relations, and especially of the third law, 
has been questioned Kegel et al. 1989 - Ballesteros-Paredes et al.2006.
In particular, it has been suggested that this law is merely the result
of the limited dynamic range of observations, and that in reality mass
surface densities of molecular clouds span at least two orders of
magnitude.

We re-examine the validity of Larson's third law using extinction as a
tracer of molecular gas Lada et al. 1994. The use of this
tracer, in combination with advanced techniques Lombardi et al.
2001 and 2009, allows us to probe
clouds over a large dynamical range (typically more than two order of
magnitudes in extinction); additionally, the column density
measurements use a simple and reliable tracer, dust.

We consider first the following version of Larson's third law.  Since
we have at our disposal complete extinction maps, we can consider the
area $S$ of a cloud \textit{above a given extinction threshold\/}
$A_0$ (unless otherwise noted, we will refer to extinction
measurements in the $K$ band, $A_K$, and drop everywhere the index
$K$).  We then define the cloud size implicitly from $S = \pi (L/2)^2$
(or the cloud radius as $R = L/2$).  Similarly, we can consider the
cloud mass $M$ above the same extinction threshold.

\begin{figure}
  \centering
  \includegraphics[width=0.445\hsize]{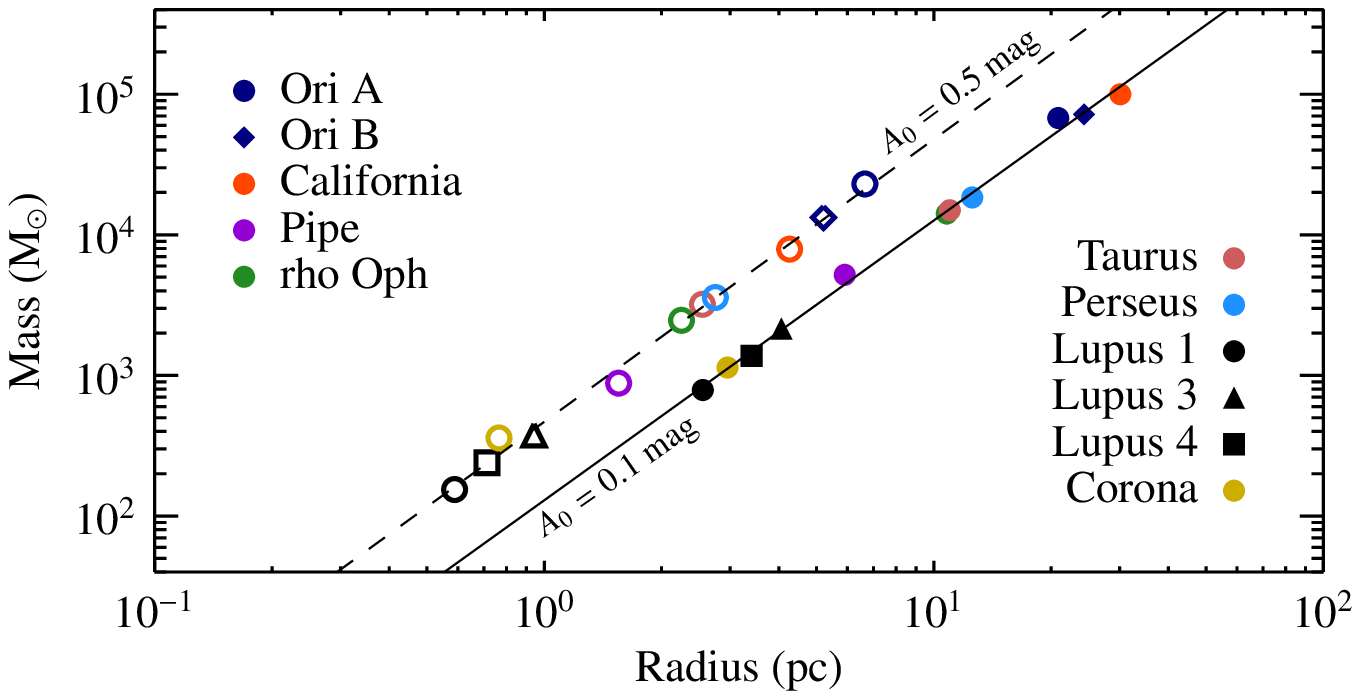}
  \includegraphics[width=0.54\hsize]{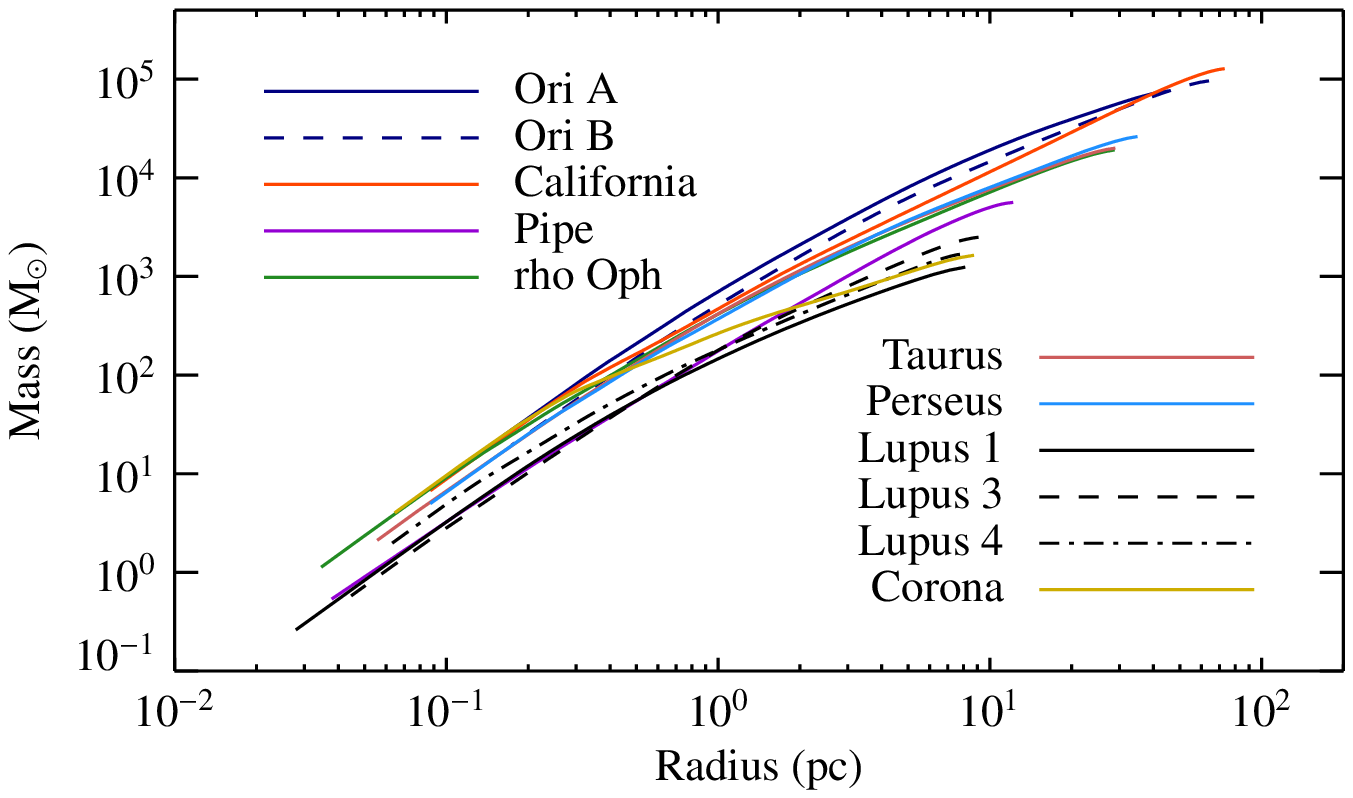}
  \caption{{\bf Left: } Cloud masses above extinction thresholds of
    $A_0 = 0.1 \mbox{ mag}$ (filled symbols) and $A_0 = 0.5 \mbox{
      mag}$ (open symbols) as a function of their size.  The two line
    shows the best constant surface density fits, which correspond to
    $\Sigma = 41 \mbox{ M}_\odot \mbox{ pc}^{-2}$ and $\Sigma = 149
    \mbox{ M}_\odot \mbox{ pc}^{-2}$ respectively. {\bf Right: } Mass
    vs.\ radius relationship for the same clouds.}
  \label{fig:1}
\end{figure}

Figure~\ref{fig:1} left shows the amount of mass different clouds have
above extinction thresholds of $A_K = 0.1 \mbox{ mag}$ and $A_K = 0.5
\mbox{ mag}$ as a function of the cloud ``radii'', together with the
best power-law fit.  As apparent from this plot, all clouds follow
exquisitely well a Larson-type relationship, with $M \propto R^2$, and
have therefore very similar projected mass densities \textit{at each
  extinction threshold}.  The exceptionally small scatter observed in
Fig.~\ref{fig:1} is also confirmed by a quantitative analysis:
\textit{at all extinctions considered, data follow the best-fit
  power-laws with relative standard deviations always below $15\%$.}

Figure~\ref{fig:1} right shows the second version of Larson's third
law considered here, i.e.\ the mass vs.\ radius relationship.  As
apparent from this figure, the tracks for the various clouds have
similar trends, but span a relatively large range of masses.  In the
range $R \in [0.1,1] \mbox{ pc}$ we can fit a power-law of the form
$M(R) = 380 \mbox{ M}_\odot \, (R / \mathrm{pc})^{1.6}$, a result that
compares well with the one obtained by Kauffmann et al. 2010,
$M(R) = 400 \mbox{ M}_\odot \, (R / \mathrm{pc})^{1.7}$.  Different
clouds have quite similar exponents (the standard deviation of the
power-law index is $\sim 0.18$), but rather different masses (the
best-fit scale parameter for the mass ranges from $150$ to $710 \mbox{
  M}_\odot$).  Note, however, that since the power-law index is
significantly different from two, errors on the assumed distances of
the clouds would affect the scale parameter for the mass.  From this
analysis we conclude that Larson's third law is not an accurate
description of the mass vs.\ radius relationship for single clouds.
Specifically, at larger scales all clouds show a flattening of the
curves and deviates significantly from a power-law, while at smaller
scales clouds follow power-laws, but with an exponent significantly
different than two.

Our results can be summarized as follow:
\begin{enumerate}
\item Using near-infrared extinction maps of a set of nearby clouds we
  tested Larson's third law for molecular clouds, the constancy of
  average mass surface densities above a given extinction threshold.
  We verified this scaling law to a relatively high degree of
  precision.  We found a very small ($<15\%$) relative scatter for the
  measured column densities independent of the adopted extinction
  thresholds over a very large range, from $A_K = 0.1 \mbox{ mag}$ to
  $A_K = 1.5 \mbox{ mag}$.  Additionally, we found the value of the
  average mass surface density to be a function of the adopted
  extinction threshold.
\item We verified that Larson's third law does not hold when
  considering the mass-radius relation within single clouds.  In the
  range $R \in [0.1, 1] \mbox{ pc}$ we find that the mass scales as
  $M(R) \propto R^{1.6}$, and is therefore significantly shallower
  than what was predicted by Larson; at larger radii, the relation
  appears to flatten even more.
\item We interpret these results, and in particular item~1 above, as
  the effects of a universal physical structure shared among the
  different clouds.  This universal structure is represented by a
  uniformity in the cloud density distributions.  We find that a
  log-normal model is able to account for this uniformity, provided
  that the log-normal parameters are restricted to relatively narrow
  ranges.  This suggests that Larson's third law might be a
  consequence of this special property of cloud structure.
\end{enumerate}


{\bf References}

\begin{description}
\expandafter\ifx\csname natexlab\endcsname\relax\def\natexlab#1{#1}\fi
\expandafter\ifx\csname bibnamefont\endcsname\relax
  \def\bibnamefont#1{#1}\fi
\expandafter\ifx\csname bibfnamefont\endcsname\relax
  \def\bibfnamefont#1{#1}\fi
\expandafter\ifx\csname citenamefont\endcsname\relax
  \def\citenamefont#1{#1}\fi
\expandafter\ifx\csname url\endcsname\relax
  \def\url#1{\texttt{#1}}\fi
\expandafter\ifx\csname urlprefix\endcsname\relax\def\urlprefix{URL }\fi
\providecommand{\bibinfo}[2]{#2}
\providecommand{\eprint}[2][]{\url{#2}}
\item[{\citenamefont{{Larson}}(1981)}]{1981MNRAS.194..809L}
\bibinfo{author}{\bibfnamefont{R.~B.} \bibnamefont{{Larson}}},
  \bibinfo{journal}{\mnras} \textbf{\bibinfo{volume}{194}},
  \bibinfo{pages}{809} (\bibinfo{year}{1981}).

\item[{\citenamefont{{Solomon} et~al.}(1987)\citenamefont{{Solomon},
  {Rivolo}, {Barrett}, and {Yahil}}}]{1987ApJ...319..730S}
\bibinfo{author}{\bibfnamefont{P.~M.} \bibnamefont{{Solomon}}},
  \bibinfo{author}{\bibfnamefont{A.~R.} \bibnamefont{{Rivolo}}},
  \bibinfo{author}{\bibfnamefont{J.}~\bibnamefont{{Barrett}}},
  \bibnamefont{and} \bibinfo{author}{\bibfnamefont{A.}~\bibnamefont{{Yahil}}},
  \bibinfo{journal}{\apj} \textbf{\bibinfo{volume}{319}}, \bibinfo{pages}{730}
  (\bibinfo{year}{1987}).

\item[{\citenamefont{{Sanders} et~al.}(1986)\citenamefont{{Sanders},
  {Clemens}, {Scoville}, and {Solomon}}}]{1986ApJS...60....1S}
\bibinfo{author}{\bibfnamefont{D.~B.} \bibnamefont{{Sanders}}},
  \bibinfo{author}{\bibfnamefont{D.~P.} \bibnamefont{{Clemens}}},
  \bibinfo{author}{\bibfnamefont{N.~Z.} \bibnamefont{{Scoville}}},
  \bibnamefont{and} \bibinfo{author}{\bibfnamefont{P.~M.}
  \bibnamefont{{Solomon}}}, \bibinfo{journal}{\apjs}
  \textbf{\bibinfo{volume}{60}}, \bibinfo{pages}{1} (\bibinfo{year}{1986}).

\item[{\citenamefont{{Heyer} et~al.}(2009)\citenamefont{{Heyer}, {Krawczyk},
  {Duval}, and {Jackson}}}]{2009ApJ...699.1092H}
\bibinfo{author}{\bibfnamefont{M.}~\bibnamefont{{Heyer}}},
  \bibinfo{author}{\bibfnamefont{C.}~\bibnamefont{{Krawczyk}}},
  \bibinfo{author}{\bibfnamefont{J.}~\bibnamefont{{Duval}}}, \bibnamefont{and}
  \bibinfo{author}{\bibfnamefont{J.~M.} \bibnamefont{{Jackson}}},
  \bibinfo{journal}{\apj} \textbf{\bibinfo{volume}{699}}, \bibinfo{pages}{1092}
  (\bibinfo{year}{2009}), \eprint{0809.1397}.

\item[{\citenamefont{{Jackson} et~al.}(2006)\citenamefont{{Jackson},
  {Rathborne}, {Shah}, {Simon}, {Bania}, {Clemens}, {Chambers}, {Johnson},
  {Dormody}, {Lavoie} et~al.}}]{2006ApJS..163..145J}
\bibinfo{author}{\bibfnamefont{J.~M.} \bibnamefont{{Jackson}}},
  \bibinfo{author}{\bibfnamefont{J.~M.} \bibnamefont{{Rathborne}}},
  \bibinfo{author}{\bibfnamefont{R.~Y.} \bibnamefont{{Shah}}},
  \bibinfo{author}{\bibfnamefont{R.}~\bibnamefont{{Simon}}},
  \bibinfo{author}{\bibfnamefont{T.~M.} \bibnamefont{{Bania}}},
  \bibinfo{author}{\bibfnamefont{D.~P.} \bibnamefont{{Clemens}}},
  \bibinfo{author}{\bibfnamefont{E.~T.} \bibnamefont{{Chambers}}},
  \bibinfo{author}{\bibfnamefont{A.~M.} \bibnamefont{{Johnson}}},
  \bibinfo{author}{\bibfnamefont{M.}~\bibnamefont{{Dormody}}},
  \bibinfo{author}{\bibfnamefont{R.}~\bibnamefont{{Lavoie}}},
  \bibnamefont{et~al.}, \bibinfo{journal}{\apjs}
  \textbf{\bibinfo{volume}{163}}, \bibinfo{pages}{145} (\bibinfo{year}{2006}),
  \eprint{arXiv:astro-ph/0602160}.

\item[{\citenamefont{{Kegel}}(1989)}]{1989A\&A...225..517K}
\bibinfo{author}{\bibfnamefont{W.~H.} \bibnamefont{{Kegel}}},
  \bibinfo{journal}{\aap} \textbf{\bibinfo{volume}{225}}, \bibinfo{pages}{517}
  (\bibinfo{year}{1989}).

\item[{\citenamefont{{Scalo}}(1990)}]{1990ASSL..162..151S}
\bibinfo{author}{\bibfnamefont{J.}~\bibnamefont{{Scalo}}}, in
  \emph{\bibinfo{booktitle}{Physical Processes in Fragmentation and Star
  Formation}}, edited by \bibinfo{editor}{\bibnamefont{{R.~Capuzzo-Dolcetta,
  C.~Chiosi, \& A.~di Fazio}}} (\bibinfo{year}{1990}), vol.
  \bibinfo{volume}{162} of \emph{\bibinfo{series}{Astrophysics and Space
  Science Library}}, pp. \bibinfo{pages}{151--176}.

\item[{\citenamefont{{Vazquez-Semadeni}
  et~al.}(1997)\citenamefont{{Vazquez-Semadeni}, {Ballesteros-Paredes}, and
  {Rodriguez}}}]{1997ApJ...474..292V}
\bibinfo{author}{\bibfnamefont{E.}~\bibnamefont{{Vazquez-Semadeni}}},
  \bibinfo{author}{\bibfnamefont{J.}~\bibnamefont{{Ballesteros-Paredes}}},
  \bibnamefont{and} \bibinfo{author}{\bibfnamefont{L.~F.}
  \bibnamefont{{Rodriguez}}}, \bibinfo{journal}{\apj}
  \textbf{\bibinfo{volume}{474}}, \bibinfo{pages}{292} (\bibinfo{year}{1997}),
  \eprint{arXiv:astro-ph/9607175}.

\item[{\citenamefont{{Ballesteros-Paredes} and {Mac
  Low}}(2002)}]{2002ApJ...570..734B}
\bibinfo{author}{\bibfnamefont{J.}~\bibnamefont{{Ballesteros-Paredes}}}
  \bibnamefont{and} \bibinfo{author}{\bibfnamefont{M.}~\bibnamefont{{Mac
  Low}}}, \bibinfo{journal}{\apj} \textbf{\bibinfo{volume}{570}},
  \bibinfo{pages}{734} (\bibinfo{year}{2002}), \eprint{arXiv:astro-ph/0108136}.

\item[{\citenamefont{{Ballesteros-Paredes}}(2006)}]{2006MNRAS.372..443B}
\bibinfo{author}{\bibfnamefont{J.}~\bibnamefont{{Ballesteros-Paredes}}},
  \bibinfo{journal}{\mnras} \textbf{\bibinfo{volume}{372}},
  \bibinfo{pages}{443} (\bibinfo{year}{2006}), \eprint{arXiv:astro-ph/0606071}.

\item[{\citenamefont{{Lada} et~al.}(1994)\citenamefont{{Lada}, {Lada},
  {Clemens}, and {Bally}}}]{1994ApJ...429..694L}
\bibinfo{author}{\bibfnamefont{C.~J.} \bibnamefont{{Lada}}},
  \bibinfo{author}{\bibfnamefont{E.~A.} \bibnamefont{{Lada}}},
  \bibinfo{author}{\bibfnamefont{D.~P.} \bibnamefont{{Clemens}}},
  \bibnamefont{and} \bibinfo{author}{\bibfnamefont{J.}~\bibnamefont{{Bally}}},
  \bibinfo{journal}{\apj} \textbf{\bibinfo{volume}{429}}, \bibinfo{pages}{694}
  (\bibinfo{year}{1994}).

\item[{\citenamefont{{Lombardi} and {Alves}}(2001)}]{2001A\&A...377.1023L}
\bibinfo{author}{\bibfnamefont{M.}~\bibnamefont{{Lombardi}}} \bibnamefont{and}
  \bibinfo{author}{\bibfnamefont{J.}~\bibnamefont{{Alves}}},
  \bibinfo{journal}{\aap} \textbf{\bibinfo{volume}{377}}, \bibinfo{pages}{1023}
  (\bibinfo{year}{2001}).

\item[{\citenamefont{{Lombardi}}(2009)}]{2009A\&A...493..735L}
\bibinfo{author}{\bibfnamefont{M.}~\bibnamefont{{Lombardi}}},
  \bibinfo{journal}{\aap} \textbf{\bibinfo{volume}{493}}, \bibinfo{pages}{735}
  (\bibinfo{year}{2009}), \eprint{0809.3383}.

\item[{\citenamefont{{Kauffmann} et~al.}(2010)\citenamefont{{Kauffmann},
  {Pillai}, {Shetty}, {Myers}, and {Goodman}}}]{2010ApJ...716..433K}
\bibinfo{author}{\bibfnamefont{J.}~\bibnamefont{{Kauffmann}}},
  \bibinfo{author}{\bibfnamefont{T.}~\bibnamefont{{Pillai}}},
  \bibinfo{author}{\bibfnamefont{R.}~\bibnamefont{{Shetty}}},
  \bibinfo{author}{\bibfnamefont{P.~C.} \bibnamefont{{Myers}}},
  \bibnamefont{and} \bibinfo{author}{\bibfnamefont{A.~A.}
  \bibnamefont{{Goodman}}}, \bibinfo{journal}{\apj}
  \textbf{\bibinfo{volume}{716}}, \bibinfo{pages}{433} (\bibinfo{year}{2010}),
  \eprint{1004.1170}.
\end{description}


\newpage

\subsection{Katarina Markovi\v{c}}

\vskip -0.3cm

\begin{center}

University Observatory Munich, Ludwig-Maximillian University, Scheinerstr. 1, 81679 Munich, Germany \\
and\\
Excellence Cluster Universe, Boltzmannstr. 2, 85748 Garching, Germany
\bigskip

{\bf Warm Dark Matter with future cosmic shear data} 

\end{center}

\medskip

Free-streaming dark matter particles dampen the overdensities on small scales of the initial linear matter density field. This corresponds to a suppression of power in the linear matter power spectrum and can be modelled relatively straightforwardly for an early decoupled thermal relic dark matter particle. Such a particle would be neutrino-like, but heavier; an example being the gravitino in the scenario, where it is the Lightest Supersymmetric Particle and it decouples much before neutrinos, but while still relativistic, see [1]. Such a particle is not classified as Hot Dark Matter, like neutrinos, because it only affects small scales as opposed to causing a suppression at all scales. However its free-streaming prevents the smallest structures from gravitationally collapsing and does therefore not correspond to Cold Dark Matter. The effect of this Warm Dark Matter may be observable in the statistical properties of cosmological Large Scale Structure.

Cosmic shear is the weak gravitational lensing of the images of very distant galaxies caused by the deflection of photons from these galaxies by the gravitational potential wells of the intervening dark matter density field. Because this effect does not strongly depend on baryonic physics it is a promising probe of the statistics of the dark matter density field. Future cosmic shear data will be able to measure the cosmic shear power spectrum to a very good accuracy and will therefore provide a probe for the non-linear matter power spectrum at low redshifts (late times). Unfortunately robust modelling of gravitational non-linearities at late times is not straightforward, but is absolutely necessary in order to extract cosmological information from large scale structure probes, in particular from cosmic shear.

The suppression of the linear matter density field at high redshifts in the WDM scenario can be calculated by solving the Boltzmann equations. A fit to the resulting linear matter power spectrum in the simple thermal relic scenario is provided by [2]. This linear matter power spectrum must then be corrected for late-time non-linear collapse. This can be done using the \emph{halofit} method of [3] or using the halo model ([4]). In [5], both of these non-linear methods are used to calculate weak lensing effect from a future survey and a limit of $m_{\rm WDM} \sim 2$ keV is predicted. WDM with greater particle masses in the simple thermal relic scenario is predicted not to be detectable with the next generation of weak lensing surveys.

\begin{figure*}
\begin{center}
\includegraphics[width=16.0cm,height=6.5cm]{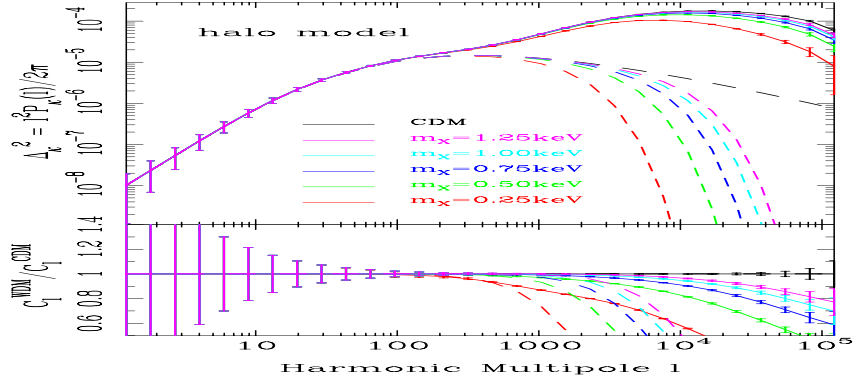}
\end{center}
\caption{\small Comparison of the weak lensing convergence power spectrum in the CDM and WDM halo models, as a function of angular multipole for source galaxies at $z_{\rm s}=1$. Top panel: Absolute power. Dashed and solid lines show the predictions from linear theory and from the nonlinear halo model. Black lines represent results for CDM. Coloured lines with red, green, blue, cyan, magenta denote results for the WDM model with particle masses $m_{WDM}=0.25,0.5,0.75,1.0,1.25$ keV respectively. Bottom panel: Ratio of the WDM model predictions to the CDM predictions. Lines styles and colours are as above.\\
This figure is from [6].}
\label{fig_wlps}
\end{figure*}

However, both of the above non-linear methods were developed assuming CDM and are therefore not necessarily appropriate for the WDM case. For this reason, [6] modify the halo model. Firstly, they treat the dark matter density field as made up of two components: a smooth, linear component and a non-linear component, both with power at all scales. Secondly, they introduce a cut-off mass scale, below which no haloes are found. Thirdly, they suppress the mass function also above the cut-off scale and finally, they suppress the centres of halo density profiles by convolving them with a Gaussian function, whose width depends on the WDM relic thermal velocity. The latter effect is shown to not be significant in the WDM scenario for the calculation of the non-linear matter power spectrum at the scales relevant to weak lensing. They make predictions for the next generation weak lensing surveys, which are consistent with the results of [5]. See Figure \ref{fig_wlps} 
for the plot of the weak lensing power spectra in different WDM scenarios.

\begin{figure*}
\begin{center}
\includegraphics[width=16.0cm,height=7cm]{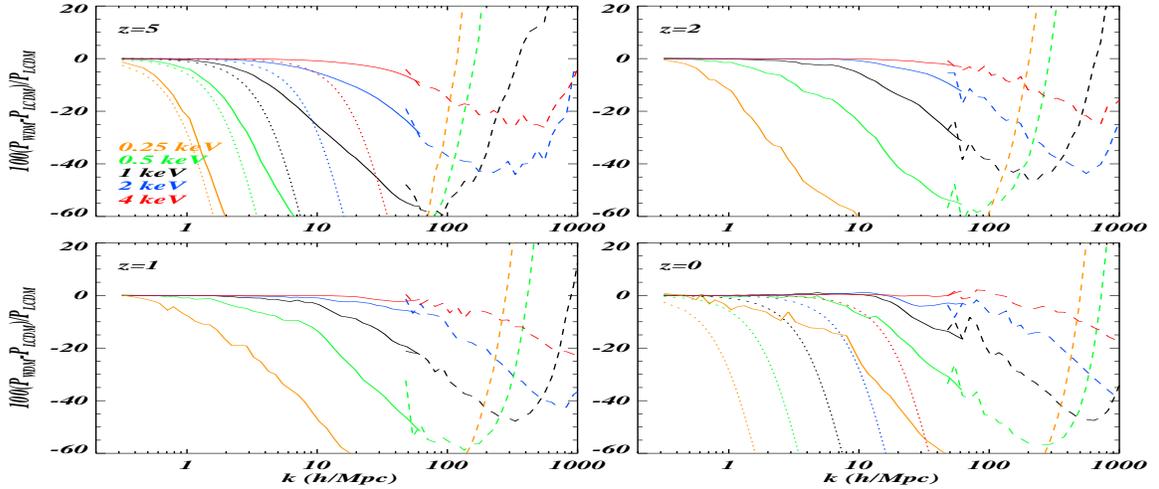}
\end{center}
\caption{\small Percentage difference between warm dark matter non-linear power and cold dark matter for the different runs. The resolution is kept fixed in this plot and only 25 Mpc/h boxes are considered. Orange, green, black, blue and red curves refer to $m_{\rm WDM}=0.25,0.5,1,2,4 $ keV, respectively. The continuous lines represent the large scale estimate of the power, while the dashed ones describe the small scale power obtained with the folding method. The four panels represent different redshifts at $z=0,1,2,5$ (bottom right, bottom left, top right and top left, respectively). The dotted coloured curves plotted at $z=0$ and $z=5$ are the redshift independent linear suppression between the different models.\\
This figure is from [7].}
\label{fig_sims}
\end{figure*}

In order to determine the validity of the different non-linear WDM models, [7] ran cosmological simulations with WDM. They provide a fitting function that can be easily applied to approximate the non-linear WDM power spectrum at redshifts $z=0.5-3.0$ at a range of scales relevant to the weak lensing power spectrum. Figure \ref{fig_sims} shows the percentage differences between the WDM and CDM non-linear matter power spectrum in the simple thermal relic scenario for different WDM masses.

{\bf References}

\begin{description}

\item[1] Bond, Szalay, and Turner, \emph{Formation of Galaxies in a Gravitino-Dominated Universe}, Phys. Rev. Lett., 48, 1636, (1982).

\vspace{-0.3cm}

\item[2] Viel, Lesgourgues, Haehnelt, Matarrese and Riotto, \emph{Constraining warm dark matter candidates including sterile neutrinos and light gravitinos with WMAP and the Lyman-$\alpha$ forest}, Phys. Rev. D, vol. 71 pp. 63534, (2005).

\vspace{-0.3cm}

\item[3] Smith, Peacock, Jenkins, White, Frenk, Pearce, Thomas, Efstathiou and Couchman, \emph{Stable clustering, the halo model and non-linear cosmological power spectra}, MNRAS, 341 pp. 1311 (2003).

\vspace{-0.3cm}

\item[4] Press and Schechter, \emph{Formation of Galaxies and Clusters of Galaxies by Self-Similar Gravitational Condensation}, Astroph. J., 187, 425 (1974).

\vspace{-0.3cm}

\item[5] Markovi\v{c}, Bridle, Slosar and Weller, \emph{Constraining warm dark matter with cosmic shear power spectra}, JCAP01(2011)022

\vspace{-0.3cm}

\item[6] Smith and Markovi\v{c}, \emph{Testing the Warm Dark Matter paradigm with large-scale structures}, arXiv:1103.2134, (2011).

\vspace{-0.3cm}

\item[7] Viel, Markovi\v{c}, Baldi and Weller, \emph{The Non-linear Matter Power Spectrum in Warm Dark Matter Cosmologies}, (2011, in preparation).

\end{description}

\newpage

\subsection{Alexander Merle}

\vskip -0.3cm

\begin{center}
Department of Theoretical Physics, School of Engineering Sciences, Royal Institute of Technology (KTH),\\
AlbaNova University Center, Roslagstullsbacken 21, 106 91 Stockholm, Sweden

\bigskip

{\bf Neutrino Model Building and keV sterile Neutrino Dark Matter} 

\end{center}

\medskip

One of the most exciting problems in astrophysics and cosmology, but maybe also one of the most attractive connection points of these fields to elementary particle physics is the explanation of the so-called {\em Dark Matter} (DM) in the Universe. DM could be {\em cold} (non-relativistic), like in the standard cosmological model, or {\em hot} (relativistic), which is, however, strongly disfavored by structure formation. Between those limiting cases there would be {\em warm} Dark Matter (WDM), favored by simulations of large scale structure formation. A particularly interesting candidate for WDM from the particle physics side is a sterile neutrino with a mass of a few keV, which can be accommodated for by a simple extension of the Standard Model (SM) of Elementary Particle Physics dubbed {\em $\nu$MSM}~[1].

Indeed, gauge extensions of the SM could easily include such a scenario~[2], while still being consistent with all phenomenological, observational, and experimental constraints. Although these frameworks can be brought into accordance with all data, they actually do not give an ``explanation'' for the values of the parameters involved, and actually they are {\em scenarios} rather than {\em models}: In a scenario, all the parameters involved can assume values such that there is no disagreement with any observation or experiment, while models give, at least to some extend, reasons for why certain parameters have their respective values. On the other hand, the experimental data from neutrino experiments seems to indicate that there is a reason for the parameter values observed~[3], which makes it tempting to develop dedicated models with certain features that enable them to relate the different observables in such a way as to explain part or all of the data.

Although the neutrino model building industry is very active~[4], and although sterile (right-handed) neutrinos most probably exist anyway due to the necessity of a non-zero active neutrino mass, up to now, there are only a few models around that can explain such a peculiar mass patterns as needed for keV sterile neutrino DM. Observational constraints require one neutrino to be at the keV-scale, while the other two sterile neutrinos (in a three-generation model) must have masses of at least GeV~[2]. From a theory point of view, one would expect them to be even heavier~[4].

The first class of models yielding an explanation of this mass pattern is the one proposed by Kusenko {\it et al.}~[5], who used the exponential warping factor present in Randall-Sundrum models of extra dimensions to suppress the lightest sterile neutrino mass down to the keV scale, while the other two can have masses of about $10^{11}$~GeV or higher. This leads to a generation dependent exponent $\alpha_i$, which suppresses the natural mass scale $M_0$ according to
\begin{equation}
M_i=\frac{M_0}{e^{\alpha_i}-1},
\nonumber
\end{equation}
such that a mild hierarchy $\alpha_1>\alpha_2>\alpha_3$ of the exponents will translate into a very strong hierarchy $M_1\ll M_2\ll M_3$ of the 4D sterile neutrino masses. A further bonus of this model is that, by the structure of the exponential suppressions involved, the seesaw mechanism to explain the light neutrino masses is not spoiled by the presence of the keV-scale sterile neutrino.

\begin{figure}[t]
\centering
\begin{tabular}{lr}\includegraphics[width=6.5cm]{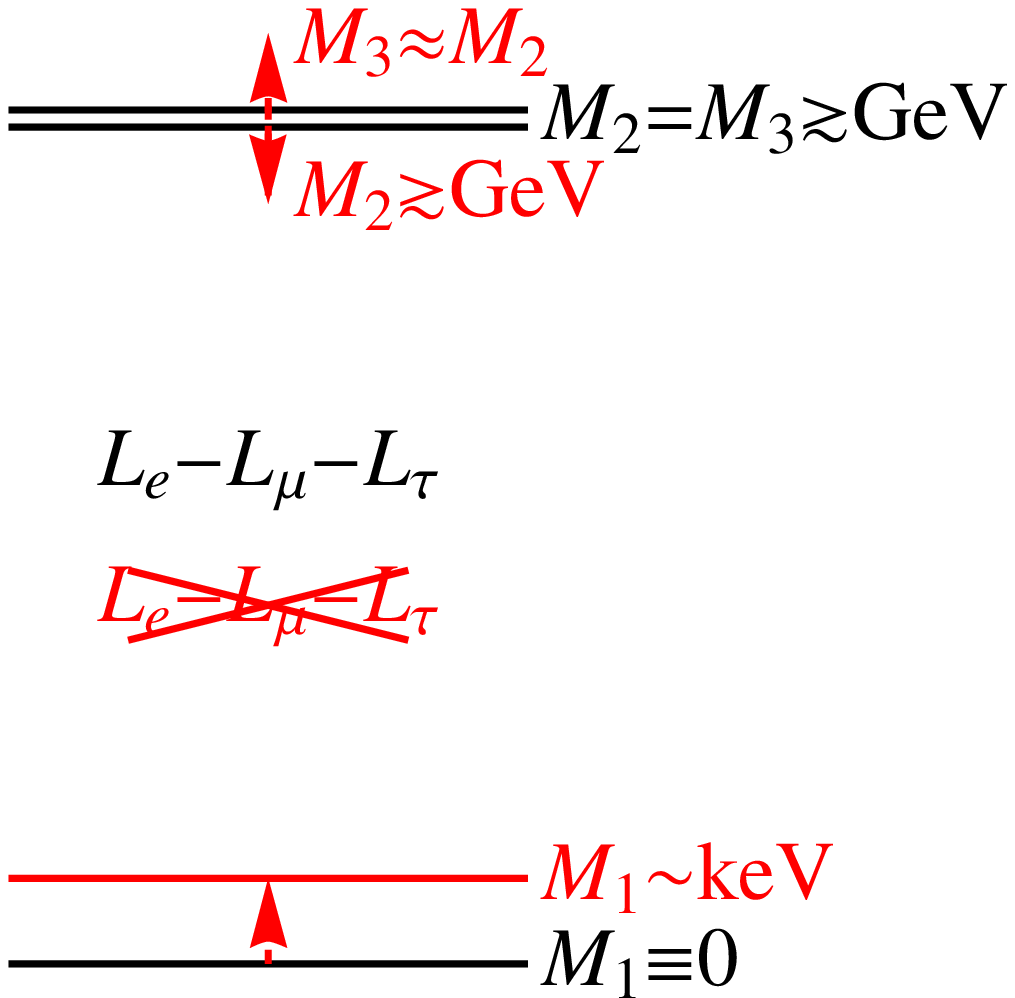}\hspace{0.5cm}  \& \includegraphics[width=8.5cm]{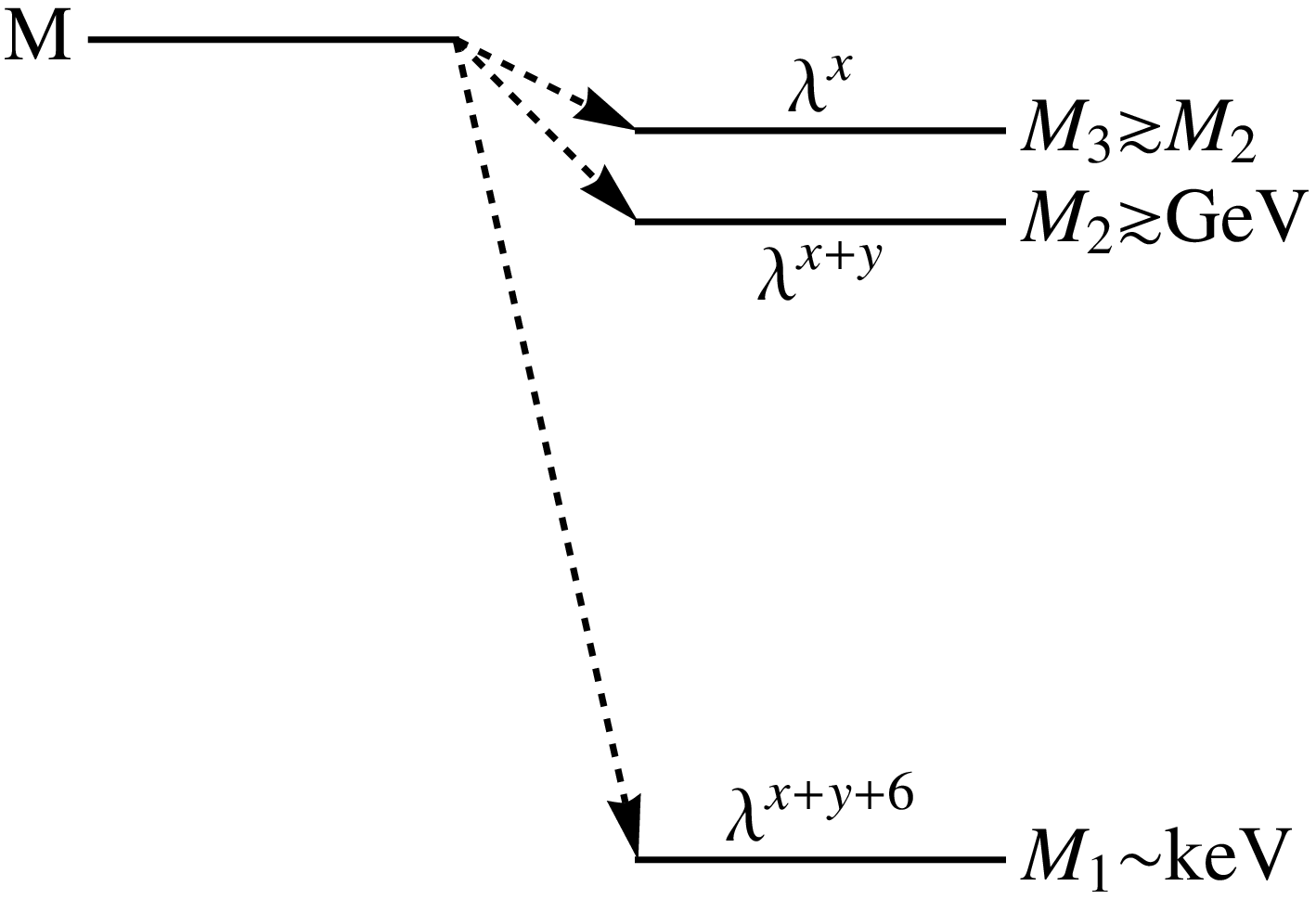}
\end{tabular}
\caption{\label{fig:schemes} The mass shifting schemes for the $L_e-L_\mu-L_\tau$ and the Froggatt-Nielsen models.}
\end{figure}

A second class yielding an explanation for the pattern required has been constructed by Lindner {\it et al.}~[6], who used an $\mathcal{F}=L_e-L_\mu-L_\tau$ {\em flavour symmetry} as basic ingredient. The mechanism is illustrated in Fig.~\ref{fig:schemes}a: The exact $\mathcal{F}$ symmetry (black) leads to a heavy (and light) neutrino mass pattern that is already quite close to the one required, by having one neutrino exactly massless while the other two are degenerate and have, let us say, the mass $M$. However, instead of one sterile neutrino being massless, one would require it to be at the keV scale. Furthermore, light neutrino data excludes the existence of two strictly degenerate light neutrinos. Accordingly, as practically any known flavour symmetry, the $\mathcal{F}$ symmetry has to be broken at some level. A convenient way to parametrize such a breaking is by introducing so-called {\em soft breaking terms}, whose magnitude $s$ is much smaller than the magnitude $M$ !
 of the symmetry-preserving terms. One could also try to construct an explicit scalar sector that leads to the correct breaking, but the resulting neutrino mass matrix will just be an explicit realization of the soft-breaking parametrization. They key point is that the terms of order $s$ will slightly alter the mass eigenvalues, in such a way that the degeneracies are broken and the otherwise massless neutrino obtains a small mass of magnitude $s\ll M$. The parameter $s$ can easily be taken to have a value of a few keV, while $M$ must be considerably larger, giving a motivation to have it at the GeV-scale or higher.

This situation is depicted by the red part of Fig.~\ref{fig:schemes}a, which shows the modification introduced by the broken symmetry. Indeed, the sterile neutrino mass eigenvalues are changed according to
\begin{equation}
(0,M,M) \to (s,M-s,M+s),
\nonumber
\end{equation}
with an analogous modification in the light neutrino sector. This model explains two of the three neutrino mixing angles in a natural way, by predicting $\theta_{13}=0$ and $\theta_{23}=45^\circ$, while unfortunately also predicting $\theta_{12}=45^\circ$, which is off compared to experiments by $6\sigma$. This problem, however, can be cured by slightly modifying the charged lepton matrices, which are naturally expected to receive corrections from renormalization group running. Summing up, this model cannot only explain the sterile neutrino mass pattern, but it can also naturally explain part of the leptonic mixing, and it does in fact predict an exact light neutrino mass pattern, to be probed within the next few years.

Finally, a third class of models has just been developed~[7], which is based on the {\em Froggatt-Nielsen} (FN) mechanism~[8]. The model makes use of the generation dependent FN-suppressions to push one sterile neutrino mass down to the keV scale, as displayed in Fig.~\ref{fig:schemes}b. The key point is that a suitable choice of FN-charges leads to a suppression of certain mass matrix elements by roughly one order of magnitude per charge unit. In order to achieve the required sterile neutrino splitting, it is necessary for one eigenvalue of the right-handed neutrino mass matrix to have an FN-charge that is larger than the one of the second-to lightest eigenvalue by at least six units.

Although models based on the FN-mechanism often suffer from a high degree of arbitrariness, and hence from a lack of predictivity, one can show that, for the keV sterile neutrino DM framework, the number of possibilities can be reduced considerably by carefully taking into account all constraints. For example, in the framework under discussion, the FN-mechanism is incompatible with Left-Right symmetry, which has been used in the scenario of~[2], it is in conflict with the bimaximal mixing predicted by the model from Ref.~[6], and it favors Grand Unified Theories based on $SU(5)$ compared to the ones based on $SO(10)$. One further feature is that, due to the construction principle of the FN-mechanism, also in this model the validity of the seesaw mechanism is guaranteed, even though one could naively expect that a keV sterile neutrino could spoil its success.

In conclusion, keV sterile neutrinos as WDM allow for very interesting connections between the Dark Matter problem and neutrino model building. In particular, the strong entanglement of the light neutrino data with the properties of the sterile neutrinos provides an interesting possibility to test models with keV sterile neutrino DM without the need to rely on direct (and rare) signals from the sterile neutrinos themselves. Currently, there are only three working models around that are able to explain the distinctive features of the keV sterile neutrino DM scenario, and the community should hunt for more such models in the future.

{\em Acknowledgements:} AM is very grateful to N.~Sanchez and H.~J.~de Vega for their kind invitation to this workshop, and to all participants for the interesting and joyful time spent together. It has been a great pleasure for him to work with his collaborators M.~Lindner and V.~Niro. The work of AM is supported by the Royal Institute of Technology (KTH), under project no.\ SII-56510.

\bigskip

{\bf References}

\begin{description}

\item[1] T.~Asaka, S.~Blanchet, and M.~Shaposhnikov, Phys.\ Lett.\  B {\bf 631} (2005) 151 [arXiv:hep-ph/0503065].
\vspace{-0.3cm}

\item[2] F.~Bezrukov, H.~Hettmansperger, and M.~Lindner, Phys.\ Rev.\  D {\bf 81} (2010) 085032, arXiv:0912.4415 [hep-ph].
\vspace{-0.3cm}

\item[3] K.~Nakamura {\it et al.}  [Particle Data Group], J.\ Phys.\ G {\bf 37} (2010) 075021.
\vspace{-0.3cm}

\item[4] R.~N.~Mohapatra {\it et al.}, Rept.\ Prog.\ Phys.\  {\bf 70} (2007) 1757, [arXiv:hep-ph/0510213].
\vspace{-0.3cm}

\item[5] A.~Kusenko, F.~Takahashi, and T.~T.~Yanagida, Phys.\ Lett.\  B {\bf 693} (2010) 144, arXiv:1006.1731 [hep-ph].
\vspace{-0.3cm}

\item[6] M.~Lindner, A.~Merle, and V.~Niro, JCAP {\bf 1101} (2011) 034, arXiv:1011.4950 [hep-ph].
\vspace{-0.3cm}

\item[7] A.~Merle and V.~Niro, arXiv:1105.5136 [hep-ph].
\vspace{-0.3cm}

\item[8] C.~D.~Froggatt and H.~B.~Nielsen, Nucl.\ Phys.\ {\bf B147}, 277 (1979).

\end{description}

\newpage

\subsection{Christian Moni Bidin}

\vskip -0.3cm

\begin{center}
Universidad de Concepci\'on, Concepci\'on, Chile

\bigskip

{\bf No evidence of dark matter in the Galactic disk} 
\end{center}

\medskip

Measuring the matter density of the Galactic disk in the proximity of the Sun through stellar kinematics
is an old art, dating nearly a century [1,2]. The comparison of the results with the expected amount
of visible matter can provide an estimate of the dark matter (DM) density in the analyzed volume. So
far, all but two estimates [3,4] have converged to the conclusion that ``there is no evidence for a
significant amount of disk DM" [e.g., 5,6]. Despite the general agreement, this result is still very
general, and only a couple of works obtained some stronger constraints on the fundamental properties of
the DM halo, such as its flattening and local density [7,8]. However, previous investigations did
not directly measure the mass density, but this was derived adjusting the Galactic model parameters to
fit the observations, with the exception of [9]. Moreover, the results are mainly limited by the
approximations used in the calculations, whose validity breaks down at increasing height from the
Galactic plane ($z$), and no study has ever exceeded $z$=1.1~kpc. In this restricted volume the expected
amount of DM is small, and within errors the presence of a classical DM halo is often not required, but
always allowed.

We propose a new formulation for the estimate of the disk mass density, based on the distribution of
stars in the six-dimentional phase space. By means of the Poisson and Jeans equation, and very basic
assumptions both theoretically and observationally confirmed, we obtain the exact (not approximated)
analytical expression of the surface mass density $\Sigma$ (M$_\odot$ pc$^{-2}$) valid at any $z$. The
information required to solve this equation is the vertical trend of the three components of the
dispersion ellipsoid in cylindrical coordinates $\sigma_U$(z), $\sigma_V$(z), $\sigma_W$(z), and of the
non-diagonal term $\sigma^2_{UW}$(z), the radial derivative of $\sigma_U$, $\sigma_V$, and $\sigma^2_{UW}$,
plus three parameters: the solar Galactocentric distance R$_\odot$, and the scale length and height of
the disk population under study (h$_{R,\rho}$ and h$_{z,\rho}$, respectively).

We applied our formulation to a sample of thick disk red giant stars with 2MASS photometry [10], SPM3
absolute proper motion [11], and radial velocity measurements [12]. We thus measured $\Sigma$(z), for the
first time, up to 4~kpc from the plane. The data provided no information about the radial behavior of the
dispersions, but the assumed radial decay is supported by both observations [13] and theoretical models
[14]. The results were first presented in [15], and a publication with all the details of the analysis is
in preparation.

\begin{figure}
\centering
\includegraphics[width=10cm]{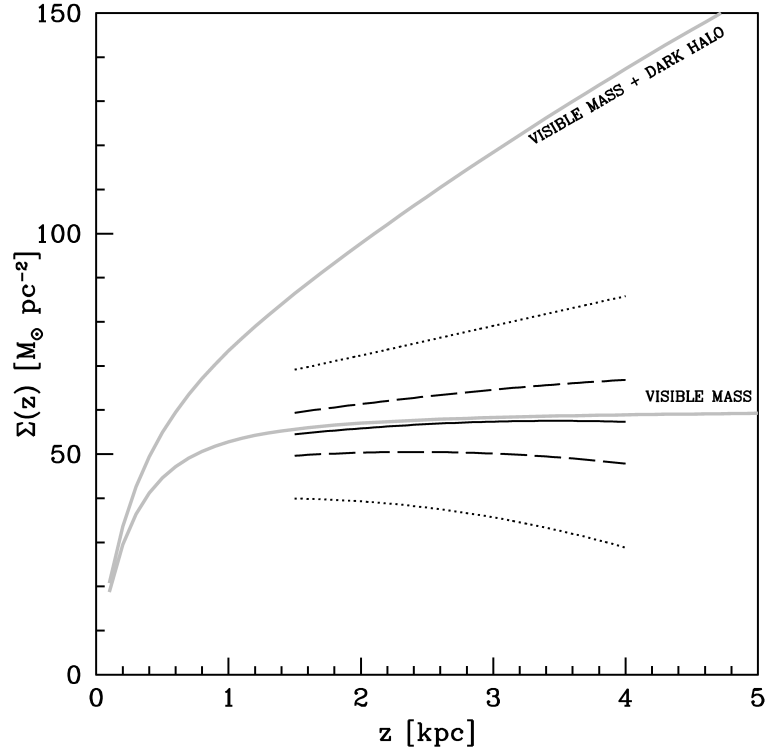}
\caption{Measured surface mass density as a function of distance from the plane (full black curve). The
1$\sigma$ and 3$\sigma$ strips are indicated by the dashed and dotted curves, respectively. The lower
grey curve shows the expectations for visible mass alone, while the upper one corresponds to a model
where a classical DM halo from [18] is added. More details about these models can be found in [15].}
\label{cmbidin}
\end{figure}

As shown in Figure~\ref{cmbidin}, we find a striking coincidence between the calculated curve and the expectations
for visible mass alone. A classical DM halo is excluded at a 6$\sigma$ level, and the amount of local
DM allowed by the 1$\sigma$ error is negligible ($\leq$1~M$_\odot$ pc$^{-3}$). We find that the results
can be forced to allow a certain amount of local DM, but only under a series non-standard assumptions,
e.g. a very thin and extended stellar thick disk (h$_{R,\rho}\geq$4.5~kpc, h$_{z,\rho}\leq$0.6~kpc)
ruled out by observations. While the break-down of a single assumption is possible, the required series
of ad-hoc alternative hypothesis is very unlikely.

Our work, while representing a noticeable improvement to the knowledge of the Galactic mass spatial
distribution, still lacks the information required to solve the equations with observed quantities only.
Present and future extensive surveys such as SDSS [16], RAVE [17], and GAIA, will provide the missing
data, allowing the study of the density distribution of the DM halo with unprecedented detail.

\newpage

{\bf References}

\begin{description}

\item[1] J.C. Kapteyn 1922, ApJ, 588, 823
\vspace{-0.3cm}
\item[2] J.H. Oort 1932, BAN, 6, 249
\vspace{-0.3cm}
\item[3] J.N. Bahcall 1984, ApJ, 276, 169
\vspace{-0.3cm}
\item[4] J.N. Bahcall, C. Flynn, \& A. Gould 1992, ApJ, 389, 234
\vspace{-0.3cm}
\item[5] K. Kuijken, \&  G. Gilmore 1989, MNRAS, 239, 605
\vspace{-0.3cm}
\item[6] J. Holmberg, \& C. Flynn 2004, MNRAS, 352, 440
\vspace{-0.3cm}
\item[7] M. Cr\'ez\'e, E. Chereul, O. Bienaym\'e, \& C. Pichon 1998, 329, 920
\vspace{-0.3cm}
\item[8] O. Bienaym\'e, C. Soubiran, T.V. Mishenina, V.V. Kovtyukh, \& A. Siebert 2006, A\&A, 446, 933
\vspace{-0.3cm}
\item[9] V.I. Korchagin, T.M. Girard, T.V. Borkova, D.I. Dinescu, \& W.F. van Altena 2003, AJ, 126, 2896
\vspace{-0.3cm}
\item[10] M.F. Skrutskie, et al. 2006, AJ, 131, 1163
\vspace{-0.3cm}
\item[11] T.M. Girard, D.I. Dinescu, W.F. van Altena, I. Platais, D.G. Monet, \& C.E. L\'opez 2004, AJ, 127, 3060
\vspace{-0.3cm}
\item[12] C. Moni Bidin 2009, PhD Thesis, U. de Chile
\vspace{-0.3cm}
\item[13] J.R. Lewis, \& K.C. Freeman 1989, AJ, 97, 139
\vspace{-0.3cm}
\item[14] P. Cuddeford, \& P. Amendt 1992, MNRAS, 256, 166
\vspace{-0.3cm}
\item[15] C. Moni Bidin, G. Carraro, R.A. M\'endez, \& W.F. van Altena 2010, ApJ, 724, L122
\vspace{-0.3cm}
\item[16] D.G. York, et al. 2000, AJ, 120, 1579
\vspace{-0.3cm}
\item[17] T. Zwitter, et al. 2008, AJ, 120, 2131
\vspace{-0.3cm}
\item[18] R.P. Olling, \& M.R. Merrifield 2001, MNRAS, 326, 164
\end{description}

\newpage

\def\Journal#1#2#3#4{{#1} {\bf #2} (#3), p. #4.}
\def\NCA{Nuovo Cimento}
\def\NIM{Nucl.\,Inst. and Meth.}
\def\NIMA{Nucl.\,Instr. and Meth. A}
\def\NIMB{Nucl.\,Instr. and Meth. B}
\def\NPB{Nucl.\,Phys. B}
\def\NDS{Nucl. Data Sheets}
\def\APP{Astropart. Phys.}
\def\NPA{Nucl.\,Phys. A}
\def\NP{Nucl.\,Phys.}
\def\NPBP{Nucl.\,Phys. B (Proc. Suppl.)}
\def\PLB{Phys.\,Lett.  B}
\def\PL{Phys.\,Lett.}
\def\JPD{J.\,Phys. D}
\def\JPG{J.\,Phys. G}
\def\JPGNP{J.\,Phys. G: Nucl. Phys.}
\def\PRL{Phys.\,Rev.\,Lett.}
\def\PRA{Phys.\,Rev. A}
\def\PRD{Phys.\,Rev. D}
\def\PRC{Phys.\,Rev. C}
\def\PR{Phys.\,Rev.}
\def\ZPC{Z.\,Phys. C}
\def\EPL{Europhys.\,Lett.}
\def\JP{J.\,Phys.}
\def\AP{Astropart.\,Phys.}
\def\RP{Phys.\,Rep.}
\def\ARNPS{Annu.\,Rev.\,Nucl.\,Part.\,Sci.}
\def\JLT{J.\,Low\,Temp.\,Phys.}

\def\ppg{\pi^+\pi^-\gamma}
\def\bbn{$\beta\beta$-0$\nu$}
\def\bb{$\beta\beta$-2$\nu$}
\def\Qbb{$Q_{\beta\beta}$}
\def\taubb{$\tau_{1/2}^{0\nu}$}
\def\meff{$\langle m_\nu\rangle$}
\def\cky{c/kg/y}
\def\ckky{c/keV/kg/y}
\def\ckty{c/keV/t/y}
\def\ky{kg$\times$y}
\def\my{moles$\times$y}
\def\Te{$^{130}$Te}
\def\Ge{$^{76}$Ge}
\def\Ca{$^{48}$Ca}
\def\Cd{$^{116}$Cd}
\def\Nd{$^{150}$Nd}
\def\Gd{$^{160}$Gd}
\def\Se{$^{82}$Se}
\def\Mo{$^{100}$Mo}
\def\Xe{$^{136}$Xe}
\def\Fe{$^{55}$Fe}
\def\Re{$^{187}$Re}
\def\Ho{$^{163}$Ho}
\def\Co{$^{60}$Co}
\def\Tl{$^{208}$Tl}
\def\teo{TeO$_2$}
\def\Tr{$^3$H}
\def\mn{$m_\nu$}
\def\mug{$\mu$g}
\def\mum{$\mu$m}
\def\mumsq{$\mu$m$^2$}
\def\mumq{$\mu$m$^3$}
\def\mus{$\mu$s}
\def\halft{$\tau_{1/2}$}
\def\de{$\Delta E$}
\def\fwhm{$_{\mathrm{FWHM}}$}
\def\b{$\beta$}
\def\g{$\gamma$}
\def\agre{AgReO$_4$}

\subsection{Angelo Nucciotti, on behalf of the MARE collaboration}

\vskip -0.3cm

\begin{center}

Dip. di Fisica ``G. Occhialini'', Universit\`a di Milano-Bicocca and
INFN - Sez. di Milano-Bicocca, Milano, Italy

\bigskip

{\bf The MARE experiment and its capabilities to measure the mass of light (active) and heavy (sterile) neutrinos} 

\end{center}

\medskip
Direct neutrino mass measurement analyzing the kinematics of electrons emitted in beta decays is the most sensitive model independent method to assess the neutrino mass absolute value. 
In practice this method consists in measuring the minimum energy carried away by the anti-neutrino -- i.e. its rest mass -- by observing the highest energy electrons emitted in the decay.

To date, the most sensitive experiments were carried out analyzing the \Tr\ decay in magnetic adiabatic collimation spectrometers with electrostatic filter, yielding an upper limit on the electron anti-neutrino mass of
2.2\,eV. Starting from 2013 or 2014 the new experiment KATRIN will  analyze the \Tr\ beta decay end-point  with a
much more sensitive electrostatic spectrometer and with an expected  statistical sensitivity of about 0.2\,eV.
However a severe limitation to these experiments resides in their configuration whereas the \Tr\ source is external to the spectrometer. 
This causes many systematic uncertainties because the measured electron energy has to be corrected for the
energy lost in exciting atomic and molecular states, in crossing the source, in scattering through the spectrometer, and more. 
Therefore, to improve the sensitivity, it is necessary to reduce both the systematic and the statistical uncertainties. 
Because of the large weight of systematics, it is inherent in this type of
measurement that confidence in the results can be obtained only through confirmation by independent
experiments. 

An alternative approach is calorimetry where the
beta source is embedded in the detector so that all the energy emitted in the decay is measured, except for that
taken away by the neutrino.
In this configuration, the systematic uncertainties arising from the electron source being external to the detector are eliminated. 
On the other hand, since calorimeters detects all the decays occurring over the entire beta energy spectrum,
the source activity must be limited to avoid spectral distortions and background at the end-point due to pulse pile-up.
As a consequence the statistics near the end-point is limited as well.
Since the fraction of decays in a given energy interval $\Delta E$ below the end-point $Q$ is only
$\propto (\Delta E/Q)^3$, the limitation on the statistics may be partially overcome by using as beta source \Re, 
the beta-active nuclide with the second lowest known transition energy ($Q\sim2.5$\,keV).

A perfect practical way to make a calorimetric measurement is to use thermal detectors. 
At thermal equilibrium, the temperature rise of the detector -- measured by a suitable thermometer -- is due to  the sum of the energy of the emitted electron and of all other initial excitations. 
The measurement is then free from the systematics induced by any possible energy
loss in the source and is not affected by problems related to decays on excited final states.

Thermal detectors with absorbers containing natural Rhenium are the most straightforward way to make a calorimetric measurement.  
\Re\ natural isotopic abundance (A.I.$=63$\%) and its half-life time ($\tau_{1/2}=43.2$\,Gy) make it in principle perfect to design small sized ($\approx$\,mg) high performance detectors with an activity $A_\beta$ of few decays per second. 
The high statistics required can be accumulated with the help of large arrays of thermal detectors.

To date, only two experiments have been carried out with thermal detectors containing \Re: 
the MANU [1] and MIBETA [2] experiments.
The two experiments, with a statistics corresponding to about 10$^7$ decays, yielded
limits on \mn\ of about 26\,eV at 95\% CL and 15\,eV at 90\% CL respectively.
The systematics affecting these experiments are still small compared with the statistical errors. The main sources
are the background, the detector response function, the theoretical spectral shape of the \Re\ \b\ decay,
the Beta Environmental Fine Structure (BEFS), and the pile-up.

\begin{figure}
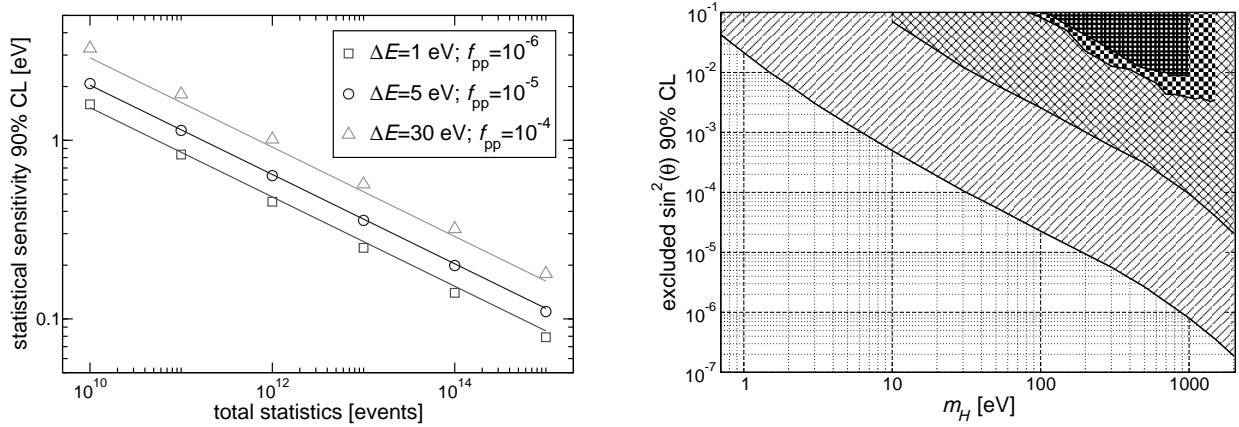

\begin{center}
\includegraphics*[bb=40 20 580 430,clip,width=0.43\linewidth]{scan-Nev_c.eps}
\hspace{0.03\linewidth}
\includegraphics*[bb=20 20 575 415,clip,width=0.445\linewidth]{sens-heavynu_mare1-2.eps}
\caption{Left: statistical sensitivity of a \Re\ neutrino mass calorimetric experiment estimated with both 
Montecarlo (points) and analytic (curves) approaches. Right: statistical sensitivity to the emission of heavy 
neutrinos with mass $m_H$. From lower to upper curve: Montecarlo symulation with $N_{ev}=10^{14}$, 
$f_{pp}=10^{-6}$ and $\Delta E = 1.5$\,eV; Montecarlo symulation with $N_{ev}=8\times10^{9}$, $f_{pp}=10^{-5}$ 
and $\Delta E = 15$\,eV; MIBETA experiment; MANU experiment [2].}
\label{fig14nuc}
\end{center}
\end{figure}

The Microcalorimeter Arrays for a Neutrino Mass Experiment (MARE) project was launched 
by a large international collaboration in 2005 with the aim of measuring the neutrino mass with a calorimetric 
approach [3].
The baseline of the MARE project consists in a large array of Rhenium based thermal detectors, but different options 
for the isotope are also being considered [5]. 
Left panel of Fig. \ref{fig14nuc} shows what are the experimental requirements for this experiment in terms of total 
statistics $N_{ev}$, energy resolution \de\ and pile-up level $f_{pp}$ -- i.e. the fraction of pile-up 
events with respect to the total decays
 given by $\tau_R A_\beta $, where $\tau_R$ is approximately the signal rise time [4].
It is apparent that, in order to reach  sub-eV sensitivities, the MARE experiment must collect  
more than $10^{13}$\,events. 
This calls for the use of large arrays  -- for a total number of channels of the order of 10000 -- coupled 
to an appropriate signal multiplexing scheme to avoid running into insurmountable cryogenic and economic problems.
Altogether these specifications are about the same as for IXO -- the next generation X-ray space observatory -- and largely proved to be technically feasible. The major open issue remains the  metallic Rhenium absorber coupling to the sensor.
In fact, in spite of the many efforts, persisting technical difficulties prevented from realizing the target performances with Rhenium absorbers so far. 
In order to have a viable alternative to the baseline MARE design using Rhenium \b\ decay, the MARE collaboration is
considering the possibility to use of \Ho\ electron capture (EC). \Ho\ decays to $^{163}$Dy with a half life of about 4570 years and, because of the low transition energy, capture is only allowed from the M shell or higher.  The EC decay may be detected only through the mostly non-radiative atomic de-excitation of the Dy atom and from the Inner Bremsstrahlung (IB) radiation.
 It has been proposed in [6] that, thanks to its very low transition energy ($Q_{EC}\approx2.5$\,keV), the analysis of the end-point of the calorimetric spectrum can yield a very high sensitivity to the neutrino mass. As for \b\ decay end-point experiments, also in \Ho\ experiments the sensitivity on the neutrino mass depends on the fraction of events at the end-point which increases for decreasing $Q_{EC}$.
Unfortunately, $Q_{EC}$ is only approximately known: its determinations span from 2.2 to 2.8 keV with a recommended value of about 2.555\,keV.
In case of the more favorable values of $Q_{EC}$, \Ho\ represents a very interesting alternative to \Re. Because of the relatively short half life, detectors may be realized by introducing only few \Ho\ nuclei (about $10^{11}$ for 1 decay/s) in low temperature microcalorimeters optimized for low energy X-ray spectroscopy, without any further modification.  

The calorimetric spectra measured by MARE with either \Re\ or \Ho\ are also suitable to investigate  the emission 
of heavy (sterile) neutrinos with a mixing angle $\theta$. Assuming the electron neutrino $\nu_e$ is a mixture of two 
mass eigenstates $\nu_H$ and $\nu_L$, with masses $m_H \gg m_L$, then $\nu_e =  \nu_L \cos \theta +   \nu_H \sin \theta$ 
and the measured energy spectrum is $N(E,m_L,m_H,\theta) = N(E,m_L) \cos^2 \theta + N(E,m_H) \sin^2 \theta$. 
The emission 
of heavy neutrinos would manifest as a kink in the spectrum at an energy of $ Q - m_H $ for heavy neutrinos with masses 
between about 0 and $Q-E_{th}\le\approx2.5$\,keV, where $ E_{th} $ is the experimental energy threshold. Right panel in 
Fig. \ref{fig14nuc}  
shows the statistical sensibility achievable by the MARE experiment with \Re\ as estimated by Montecarlo simulations. 
Similar curves can be obtained for \Ho. The two curves in the upper right corner of Fig. \ref{fig14nuc} are the 
experimental limits set by the MANU [2] and MIBETA experiments.
The search for heavy neutrinos may be affected by systematic uncertainties due to the background and to the 
ripple observed in the \Re\ spectrum and caused by the BEFS. 

The MARE project is subdivided into two phases. The second phase -- MARE-2 -- is the final large scale experiment with sub-electron sensitivity. The first one -- MARE-1 -- is a collection of activities with the aim of sorting out both the best isotope and 
the most suited detector technology to be used for the final experiment [5].
On the short term, intermediate scale experiments are being prepared, with the purpose of both reaching a neutrino mass sensitivity of few electronvolts and studying in depth the possible sources of systematic uncertainties. The first of these experiments is starting in Milano using two $6\times6$ silicon implanted arrays with \agre\ absorbers. The experiment may be later expanded to up to 8 arrays, for a total of 288 crystals. The potential statistical sensitivity to heavy neutrinos for the full scale experiment is shown by the cross-hatched curve in Fig. \ref{fig14nuc} (right panel).

 \medskip

{\bf References}

\begin{description}

\item[1]  M. Sisti et al., \Journal{\NIMA}{520}{2004}{125}
\item[2]  F. Gatti et al., Nucl.Phys. B, 91 (2001) 293; M. Galeazzi et al., \Journal{\PRL}{86}{2001}{1978}
\item[3] the MARE proposal, http://mare.dfm.uninsubria.it
\item[4]  A. Nucciotti et al., \Journal{\APP}{34}{2010}{80}
\item[5]  A. Nucciotti, proceeding of Neutrino 2010, Athens, Greece, June 14-19, 2010, arXiv:1012.2290v1
\item[6]  A. De Rujula and M. Lusignoli, \Journal{\PLB}{118}{1982}{429}
\vspace{-0.3cm}

\end{description}

\newpage

\subsection{Sinziana Paduroiu, Andrea Macci\`o, Ben Moore, Joachim Stadel, Doug Potter, George Lake, 
Justin Read \& Oscar Agertz}



\vskip -0.3cm

\begin{center}

S. P.: Observatory of Geneva, CH-1290, Geneva, Switzerland.
A. M. : Max-Planck-Institute for Astronomy, K\"onigstuhl 17, D-69117
Heidelberg, Germany.
B. M., J. S., D. P., \& G. L.: Institute for Theoretical Physics, University of Z\"urich, CH-8057 Z\"urich, Switzerland.
J. R.: Department of Physics \& Astronomy, University of Leicester, Leicester LE17RH, United Kingdom.
and Institute for Astronomy, Department of Physics, ETH Z\"urich, Wolfgang-Pauli-Strasse 16, CH-8093 Z\"urich, Switzerland.
O. A.: Kavli Institute for Cosmological Physics, University of
Chicago, 5640 South Ellis Avenue, Chicago, IL, USA 60637

\bigskip

{\bf The effects of free streaming on warm dark matter haloes: a test of the Gunn-Tremaine limit} 

\end{center}

\medskip

The free streaming of warm dark matter particles dampens the 
fluctuation spectrum, flattening the mass function
of haloes and imprinting a fine grained phase density (PSD) limit for dark matter structures.
The Gunn-Tremaine limit is expected to
imprint a constant density core at the halo center. 
In a purely cold dark matter model the fine grained phase space density is effectively infinite in the initial conditions and would therefore be infinite everywhere today. However when the phase space density profiles are computed using coarse grained averages that can be measured from N-body simulations, the value is finite everywhere and even falls with radius with a universal power law slope within virialised structures (Taylor \& Navarro 2001). The coarse grained phase space is an average over mixed regions of fine grained phase space, so this behaviour is as expected (Tremaine \& Gunn 1979).

Using high resolution simulations of structure
formation in a warm dark matter universe (movies available: http://obswww.unige.ch/$\sim$paduroiu ) we explore these effects on structure formation and the properties of warm dark matter halos.
\begin{itemize}
\item{The finite initial fine grained PSD is a also a maximum of the coarse grained PSD, resulting in PSD profiles of WDM haloes that are similar to CDM haloes in the outer regions, however they turn over to a constant value set by the initial conditions}
\item{The turn over in PSD results in a constant density core with characteristic size that is 
in agreement with the simplest expectations } 
\item{We demonstrated that if the primordial velocities are large enough to produce a significant core in dwarf 
galaxies i.e. $\sim$ kpc, then the free streaming erases all perturbations on that scale and the haloes cannot form}
\item{Halo formation occurs top down on all scales with the most massive haloes collapsing first}
\item{The concentration - mass relation for WDM haloes is reversed with respect to that found for CDM}
\item{Warm dark matter haloes contain visible caustics and shells}

\end{itemize}

\begin{figure}   
\psfig{file=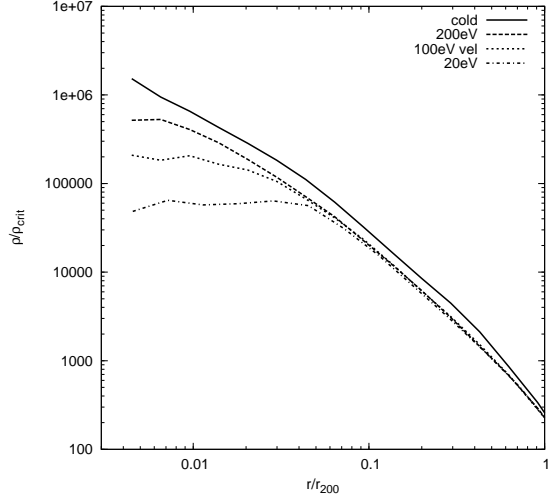,angle=-90,width=280pt}
\caption[]{The spherically averaged 
density profiles for CDM, WDM1, WDM2 \& WDM3 haloes.
The resolution limit is at approximately 0.5\% of the virial
radius (the softening radii are a 0.26\% of the virialized radius).}
\label{fig:11}
\end{figure}

We ran two suites of simulations, first with a $160^{3}$ particles in
40 Mpc box, and the second one with $300^3$ in a 42.51 Mpc box.
We adopt a flat $\Lambda$CDM  cosmology with parameters from the first
year  WMAP results  (Spergel et al. 2003). The transfer function in WDM model has been computed using the fitting formula suggested by Bode, Turok and Ostriker (2001) where $\alpha$, the scale of the break, is a function of the WDM parameters (Viel et al (2005)), while the index $\nu$ is fixed. From these simulations several galaxy mass haloes were re-simulated at higher resolution, with and without thermal velocities. 
The density profile for a  $7\times10^{11}M_{\odot}$ is shown in Figure 
\ref{fig:11} in four different
contexts, the CDM case, the WDM case with just the cutoff in the
power spectrum as expected for a 200eV particle (WDM1), with velocities corresponding to the 200eV particle (WDM2), and with velocities artificially increased such that they correspond to a 20eV particle but with the power spectrum
of the 200eV case (WDM3).

\begin{figure}
\psfig{file=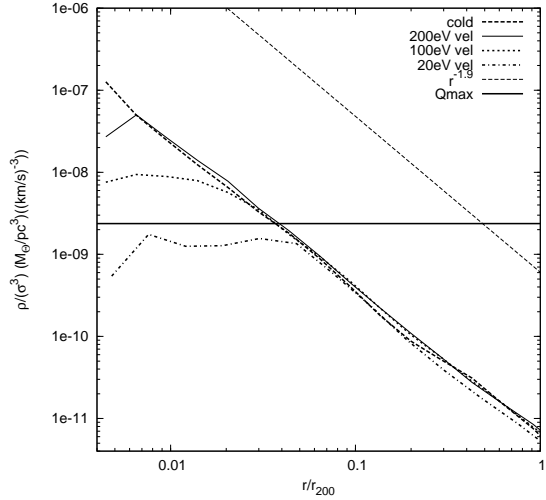,angle=-90,width=280pt}
\caption[]{``Phase-space density'' (PSD) profiles of the same haloes 
shown in Figure \ref{fig:11}, calculated using $\rho/\sigma^{3}$.} 
\label{fig:2}
\end{figure}

Figure \ref{fig:2} shows the corresponding coarse grained PSD 
profiles calculated by spherical averaging the quantity
$\rho/\sigma^{3}$.

In the case of our simulations, for a constant density in the initial 
conditions, the phase space density is:
\vspace{-0.3cm}
\begin{equation}
\label{eq=q0}
Q_0=\frac{\rho}{\sigma^{3}}=
\rho_{crit}\Omega\left({\frac{m_{\nu}c^2}{KTc}}\right)^3
\end{equation}
\vspace{-0.3cm}
For a critical density $\rho_{crit}=1.4\times10^{-7}M_\odot/pc^3$ and $\Omega=0.268$ we find the phase space density:
\begin{equation}
\label{eq=q01}
Q_0=3\times10^{-4}M_\odot{\rm pc}^{-3}\left({\rm km/s}\right)^{-3}\left(\frac{m_{x}}{1keV}\right)^{3}
\end{equation} 
\vspace{-0.3cm}
The minimum core radius is described by
\vspace{-0.3cm}
\begin{equation}
\label{eq=core}
r_{c,min}^{2}=\frac{\sqrt{3}}{4\pi
  GQ_{0}}\frac{1}{ \langle {\sigma^{2} \rangle }^{1/2}},
\end{equation}

Thus for a particle with $m_{x}=20$ eV and $\sigma=150$ km/s the 
phase space density is $2.4\times 10^{-9}M_\odot{\rm pc}^{-3}\left({\rm km/s}\right)^{-3}$  
which gives a theoretical value for the core radius of 9.3 kpc. If we consider the core radius to be the radius where the density starts decreasing from the constant value, a ''by eye'' fit gives a slightly larger radius of ~6.5 kpc, while the radius where the density drops by a factor of 2 is ~9.4 kpc in agreement with the theoretical predictions.

\bigskip

{\bf References}

\begin{description}

\item[1] Bode, P., Ostriker, J. P., \& Turok, N. 2001, \apj, 556, 93

\vspace{-0.3cm}

\item[2] {Dalcanton}, J.~J. \& {Hogan}, C.~J. 2001, \apj, 561, 35

\vspace{-0.3cm}

\item[3] {Navarro}, J.~F., {Frenk}, C.~S., \& {White}, S.~D.~M. 1996, \apj, 462, 563

\vspace{-0.3cm}

\item[4]{Strigari}, L.~E., {Bullock}, J.~S., {Kaplinghat}, M.,  {Kravtsov}, A.~V.,
  {Gnedin}, O.~Y., {Abazajian}, K. \& {Klypin}, A.~A. 2006, \apj, 652, 306 

\vspace{-0.3cm}

\item[5]{Taylor}, J.~E. \& {Navarro}, J.~F. 2001, \apj, 563, 483

\vspace{-0.3cm}

\item[6]{Tremaine}, S. \& {Gunn}, J.~E. 1979, Phys.Rev.Let., 42, 407

\vspace{-0.3cm}

\item[7]Viel, M., Lesgourgues, J.,
Haehnelt, M.~G., Matarrese, S., \& Riotto, A.\ 2006, Physical Review
Letters, 97, 071301

\vspace{-0.3cm}

\item[8]{Wang}, J.,{White}, S.~D.~M. 2007 (astroph-0702575)

\end{description}
\newpage

\subsection{Henri Plana}

\vskip -0.3cm

\begin{center}

Laborat\'orio de Astrof\'{i}sica Observacional Te\'orica e Observacional  \\
Universidade Estadual  de Santa Cruz - Ilh\'eus - Brazil

\bigskip

{\bf Mass distribution of Galaxies in Hickson Compact Groups} 

\end{center}

\medskip

This study presents the mass distribution for a sample of 18 late-type galaxies in nine Hickson Compact Groups [1]. We used rotation curves from high resolution 2D velocity fields of Fabry-Perot observations and J-band photometry from the 2MASS survey, in order to determine the dark halo and the visible matter distributions.   The study compares two halo density profile, an isothermal core-like distribution and a cuspy one. We also compare their visible and dark matter distributions with those of galaxies belonging to cluster and field galaxies coming from two samples: 40 cluster galaxies of [1] and 35 field galaxies of [3]. The central halo surface density is found
to be constant with respect to the total absolute magnitude similar to what is found for the isolated galaxies. 
This suggests that the halo density is independent to galaxy type and environment. 
 We have found that core-like density profiles fit better the rotation curves than cuspy-like ones. No major differences have been found between field, cluster and compact group galaxies with respect to their dark halo density profiles.

\medskip

Based on a sample of $\approx$ 100 velocity fields of spiral galaxies belonging to 25 HCGs, our previous study has demonstrated that:  40\% (38 galaxies) of the HCG galaxies analysed in our sample do not allow  the computation of the RC. This is because these galaxies are strongly perturbed by interaction and mergers. 33\% (31 galaxies) of HCG galaxies are regular enough to calculate a RC but the RC is not symmetric to allow the derivation of a mass model. These perturbed galaxies are mildly interacting. The remaining 27\% (25 galaxies) of them are suitable to derive mass models. For a variety reasons only 18 galaxies have been studied  in the present work. The RCs have been combined with 2MASS J-band surface brightness profiles to derive their mass models. Two shapes of dark halos have been considered: a core-like density profile (isothermal sphere or ISO) and a cuspy-like one (NFW [4]). Best fit models and maximum disk models have been computed for the two dark halo profil!
 es. We have compared the HCG galaxies with two samples of galaxies in different environments: field and clusters galaxies. The three samples (HCG, isolated and cluster) have been analysed using the same tools.

The results are summarized below:

\begin{itemize}

\item No obvious differences can be found between the halo parameters for the HCG galaxies and galaxies in other environments. The strong correlation between $\rho_0$ and $r_0$ is present for the three samples and the slopes of the linear
regressions between them are very similar (between -1.62 and -1.00), well inside the uncertainties. In a study based on late-type and dwarf spheroidal galaxies, [5] confirms this correlation giving a slope of  -1.038. On the other hand, the linear regression constant found by [5] is different in comparison with those found with the three samples.
 This scale difference is may be due to the fact that [5] compiled his sample using several sources using different methods to obtain halo parameters.

\item The use of the NFW model gives less satisfactory results than the ISO model. As mentioned by S08 [3], the $\chi^2$ coefficient is usually larger when using the NFW. NFW model gives a worse correlation between the halo central density and the core radius than ISO. This is also true if we use the field galaxies sample. In contrast, the NFW model gives a consistent result for the cluster galaxies sample. Halo profiles are closer to isothermal spheres than NFW profile as already found by other authors [3], [6], [7].
    The slope of the linear regression between $\rho_0$ and $r_0$ is higher for the cluster galaxies sample than for the two other samples when the NFW model is used.
The disagreement between ISO and NFW is smaller with cluster galaxies sample.  61\% of HCG galaxies shows higher disk M/L using the ISO model compared to NFW.
No relation between the disk scale length and the halo central density is seen using either ISO or NFW models.

\item We explored the possible connection between the halo parameters, the halo mass fraction and the disk M/L. The halo mass is high for both field galaxies
and compact groups galaxies (75 to 95\% of the total mass), leaving only modest room for the disk mass.
The halo surface density is independent of the absolute B magnitude and no clear relation is seen between the disk M/L and $M_B$, meaning that the halo is independent of the galaxy luminosity.

\end{itemize}

Recently, [8] suggest that core halos observed using high-resolution velocity fields in real dark matter-dominated galaxies are genuine and cannot be ascribed to systematic errors, halo triaxiality, or non-circular motions.

\begin{figure}[h]
\includegraphics[width=10cm,height=7cm]{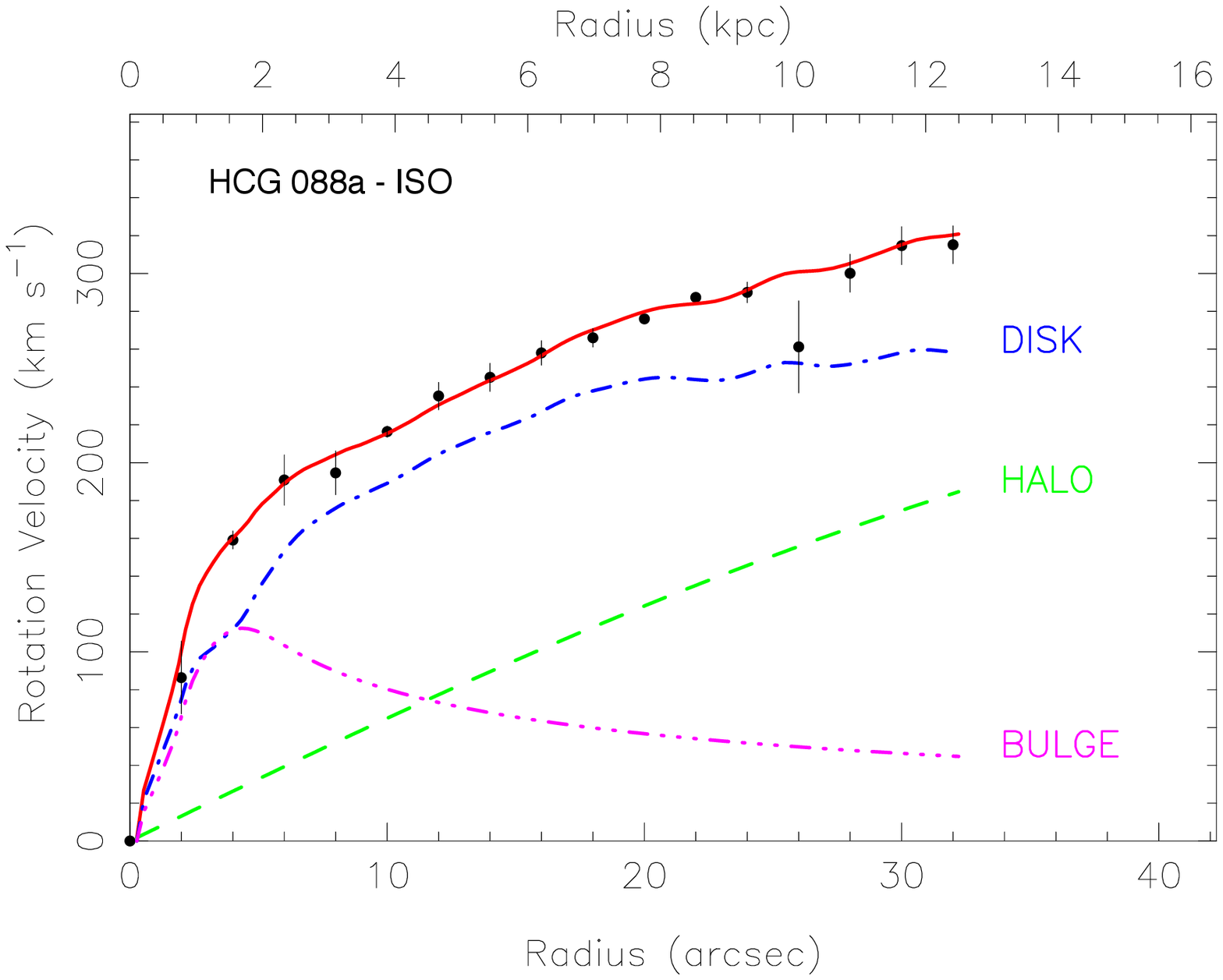}
\caption{\label{fig:epsart} Example of rotation curve and the fit of the different components: bulge, disk and isothermal halo.}
\end{figure}

\begin{figure}[h]
\includegraphics[scale=0.50]{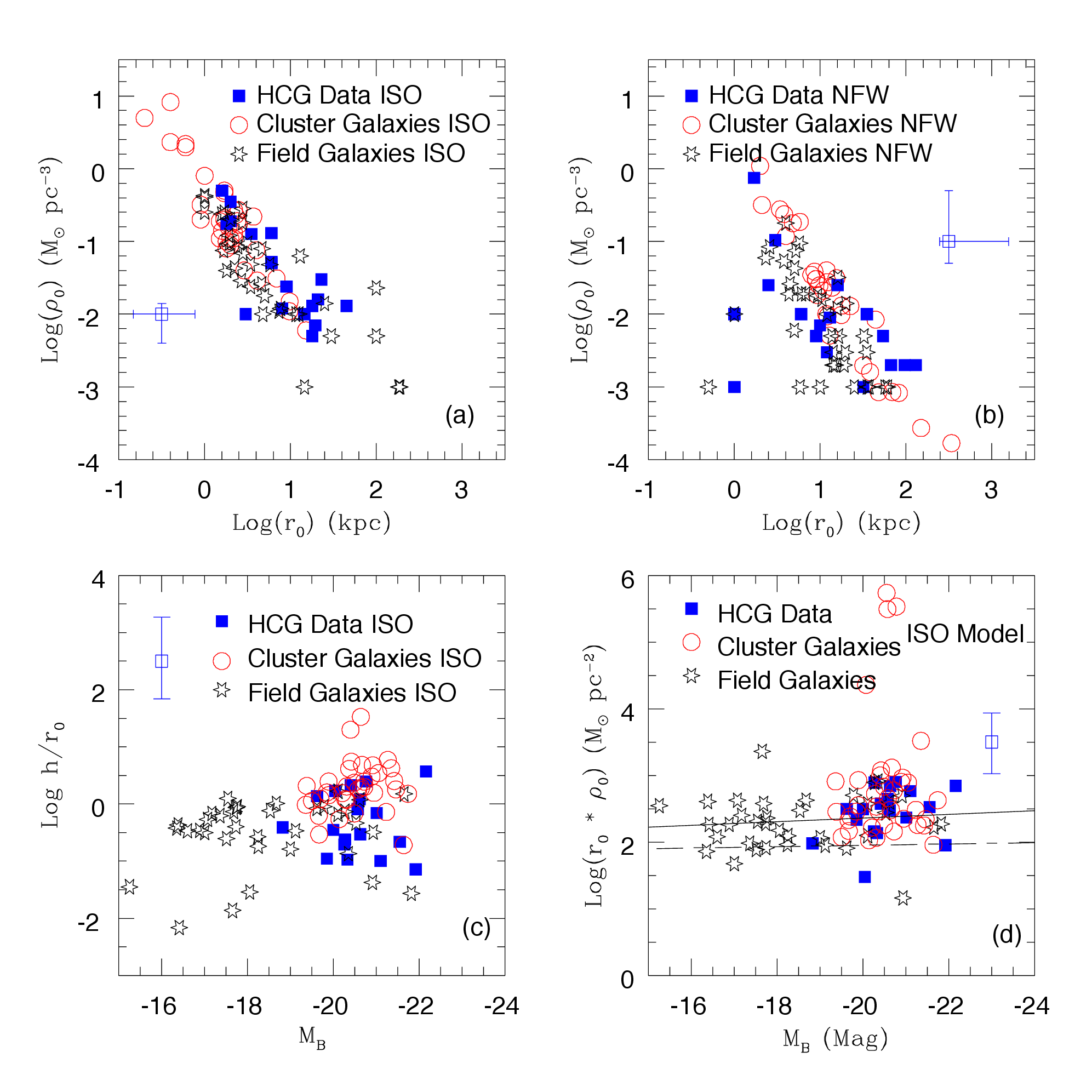}
\caption{\label{fig:epsart2} Different correlation found with halos parameters and magnitude and disk scale length.}
\end{figure}

\bigskip

{\bf References}

\begin{description}

\item[1] Plana, H., Amram, P., Mendes de Oliveira, C., Balkowski, C. 2010, Astronomical Journal, 139, 1
\vspace{-0.3cm}

\item[2] Barnes E.I., Sellwood J.A., Kosowsky A. 2004, Astronomical Journal, 128, 2724
\vspace{-0.3cm}

\item[3] Spano M., Marcelin M., Amram P., Carignan C., Epinat B., Hernandez O., 2008 MNRAS, 383, 297
\vspace{-0.3cm}

\item[4] Navarro, J.F., Frenk, C.S., White, S.D. 1996, Astrophysical Journal, 462, 563 - NFW
\vspace{-0.3cm}

\item[5] Kormendy J., Freeman K.C. 2004, IAU Symp. 220. S. Ryder, D.J. Pisano, M. Walker \& K.C. Freeman ed. p. 377
\vspace{-0.3cm}

\item[6] de Block W.J.G. \& Bosma A., 2002, Astronomy  \& Astrophysics, 385, 816
\vspace{-0.3cm}

\item[7] Gentile, G.,  Salucci, P., Klein, U., Vergani, D., Kalberla, P. 2004, MNRAS, 351, 903
\vspace{-0.3cm}

\item[8] Kuzio de Naray, R.; Kaufmann, T. 2011, MNRAS, tmp, 659K

\end{description}

\newpage

\subsection{Norma G. Sanchez and Hector J. de Vega}

\vskip -0.3cm

\begin{center}

HJdV: LPTHE, CNRS/Universit\'e Paris VI-P. \& M. Curie \& Observatoire de Paris.\\
NGS: Observatoire de Paris, LERMA \& CNRS

\bigskip

{\bf Warm dark matter: the linear Boltzmann-Vlasov equation vs.  observations}

\end{center}

The distribution function of the decoupled dark matter (DM) particles can be written as
$$
f(\vec{x},\vec{p};t) = f_0^{DM}(p)+ F_1(\vec{x},\vec{p};t) \quad , 
$$ 
where $ f_0^{DM}(p) $ is the zeroth order  freezed-out
DM distribution function in or out of thermal equilibrium and $ F_1(\vec{x},\vec{p};t) $
the  primordial fluctuations around it. The evolution of $ F_1 $ is governed by 
the Boltzmann-Vlasov equation linearized around  $ f_0^{DM}(p) $.
We evolve $ F_1(\vec{x},\vec{p};t) $ according to the 
linearized Boltzmann-Vlasov equation since the end of inflation till today.
The DM density fluctuations are given by
$$
\Delta(t,\vec{k}) \equiv m \; \int \frac{d^3p}{(2\pi)^3} \; 
\int d^3x \; e^{-i \, \vec{x} \cdot \vec{k}} \; F_1(\vec{x},\vec{p};t) \; .
$$
The initial data for the evolution are the primordial inflationary fluctuations
$$
|\phi_k| = \sqrt2 \; \pi \; \frac{| \Delta_0 |}{k^{\frac32}} \; 
\left(\frac{k}{k_0}\right)^{\frac{n_s-1}2} \;  g(\vec{k})  \quad ,  \quad
|\Delta_0 | \simeq 4.94 \; 10^{-5} , \; n_s \simeq 0.964 , \; k_0 = 2 \; {\rm Gpc}^{-1} \; , 
$$ 
where $ g(\vec{k}) $ is a random gaussian field. It is convenient to factorize 
the initial data as $ \Delta(z,\vec{k}) = \rho_{DM} \; {\bar \Delta}(z,k)  \; |\phi_k| \; g(\vec{k}) \; .$
We show that the linearized Boltzmann-Vlasov equation can be recasted as a pair of Volterra
integral equations for Warm Dark Matter (WDM) plus the neutrino density fluctuations $ {\bar \Delta}(z,k) $ [1].
When the WDM becomes non-relativistic, the pair of integral equations reduces to the single equation:
\be\label{volt}
{\bar \Delta}(z,k) = h(z,k) + \frac6{(z+1) \; k \; r_{lin}} \; \int^s_{s_0} ds' \; \Pi\{k \; r_{lin}[s(z) - s']\}
\; {\bar \Delta}(z(s'),k) , 
\ee
where $ z(s) + 1 = (z_{eq} + 1) \, \sinh^2 s \quad , \quad z_{eq} + 1 \simeq 3200 \quad , \quad
{\bar \Delta}({\rm initial},k)=1 $. Here $ h(z,k) $ is a known function: it contains the memory from 
previous ultra-relativistic (UR) evolution of the DM and
$
\Pi(x) \equiv \int_0^{\infty} Q \; dQ \; f_0^{DM}(Q) \; \sin(Q \; x) \; .
$
The integral equation (\ref{volt}) is valid both in the radiation RD and matter MD dominated eras. Eq.(\ref{volt})
becomes the Gilbert equation in the MD era (plus memory terms).

The physical characteristic length scale in this linear evolution 
is the free streaming scale (or Jeans' scale) 
$$
r_{lin} = 2 \, \sqrt{1 + z_{eq}} \, \left(\frac{3 \, M_{Pl}^2}{H_0 \, \sqrt{\Omega_{DM}} \; Q_{prim}}\right)^\frac13  
= 21.1 \,  q_p^{-\frac13} \; {\rm kpc}  \quad , \quad q_p^\frac14 \simeq 0.4 \; m/{\rm keV} \; ,
$$
where $ q_p \equiv \! Q_{prim}/{\rm (keV)}^4  \; $ is the primordial phase space density.
DM particles can freely propagate over distances less or equal the free streaming scale.

\begin{figure}
\begin{turn}{-90}
\psfrag{"dhoyRgam.dat"}{}
\psfrag{"trucho.dat"}{}
\psfrag{"edhoygam.dat"}{}
\includegraphics[height=9.cm,width=3.5cm]{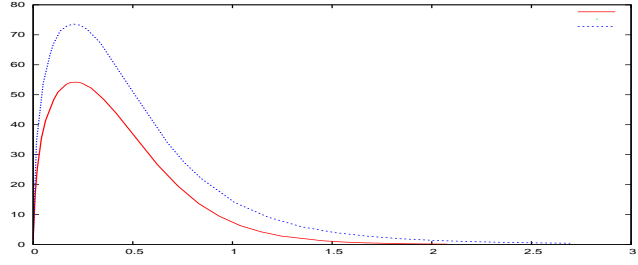}
\end{turn}
\caption{The WDM transfer function $ {\bar \Delta}(z=0,k)/{\bar \Delta}({\rm initial},k) $ 
vs. $ k  \; r_{lin} $ computed from eq.(\ref{volt}) for 1 keV particles. Red: thermal equilibrium
decoupling. Blue: out of equilibrium decoupling.}
\label{funtra}
\end{figure}

We plot in fig. \ref{funtra} the transfer function $ {\bar \Delta}(z=0,k)/{\bar \Delta}({\rm initial},k) $ 
vs. $ k  \; r_{lin}$ computed from  eq.(\ref{volt}) for particles of mass $m$ = 1 keV. The curve 
in red corresponds to fermions decoupling at thermal equilibrium and the curve in
blue to sterile neutrinos in the  $\chi$-model decoupling out of thermal equilibrium. 

\begin{figure}
\begin{turn}{-90}
\psfrag{"perfd.dat"}{}
\psfrag{"eperfd.dat"}{}
\includegraphics[height=9.cm,width=3.cm]{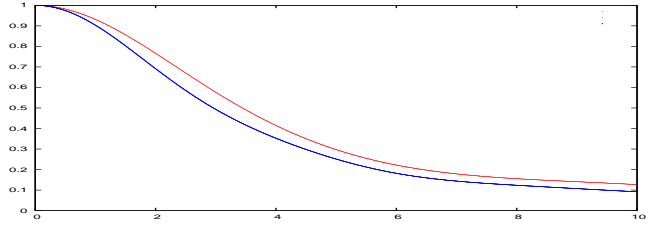}
\end{turn}
\caption{Linear density profiles $ \rho_{lin}(r)/ \rho_{lin}(0) $ vs. $ x \equiv r/r_{lin} $.
Same colur code as in fig. \ref{funtra}.}
\label{perlin}
\end{figure}
\begin{figure}
\begin{turn}{-90}
\includegraphics[height=9.cm,width=3.cm]{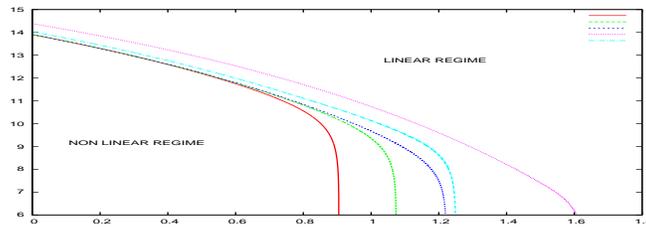}
\end{turn}
\caption{$ \log M/M_{\odot} $ vs. $ \log (z+1) $ for CDM (violet) and WDM:
1 keV (red),  2 keV (green), 4  keV (blue) 
DM particles decoupling in equilibrium
and 1 keV  (light-blue) sterile neutrinos}
\label{sigmlnl}
\end{figure}
The linear density profile $ \rho_{lin}(r,z) $  at redshift $z$ is given by
$$
\rho_{lin}(r,z) = \frac1{2 \, \pi^2 \; r} \; \int_0^{\infty} k \; dk \; \sin(k \, r) \;  
\Delta(k,z) \; \quad {\rm for}  \quad \; g(\vec{k})=1 
$$
Notice that $ \quad {\bar \Delta}(z,\gamma) \simeq \frac1{z+1} \; {\bar \Delta}(0,\gamma) $.
Therefore, the profile shape turns to be redshift independent in the MD/$\Lambda$ era.
We plot in fig. \ref{perlin} the linear density profiles $ \rho_{lin}(r)/ \rho_{lin}(0) $ vs. $ x \equiv r/r_{lin} $.
WDM linear density profiles turn to be cored for $ r \to 0 $.
We find in the intermediate regime $ r \gtrsim r_{lin} $:
$$
\rho_{lin}(r)\buildrel{r \gtrsim r_{lin}}\over= c_0 \;
 \left(\frac{r_{lin}}{r}\right)^{1+n_s/2} \; 
\rho_{lin}(0) \quad , \quad 1+n_s/2 = 1.482
$$
Namely, $ \rho_{lin}(r) $ scales with the  primordial spectral index $ n_s $.
These theoretical linear results agree with the universal empirical behaviour 
$r^{-1.6\pm 0.4}$: M. G. Walker et al.  (2009) (observations) and with
I. M. Vass et al. (2009) (simulations).
In the asymptotic regime $ r \gg  r_{lin} $, the small $ k $ behaviour of 
$ \Delta(k,t_{\rm today}) \buildrel{k \to 0}\over= c_1 \; (k \;  r_{lin})^s $
with $ s \simeq 0.5 $ implies the presence of a tail:
$  \quad \rho_{lin}(r)  \buildrel{r \gg r_{lin}}\over\simeq c \; \left( r_{lin}/r \right)^2 $.
For WDM, the agreement between the linear theory and the observations is {\it remarkable}.
Linear CDM profiles turn to be cusped: 
$ \rho_{lin}(r)_{CDM} $ vs. $ r $ exhibits a cusp behaviour for $ r \gtrsim 0.03 $ pc, 
(in agreement with CDM simulations, which show the predictivity of our results).
Notice that observations in DM dominated galaxies always find {\it cores} as predicted by WDM.

The expected overdensity within a radius $ R $ at redshift $ z $ in the linear regime is given by
$$
\sigma^2(R,z) = \int_0^{\infty} \frac{dk}{k} \; \Delta^2(z,k) \; W^2(kR) 
\quad , \quad  W(kR) = {\rm window ~ function} \; .
$$
$ \sigma^2(R,z) \sim 1 $ is the  borderline between linear 
and non-linear regimes (see fig. \ref{sigmlnl}). Objects (galaxies) of scale $ R $
and mass $ \sim R^3 $ start to form when this scale becomes non-linear.
Our linear claculation shows that smaller objects form earlier in agreement
with astronomical observations [2]. 

\medskip

In short, WDM is characterized by : (i) its initial power spectrum 
cutted off for scales below $ r_{lin} \sim 50 $ kpc as shown by Fig. \ref{funtra}.
Thus, structures are not formed in WDM for scales below
$ r_{lin} \sim 50 $ kpc. (ii) its initial velocity dispersion. However, this is negligible
for $ z < 20 $ where the non-linear regime starts.

{\bf References}

\begin{description}

\item[1] H. J. de Vega, N. G. Sanchez, to appear september 2011.

\item[2] C. Destri, H. J. de Vega, N. G. Sanchez, in preparation.

\item[3]  H. J. de Vega, N. G. Sanchez,  arXiv:0901.0922, 
Mon. Not. R. Astron. Soc. 404, 885 (2010).
D. Boyanovsky, H. J. de Vega, N. G. Sanchez, 	
arXiv:0710.5180, Phys. Rev. {\bf D 77}, 043518 (2008).
H. J. de Vega, N. G. Sanchez, Int. J. Mod. Phys. A26: 1057 (2011),
arXiv:0907.0006.

\end{description}

\newpage

\subsection{Patrick Valageas}

\vskip -0.3cm

\begin{center}

Institut de Physique Th\'eorique, CEA Saclay, 91191 Gif-sur-Yvette, France.

\bigskip

{\bf Perturbation theories for large-scale structures} 

\end{center}

\medskip

In the standard cosmological scenario, the large-scale structures we observe in
the recent Universe (galaxies, clusters, filaments, voids, ..) have formed
through the amplification by gravitational instability of small primordial
perturbations generated in the primordial Universe by quantum fluctuations.
Since the initial power increases on small scales (for popular models such as
CDM), on large scales or at early times it is possible to use
linear theory or, more generally, perturbation theory, while on small scales or at late
times one must use numerical simulations or phenomenological models such as the
halo model.

In this talk, we have described some properties of perturbative approaches, in particular the
different behaviors found in Eulerian and Lagrangian frameworks, and we have shown
how to combine these schemes with phenomenological halo models, to build unified models
that can describe all scales.

\medskip

Focusing on the power spectrum $P(k)$, we have first recalled that standard perturbation
theory is not well-behaved, because higher-order terms grow increasingly fast at high
$k$ with strong cancellations between various orders. On the other hand, it can be checked
that many orders of perturbation theory (typically up to order $30$ at $z=2$) are relevant
before non-perturbative corrections associated with shell crossing dominate [1]. Therefore,
it is useful to build other perturbative schemes (i.e. partial resummations) that are better
behaved and more efficient.

\begin{figure*}[h]
\begin{center}
\epsfxsize=7.5 cm \epsfysize=6 cm {\epsfbox{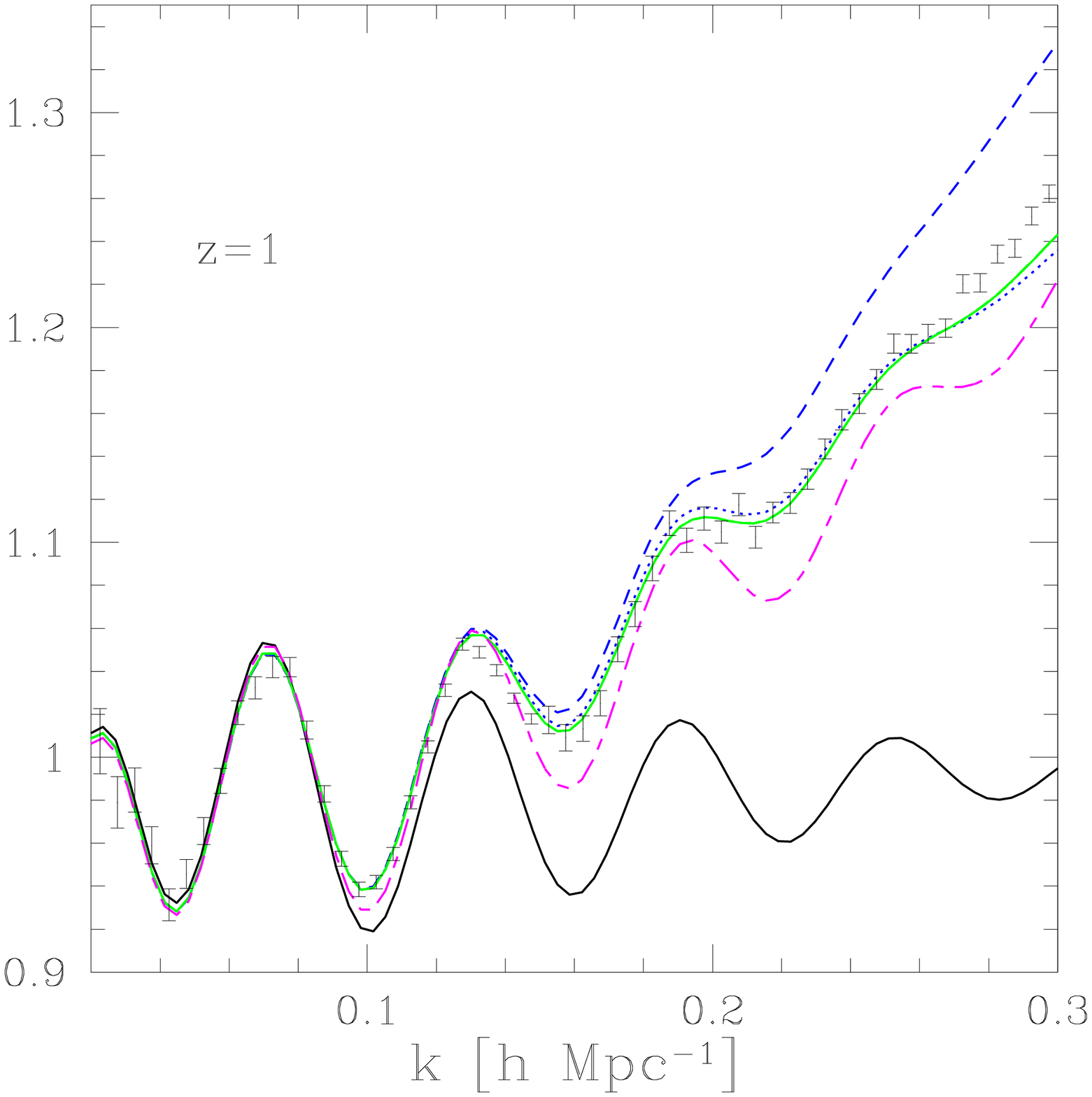}}
\epsfxsize=7.5 cm \epsfysize=6 cm {\epsfbox{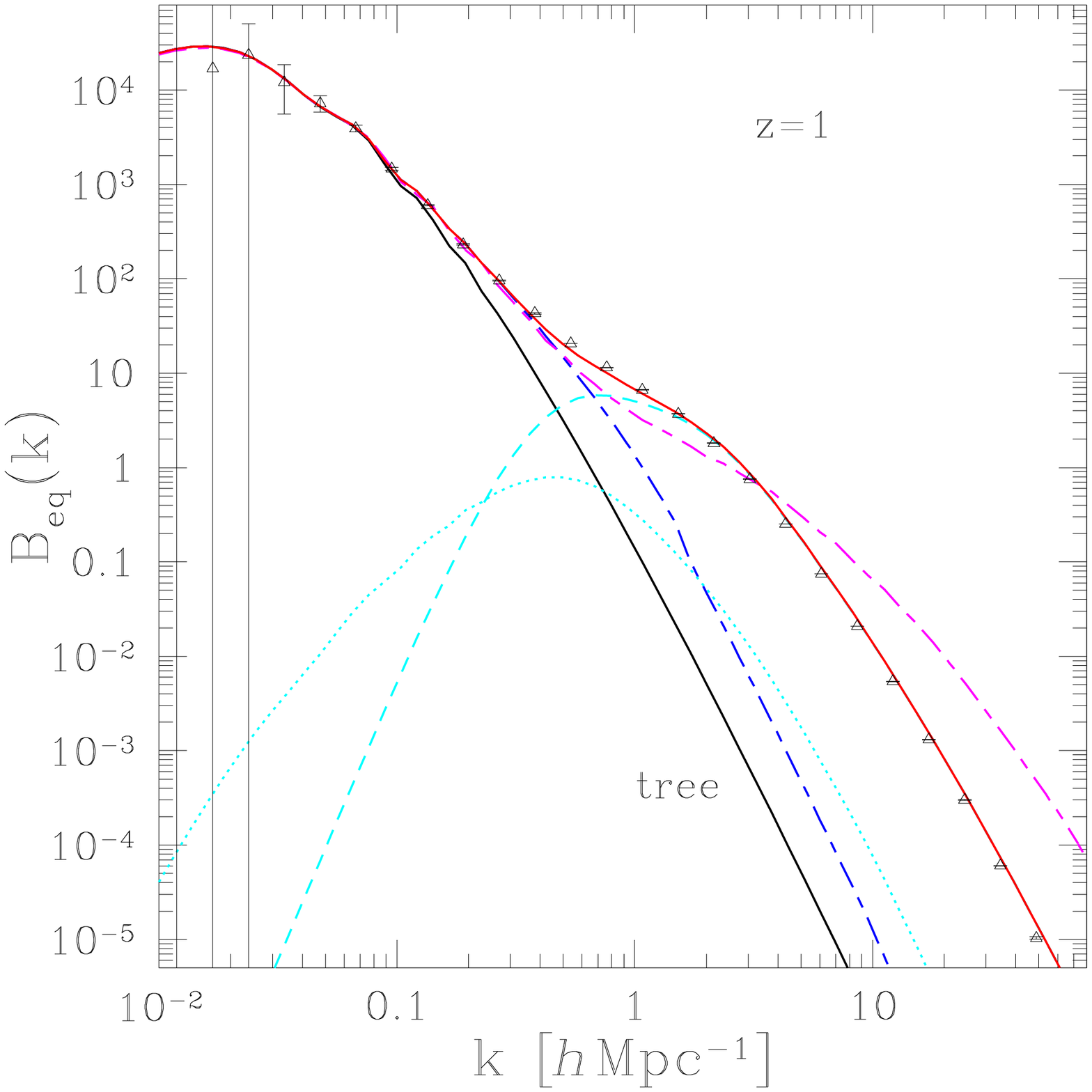}}
\end{center}
\caption{{\it Left panel:} ratio of the nonlinear power spectrum $P(k)$ to a smooth linear
power spectrum $P_{Ls}$ without acoustic baryonic oscillation, at $z=1$.
The points with error bars are the results from N-body simulations,
the black solid line is the linear power spectrum, the upper blue
dashed line is the prediction of standard perturbation theory up to 1-loop order,
the green solid line is the theoretical prediction (combining the
``large-N'' perturbative resummation and the halo model), the magenta dot-dashed line
is a popular fit to simulations [9].
{\it Right panel:} bispectrum for equilateral configurations,
at $z=1$.  The points are the results from numerical
simulations, the black solid line is the tree-order result, the blue dot-dashed line
is the standard 1-loop order prediction, the red solid line is the full theoretical
prediction that also includes the non-perturbative ``1-halo'' contribution.
The magenta dot-dashed line is a simple phenomenological model [10].}
\label{fig_Pk-200}
\end{figure*}

As a first example, we have described a ``large-N'' perturbative approach, based on
a path-integral formalism. At one-loop order this provides a significantly higher accuracy
on large scales than the standard perturbation theory, while being well-behaved at
high $k$ [2].
Next, we have explained how an alternative approach [3], based on a high-$k$ resummation
and the assumption of a wide separation of scales, corresponds to replacing the equations
of motion by linear equations with random coefficients. This yields Eulerian-space
propagators that show a Gaussian decay at high $k$, in good agreement with simulations.
However, this behavior is not a true loss of memory of the density field, but a ``sweeping
effect'' due to long wavelength modes of the velocity field that coherently move density
structures [4].
In particular, applying the same approach in Lagrangian-space, which is not
sensitive to this ``sweeping effect'', one obtains a propagator that no longer decays [5].

\medskip

Then, we have shown how these two distinct behaviors can be understood from the
study of the ``adhesion model'', where exact results can be derived. This model,
introduced by [6], extends the Zeldovich dynamics by making particles
glue together at collisions. This dynamics builds a cosmic web and density fields that are
very similar to those found in cosmological simulations, except that halos are pointlike
masses. Then, one can show that the Eulerian-space propagator is actually equal to the
velocity probability distribution, with an exponential-like high-$k$ decay, whereas the
Lagrangian propagator can be written in terms of the halo mass function, with a
power-law high-$k$ decay [7].
This confirms the explanation of these two different behaviors and explicitly shows that
Lagrangian propagators should be more sensitive probes of the density field than their
Eulerian counterparts.

\medskip

Next, we have described how to build unified models than combine perturbation theories
with halo models. More precisely, focusing on the power spectrum $P(k)$, the ``2-halo'' 
contribution can be written in terms of the perturbative prediction, whereas the
non-perturbative ``1-halo'' term involves the halo mass function and density profile.
Moreover, we have pointed out that the latter contains a specific counterterm that
ensures a physically meaningful behavior on large scales and is needed to reach
a high accuracy [2].

Then, we have described how the same approach allows us to obtain the density
bispectrum, from large to small scales [8].

\medskip

Finally, we have briefly reviewed a few other perturbative approaches and discussed
the Vlasov-Poisson system, which may allow going beyond the fluid approximation and
taking into account some shell-crossing effects.

\bigskip

{\bf References}

\begin{description}

\item[1] P. Valageas, A\&A, 526A, 67 (2011)

\vspace{-0.3cm}

\item[2] P. Valageas, T. Nishimichi, A\&A, 527A, 87 (2011)

\vspace{-0.3cm}

\item[3] M. Crocce, R. Scoccimarro, Phys. Rev. D, 73, 063519 (2006)

\vspace{-0.3cm}

\item[4] P. Valageas, A\&A, 476, 31 (2007)

\vspace{-0.3cm}

\item[5] F. Bernardeau, P. Valageas, Phys. Rev. D, 78, 083503 (2008)

\vspace{-0.3cm}

\item[6] S. N. Gurbatov, A. I. Saichev, S. F. Shandarin, Mon. Not. R. Astron. Soc., 236, 385 (1989)

\vspace{-0.3cm}

\item[7] F. Bernardeau, P. Valageas, Phys. Rev. D, 81, 043516 (2010)

\vspace{-0.3cm}

\item[8] P. Valageas, T. Nishimichi, arXiv:1102.0641

\vspace{-0.3cm}

\item[9] R. E. Smith, J. A. Peacock, et al., Mon. Not. R. Astron. Soc., 341, 1311 (2003)

\vspace{-0.3cm}

\item[10] J. Pan, P. Coles, I. Szapudi, Mon. Not. R. Astron. Soc., 382, 1460 (2007)

\end{description}

\newpage

\subsection{Shun Zhou}

\vskip -0.3cm

\begin{center}

Max-Planck-Institut f\"{u}r Physik (Werner-Heisenberg-Institut),
F\"{o}hringer Ring 6, D-80805 M\"{u}nchen, Germany

\bigskip

{\bf Supernova bound on keV-mass sterile neutrinos}

\end{center}

\medskip

Sterile neutrinos with masses in the keV range, which are a
promising candidate for warm dark matter, can be copiously produced
in the supernova (SN) core. For $m_s \gtrsim 100~{\rm keV}$, the
vacuum mixing angle of sterile neutrinos is stringently constrained
$\sin^2 2\theta \lesssim 10^{-9}$ in order to avoid excessive energy
loss [1--5]. For smaller masses, however, the
Mikheyev-Smirnov-Wolfenstein (MSW) matter effect on active-sterile
neutrino mixing becomes very important and the SN bound on vacuum
mixing angle is not that obvious. Note that the bounds on mixing
angles depend on which flavor the sterile neutrino mixes with. We
concentrate on the SN bound in $\nu_\tau$-$\nu_s$-mixing case for
simplicity, because $\nu_\tau$ and $\overline{\nu}_\tau$ only have
neutral-current interactions and essentially stay in thermal
equilibrium with ambient matter.

The matter density in the SN core is so high that the incoherent
scattering of active neutrinos on matter particles may even dominate
over flavor oscillations as the production mechanism for keV-mass
sterile neutrinos. An elegant formalism to deal with both incoherent
scattering and flavor oscillations have been developed in Refs.
[6--8], where the evolution equations for the occupation numbers of
different neutrino species have been derived. In the weak-damping
limit, which is always valid for supernova neutrinos mixing with
keV-mass sterile neutrinos, the evolution of $\nu_\tau$ number
density is determined by [9]
\begin{equation}
\dot{N}_{\nu_\tau} = -\frac{1}{4} \sum_a \int \frac{E^2 {\rm
    d}E}{2\pi^2} s^2_{2\theta_\nu} \int \frac{{E^\prime}^2 {\rm
    d}E^\prime}{2\pi^2} W^a_{E^\prime E} f^\tau_{E^\prime} \; ,~~
\end{equation}
where $s_{2\theta_\nu} \equiv \sin 2\theta_\nu$ with $\theta_\nu$
being the neutrino mixing angle in matter, $f^\tau_E$ the occupation
number of $\nu_\tau$, and $W^a_{E^\prime E}$ the transition
probability for $\nu(E^\prime) + a \to \nu(E) + a$ with $a$ being
background particles in the SN core. In a similar way, we can derive
the evolution equation of the $\bar\nu_\tau$ number density,
involving the mixing angle $\theta_{\bar\nu}$, the occupation number
$f^{\bar\tau}_E$ and the transition probability $\bar{W}^a_{E^\prime
E}$. Due to the MSW effect, the mixing angle of neutrinos in matter
is different from that of antineutrinos
\begin{equation}
\sin^2 2\theta_{\nu,\bar\nu} = \frac{\sin^2 2\theta}{\sin^2 2\theta
+ (\cos
  2\theta \pm E/E_{\rm r})^2} \;,
\end{equation}
where $\theta$ denotes the vacuum mixing angle, and the upper sign
refers to $\nu$ and the lower to $\bar\nu$. The resonant energy
$E_{\rm r} \equiv \Delta m^2/2|V_{\nu_\tau}|$ is defined as
\begin{equation}
E_{\rm r} = 3.25~{\rm MeV} \left(\frac{m_s}{10~{\rm
    keV}}\right)^2 \rho^{-1}_{14} \left|Y_0 - Y_{\nu_\tau}\right|^{-1} \; ,
\end{equation}
where $\rho_{14}$ is the matter density $\rho$ in units of
$10^{14}~{\rm g}~{\rm cm}^{-3}$ and $Y_0 \equiv (1 - Y_{e} -
2Y_{\nu_e})/4$. Note that $Y_x \equiv (N_x - N_{\bar x})/N_{\rm B}$
with $N_{\rm B}$ being the baryon number density, $N_x$ and
$N_{\bar{x}}$ being the number densities of particle $x$ and its
antiparticle $\bar x$. For tau neutrinos, the matter potential
$V_{\nu_\tau} = - (G_{\rm F}/\sqrt{2}) N_{\rm B} \left(1 - Y_e -
2Y_{\nu_e} - 4Y_{\nu_\tau}\right)$ is negative if the typical values
of $Y_e = 0.3$, $Y_{\nu_e} = 0.07$ and $Y_{\nu_\tau} = 0$ for a SN
core are taken. Therefore, the mixing angle for $\bar\nu_\tau$ is
enhanced by matter effects, and the emission rate for $\bar\nu_\tau$
exceeds that for $\nu_\tau$, indicating that a
$\nu_\tau$-$\bar\nu_\tau$ asymmetry (i.e., $Y_{\nu_\tau} \neq 0$)
will be established. An interesting feedback effect emerges: (i) The
chemical potential for tau neutrinos develops and thus changes the
occupation numbers of $\nu_\tau$ and $\bar\nu_\tau$; (ii) The
$\nu_\tau$-$\bar\nu_\tau$ asymmetry shifts the resonant energy
$E_{\rm r}$, and thus modifies the mixing angles $\theta_\nu$ and
$\theta_{\bar\nu}$; (iii) Both effects in (i) and (ii) will feed
back on the emission rates. Hence a stationary state of this
active-sterile neutrino system will be achieved if the emission
rates for neutrinos and antineutrinos become equal to each other
[9].

\begin{figure}[t]
\includegraphics[scale=0.8]{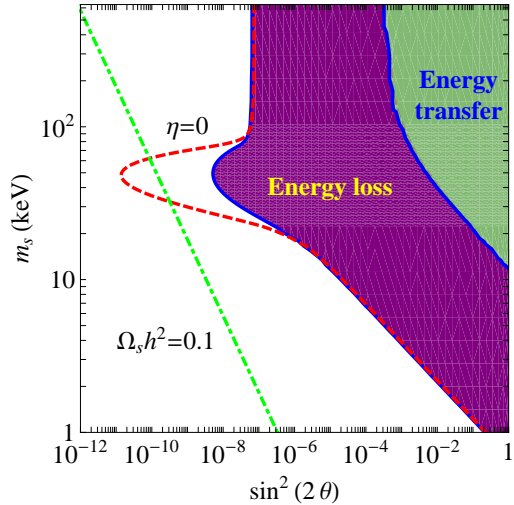}
\vspace{-0.1cm} \caption{Supernova bound on sterile neutrino masses
$m_s$ and mixing angles $\theta$, where the purple region is
excluded by the energy-loss argument while the green one by the
energy-transfer argument [9]. The excluded region will be extended
to the dashed (red) line if the build-up of degeneracy parameter is
ignored, i.e., $\eta(t) = 0$. The dot-dashed (green) line represents
the sterile neutrinos as dark matter with the correct relic
abundance $\Omega_s h^2 = 0.1$. }
\label{supernova}
\end{figure}

Given the sterile neutrino mass $m_s$ and vacuum mixing angle
$\theta$, the energy loss rate ${\cal E}(t)$ due to sterile neutrino
emission can be calculated by following the evolution of
$\nu_\tau$-$\bar\nu_\tau$ asymmetry $Y_{\nu_\tau}(t)$. It has been
found that the stationary state can be reached within one second and
the feedback effect is very important for $20~{\rm keV} \lesssim m_s
\lesssim 80~{\rm keV}$ and $10^{-9} \lesssim \sin^2 2\theta \lesssim
10^{-4}$. To avoid excessive energy losses, we require that the
average energy-loss rate $\langle {\cal E} \rangle \equiv
\int^{\tau_{\rm d}}_0 {\cal E}(t)~{\rm d}t$ with $\tau_{\rm d} =
1~{\rm s}$ should be $\langle {\cal E} \rangle \lesssim 3.0\times
10^{33}~{\rm erg}~{\rm cm}^{-3}~{\rm s}^{-1}$. Otherwise, the
duration of neutrino burst from Supernova 1987A would have been
significantly reduced. In Fig. \ref{supernova}, we show the contours of
energy-loss rates in the $(\sin^2 2\theta, m_s)$-plane, where we
have assumed a homogeneous and isotropic core with matter density
$\rho = 3.0\times 10^{-14}~{\rm g}~{\rm cm}^{-3}$ and temperature $T
= 30~{\rm MeV}$. Based on the energy-loss argument, the purple
region has been excluded. The most stringent bound $\sin^2 2\theta
\lesssim 10^{-8}$ arises for $m_s = 50~{\rm keV}$. For the
large-mixing angle region, the energy-loss rate is actually small,
because sterile neutrinos have been trapped in the core and cannot
carry away energies. However, the mean free path of sterile
neutrinos is comparable to or even longer than that of ordinary
neutrinos, indicating that they may transfer energies in a more
efficient way. As a consequence, the duration of neutrino burst will
be shortened by emitting neutrinos more rapidly. In this sense, the
excessive energy transfer should be as dangerous as the excessive
energy loss. Hence the large-mixing angle region is indeed excluded
if the energy-transfer argument is applied. The green line in Fig. \ref{supernova}
indicates the relic abundance of dark matter $\Omega_s h^2 = 0.1$,
where keV-mass sterile neutrinos are warm dark matter and the
non-resonant production mechanism is assumed. If we ignore the
feedback effect (i.e., a vanishing chemical potential for tau
neutrinos $\eta = \mu_{\nu_\tau}/T = 0$), the excluded region will
extend to the red line, which overlaps the relic-abundance line. As
shown in Fig. \ref{supernova}, however, the mixing angles are essentially
unconstrained in the favored warm-dark-matter mass range $1~{\rm
keV} \lesssim m_s \lesssim 10~{\rm keV}$ [10,11].

As for the $\nu_\mu$-$\nu_s$-mixing case, our discussions about the
feedback effects are essentially applicable. However, the
charged-current interactions of $\nu_\mu$ and $\bar\nu_\mu$ should
be taken into account, and the change of $\nu_\mu$-$\bar\nu_\mu$
asymmetry will be redistributed between muon neutrinos and charged
muons. The $\nu_e$-$\nu_s$ mixing in SN cores is more involved
because of the large trapped electron number and high $\nu_e$
degeneracy. Besides energy loss, deleptonization by sterile neutrino
emission is an effect to be taken into account. It has been
suggested that keV-mass sterile neutrinos mixing with electron
neutrinos may help supernova explosions [12,13]. However, this case
requires a dedicated investigation.

\bigskip

{\bf References}

\begin{description}

\item[1] K. Kainulainen, J. Maalampi, J. T. Peltoniemi,
Nucl. Phys. B {\bf 358}, 435 (1991).

\vspace{-0.3cm}

\item[2] G. Raffelt, G. Sigl, Astropart. Phys. {\bf 1}, 165 (1993).

\vspace{-0.3cm}

\item[3] J. T. Peltoniemi, Astron. Astrophys. {\bf 254}, 121 (1992).

\vspace{-0.3cm}

\item[4] X. Shi, G.~Sigl, Phys. Lett. B {\bf 323}, 360 (1994).

\vspace{-0.3cm}

\item[5] K. Abazajian, G. M. Fuller, M. Patel,
Phys. Rev.  D {\bf 64}, 023501 (2001).

\vspace{-0.3cm}

\item[6] L. Stodolsky, Phys. Rev. D {\bf 36}, 2273 (1987).

\vspace{-0.3cm}

\item[7] G. Raffelt, G.~Sigl, L.~Stodolsky, Phys. Rev. Lett. {\bf 70}, 2363
(1993).

\vspace{-0.3cm}

\item[8] G.~Sigl, G.~Raffelt, Nucl. Phys. B {\bf 406}, 423 (1993).

\vspace{-0.3cm}

\item[9] G. G. Raffelt, S. Zhou, Phys. Rev. D {\bf 83}, 093014
(2011).

\vspace{-0.3cm}

\item[10]  A. Kusenko, Phys. Rept. {\bf 481}, 1 (2009).

\vspace{-0.3cm}

\item[11] A. Boyarsky, O. Ruchayskiy, M. Shaposhnikov,
Ann. Rev. Nucl. Part. Sci. {\bf 59}, 191 (2009).

\vspace{-0.3cm}

\item[12] J. Hidaka, G. M. Fuller, Phys. Rev. D {\bf 74}, 083516
(2007).

\vspace{-0.3cm}

\item[13]  J. Hidaka, G. M. Fuller, Phys. Rev. D {\bf 76}, 125015
(2006).

\end{description}

\newpage

\newpage

\section{Summary and Conclusions of the Workshop 
by H. J. de Vega and \qquad \qquad \qquad \qquad N. G. Sanchez}

\subsection{General view and clarifying remarks}

Participants came from Europe, North and South America, Russia, Japan, Korea.
Discussions and lectures were outstanding.
WDM research evolves fastly in both astronomical, numerical, theoretical, particle and experimental research. 
The Workshop allowed to make visible the work on WDM made by different groups over the world and cristalize 
WDM as the viable component of the standard cosmological model in agreement with CMB + Large Scale
Structure (LSS) + Small Scale Structure (SSS) observations, $\Lambda$WDM, in contrast to 
 $\Lambda$CDM which only agree with CMB+LSS observations and is plagued with SSS problems.  

The participants and the programme represented the different communities doing
research on dark matter:

\begin{itemize}
\item{Observational astronomers}
\item{Computer simulators}
\item{Theoretical astrophysicists not doing simulations}
\item{Physical theorists}
\item{Particle experimentalists}
\end{itemize}

\medskip

WDM refers to keV scale DM particles. This is not Hot DM (HDM). (HDM refers to eV scale DM particles, 
which are already ruled out). CDM refers to heavy DM particles (so called wimps of GeV scale or any scale 
larger than keV). 

\medskip 

It should be recalled that the connection between 
small scale structure features and the mass of the DM particle 
follows mainly from the value of the free-streaming
length $ l_{fs} $. Structures 
smaller than $ l_{fs} $ are erased by free-streaming.
WDM particles with mass in the keV scale 
produce $ l_{fs} \sim 100 $ kpc while 100 GeV CDM particles produce an
extremely small $ l_{fs} \sim  0.1 $ pc. While the keV WDM $ l_{fs} \sim 100 $ kpc
is in nice agreement with the astronomical observations, the GeV CDM $ l_{fs} $ 
is a million times smaller and produces the existence of too many
small scale structures till distances of the size of the Oort's cloud
in the solar system. No structures of such type have ever been observed.

\medskip
 
Also, the name CDM  precisely refers to simulations with heavy DM particles in the GeV scale.
Most of the literature on CDM simulations do not make explicit
the relevant ingredient which is the mass of the DM particle (GeV scale wimps in the CDM case). 

\medskip
 
The mass of the DM particle with the free-streaming length naturally enters in the initial power spectrum 
used in the N-body simulations and in the initial velocity. The power spectrum for large scales 
beyond 100 kpc is identical for WDM and CDM particles,
while the WDM spectrum is naturally cut off at scales below 100 kpc, 
corresponding to the keV particle mass free-streaming length. In contrast, the CDM spectrum 
smoothly continues for smaller and smaller scales till $\sim$ 0.1 pc, which gives rise
to the overabundance of predicted CDM structures at such scales.

\medskip
 
CDM particles are always non-relativistic, the initial velocities
are taken zero in CDM simulations, (and phase space density is unrealistically infinity 
in CDM simulations), while all this is not so for WDM.

\medskip

Since keV scale DM particles are non relativistic for $ z < 10^6 $
they could also deserve the name of cold dark matter, although for historical reasons the name WDM
is used. Overall, seen in perspective today, the reasons why CDM does not work are simple: the heavy wimps 
are excessively non-relativistic (too heavy, too cold, too slow), and thus frozen, which preclude them to 
erase the structures below the kpc scale, while 
the eV particles (HDM) are excessively relativistic, too light and fast, (its free streaming length is too large), 
which erase all structures below the Mpc scale; in between, WDM keV particles produce the right answer. 

\medskip

\subsection{Conclusions}

Some conclusions are:

\begin{itemize}

\item{Sterile neutrinos with mass in the keV scale (1 to 10 keV) 
emerge as leading candidates for the dark matter (DM)
particle from theory combined with astronomical observations.

DM particles in the keV scale (warm dark matter, WDM)
naturally reproduce (i) the observed
galaxy structures at small scales (less than 50 kpc), (ii) the observed
value of the galaxy surface density and phase space density
(iii) the cored profiles of galaxy density profiles seen in
astronomical observations.

Heavier DM particles (as wimps in the GeV mass scale) do not
reproduce the above important galaxy observations and run into
growing and growing serious problems (they produce satellites problem,
voids problem, galaxy size problem, unobserved density cusps and other
problems).}


\item {Sterile neutrinos are serious WDM candidates:
Minimal extensions of the Standard Model of particle physics 
include keV sterile neutrinos which are very weakly coupled to the standard model particles
and are produced via the oscillation of the light (eV) active neutrinos, with their mixing
angle governing the amount of generated WDM. The mixing angle theta between active and sterile neutrinos
should be in the $ 10^{-4} $ scale to reproduce the average DM density in the
Universe.\\  
Sterile neutrinos are usually produced out of thermal equilibrium. 
The production can be non-resonant (in the absence of lepton asymmetries) 
or resonantly ennhanced (if lepton asymmetries are present). The usual X ray bound together
with the Lyman alpha bound forbids the non-resonant mechanism in the $\nu$MSM model. } 


\item{ Sterile neutrinos can decay into an active-like neutrino and an X-ray photon. Abundance and phase 
space density of dwarf spheroidal galaxies constrain the mass to be in the 
$ \sim $ keV range.  
Small scale aspects of sterile neutrinos and different mechanisms of their production 
were presented: The transfer function and power spectra obtained by solving the 
collisionless Boltzmann equation during the radiation and matter dominated eras feature new WDM acoustic
oscillations on mass scales $ \sim 10^8-10^{9} \, M_{\odot} $.} 


\item{Lyman-alpha constraints
have been often misinterpreted or superficially invoked in the past
to wrongly suggest a tension with WDM, but those constraints have been by now clarified
and relaxed, and such a tension does not exist: keV sterile neutrino dark matter (WDM) 
is consistent with Lyman-alpha constraints within a 
{\it wide range} of the sterile neutrino model parameters. {\bf Only} for sterile
neutrinos {\bf assuming} a {\bf non-resonant} (Dodelson-Widrow model) 
production mechanism, Lyman-alpha constraints provide a lower bound for the 
mass of about 4 keV . For thermal WDM relics (WDM particles decoupling at
thermal equilibrium) the Lyman-alpha lower particle mass bounds are 
smaller than for non-thermal WDM relics (WDM particles decoupling out of
thermal equilibrium). The number of Milky-Way satellites indicates lower bounds 
between 1 and 13 keV for different models of sterile neutrinos.}


\item {WDM keV sterile neutrinos can be copiously produced in the supernovae cores. Supernovae stringently 
constraint the neutrino mixing angle squared to be $ \lesssim 10^{-9}$ for sterile neutrino masses
$ m > 100$ keV (in order to avoid excessive energy lost) but for smaller sterile neutrino masses the 
SN bound is not so direct. Within the models 
worked out till now, mixing angles are essentially unconstrained by SN in the favoured WDM mass range, namely 
$ 1 < m  < 10 $ keV. Mixing between electron and keV sterile neutrinos could help SN explosions, case 
which deserve investigation}


\item{Signatures for a right-handed few keV sterile neutrino should be Lyman alpha emission and absorption 
at around a few 
microns; corresponding emission and absorption lines might be visible from molecular Hydrogen H$_2$  
and H$_3$  and their ions in the far infrared and sub-mm wavelength range.  The detection at very 
high redshift of massive star formation, stellar evolution and the formation 
of the first super-massive black holes would constitute the most striking and testable prediction of 
WDM sterile neutrinos.}


\item{The effect of keV WDM can be also observable in the statistical 
properties of cosmological Large Scale Structure. Cosmic shear 
(weak gravitational lensing) does not strongly depend on baryonic 
physics and is a promisingig probe. First results in a simple thermal 
relic scenario indicate that future weak lensing surveys could see a WDM 
signal for $m_{WDM} \sim 2$ keV or smaller. The predicted limit beyond 
which these surveys will not see a WDM signal is $m_{WDM} \sim 2.5$ keV 
(thermal relic) for combined 
Euclid + Planck. More realistic models deserve investigation and are 
expected to relaxe such minimal bound. With the real data, the 
non-linear WDM model should be taken into account}


\item {The possibility of laboratory detection of warm dark matter
is extremely interesting. Only a direct detection of the DM particle can give a clear-cut answer
to the nature of DM. At present,  only the {\bf Katrin and Mare experiments}
have the possibility to do that for sterile neutrinos.
{\bf Mare} bounds on sterile neutrinos have been reported in this Workshop.
Namely, bounds from the beta decay of Re187 and EC decay of Ho163.
{\bf Mare} keeps collecting data in both.}


\item {The possibility that {\bf Katrin} experiment can look to sterile neutrinos in the tritium
decay did appeared in the discussions.
Katrin experiment have the potentiality to detect warm dark matter if its set-up 
would be adapted to look to keV scale sterile neutrinos.
Katrin experiment concentrates its attention right now
on the electron spectrum near its end-point
since its goal is to measure the active neutrino mass.
Sterile neutrinos in the tritium decay will affect the electron
kinematics at an energy about $m$ below the end-point of the
spectrum ($m$ = sterile neutrinos mass). Katrin in the future
could perhaps adapt its set-up to look to keV scale sterile neutrinos.
It will be a a fantastic discovery to detect dark matter
in a beta decay}


\item{Astronomical observations strongly indicate that
{\bf dark matter halos are cored till scales below 1 kpc}. 
More precisely, the measured cores {\bf are not} hidden cusps.
CDM Numerical simulations -with wimps (particles heavier than $ 1 $ GeV)-
without {\bf and} with baryons yield cusped dark matter halos.
Adding baryons do not alleviate the problems of wimps (CDM) simulations,
on the contrary adiabatic contraction increases the central density of cups
worsening the discrepancies with astronomical observations. 
In order to transform the CDM cusps into cores, the baryon+CDM simulations
need to introduce strong baryon and supernovae feedback which produces a
large star formation rate contradicting the observations.
None of the predictions of CDM simulations at small scales
(cusps, substructures, ...) have been observed. The discrepancies of CDM 
simulations with the astronomical observations at small scales $ \lesssim 100 $ kpc {\bf is staggering}: 
satellite problem (for example, only 1/3 of satellites predicted by CDM
simulations around our galaxy are observed), the surface density problem (the value 
obtained in CDM simulations is 1000 times larger than the
observed galaxy surface density value),  the voids problem, size problem 
(CDM simulations produce too small galaxies).}


\item{The use of keV scale WDM particles in the simulations instead of the GeV CDM wimps, alleviate all the
above problems. For the core-cusp problem, setting
the velocity dispersion of keV scale DM particles seems beyond
the present resolution of computer simulations. 
However, the velocity dispersion of WDM particles is negligible
for $ z < 20 $ where the non-linear regime and structure formation starts.
Analytic work in the
linear approximation produces cored profiles for keV scale DM particles
and cusped profiles for CDM. Model-independent analysis of DM from phase-space density
and surface density observational data plus theoretical analysis
points to a DM particle mass in the keV scale.
The dark matter particle candidates with high mass (100 GeV, `wimps') 
are strongly disfavored, while cored (non cusped) dark matter halos and warm (keV scale mass) 
dark matter are strongly favoured from theory and  astrophysical observations.}

\item{An `Universal Rotation Curve' (URC) of spiral galaxies emerged from  
3200 individual observed Rotation Curves (RCs) and 
reproduces remarkably well out to the virial radius the Rotation Curve of 
any spiral galaxy. The URC is the observational counterpart of the  circular  
velocity profile from cosmological  simulations.
$\Lambda$CDM simulations give the well known NFW cuspy halo profile. A careful analysis 
from about 100 observed high quality rotation curves has now {\bf ruled out} the disk + NFW halo 
mass model, in favor of {\bf cored profiles}. 
The observed galaxy surface density (surface gravity acceleration) appears to be universal within 
$ \sim 10 \% $ with values around $ 120 \; M_{\odot}/{\rm pc}^2 $
, irrespective of galaxy morphology, luminosity and Hubble types, spanning over 14 magnitudes in 
luminosity and mass profiles determined by several independent methods.}


\item{Interestingly enough, a constant surface density  (in this case column density) with value around 
$ 120 \; M_{\odot}/{\rm pc}^2 $ similar to that found for galaxy systems  is found too for the interstellar 
molecular clouds, irrespective of size and compositions over six order of magnitude; this universal surface 
density in molecular clouds
is a consequence of the Larson scaling laws. This suggests the role of gravity on matter (whatever DM or 
baryonic) as a dominant underlying mechanism to produce such universal surface density in galaxies and molecular 
clouds. Recent re-examination of different and independent (mostly millimeter) molecular cloud data sets show 
that interestellar clouds do follow Larson law $Mass \sim (Size)^{2}$ exquisitely well, and therefore very 
similar projected mass densities at each extinction threshold. Such scaling and universality should play a key 
role in cloud structure formation.}


\item{Visible and dark matter observed distributions (from rotation curves of high resolution 2D velocity fields) 
of galaxies in compact groups, and the comparison with those of galaxies in clusters and field galaxies show 
that: (i) The central halo surface density is constant with respect to the total absolute magnitude similar to 
what is found for the isolated galaxies, suggesting that the halo density is independent of the galaxy type 
and environment. (ii) Core density profiles fit better the rotation curves than cuspy profiles. Dark halo 
density profiles are found almost the same in field galaxies, cluster and compact group galaxies. Core halos 
observed using high-resolution velocity fields in dark matter galaxies are {\bf genuine} and cannot be ascribed 
to systematic errors, halo triaxiality, or non-circular motions. (iii) The halo mass is high (75 to 95 \% of 
the total mass) for both field galaxies and compact groups galaxies, living modest room for a dark mass disk. 
No relation between the disk scale length and the halo central density is seen, 
the halo being independent of galaxy luminosity.}             


\item{Besides the DM haloes, $\Lambda$CDM models ubiquituosly predict cold dark matter disks, 
[also believed as the consequence of merging events]: CDM simulations predict that 
a galaxy such as the Milky Way should host a CDM disk 
(with a scale height of $2.1-2.4$ kpc and a local density at the solar position $0.25-1.0$ times that of the 
DM halo; some models reach a height of 4 kpc from the galactic plane), which is thicker than any visible disk 
(the scale height of the galactic old thick stellar disk is $\sim$ 0.9 kpc; young stars and ISM form even 
thinner structures of height 0.3 and 0.1 kpc respectively). However, careful astronomical observations 
performed to see such disk have found no evidence of such CDM disk in the Milky Way.}

\item{A key part of any galaxy formation process and evolution involves dark matter. Cold gas accretion and 
mergers became important ingredients of the CDM models but they have little observational evidence. DM properties 
and its correlation with stellar masses are measured today up to $z =2$; at $z > 2$ observations are much less 
certain. Using kinematics and star formation rates, all types of masses -gaseous, stellar and dark- are measured 
now up to $z =1.4$. The DM density within galaxies declines at higher redshifts.  Star formation is observed 
to be more common in the past than today. More passive galaxies are in more 
massive DM halos, namely most massive DM halos have lowest fraction of stellar mass. CDM predicts high
overabundance of structure today and under-abundance of structure in the past with respect to observations. 
The size-luminosity scaling relation is the tightest of all purely photometric correlations used to 
characterize galaxies; its environmental dependence have been highly debated but recent findings show that the 
size-luminosity relation of nearby elliptical galaxies is well defined by a fundamental line and is environmental 
independent. Observed structural properties of elliptical galaxies appear simple and with no environmental 
dependence, showing that their growth via important mergers -as required by CDM galaxy formation- is not 
plausible. Moreover, observations in brightest cluster galaxies (BCGs) show little changes in the sizes of 
most massive galaxies since $z =1$ and this scale-size evolution appears closer to that of radio galaxies over 
a similar epoch. This lack of size growth evolution, a lack of BCG stellar mass evolution is observed too, 
demonstrates that {\bf major merging is not an important process}. Again, these observations put in serious trouble 
CDM `semianalytical models' of BCG evolution which require about $70\%$ of the final BCG stellar mass to be 
accreted in the evolution and important growth factors in size of massive elliptical massive galaxies.}

\item{Recent $\Lambda$WDM N-body simulations have been performed by different groups. 
High resolution simulations for different types of DM (HDM, WDM or CDM), allow to visualize the effects of 
the mass of the corresponding DM particles: free-streaming length scale, initial velocities and associated 
phase space density properties: for masses in the eV scale (HDM), halo formation occurs top down on all scales 
with the most massive haloes collapsing first; if primordial velocities are large enough, free streaming erases 
all perturbations and  haloes  cannot form (HDM). The concentration-mass halo relation for mass of hundreds eV 
is reversed with respect to that found for
CDM wimps of GeV mass. For realistic keV WDM these simulations deserve investigation: it could be expected from 
these HDM and CDM effects that combined free-streaming and velocity effects in keV WDM simulations could produce 
a bottom-up hierarchical scenario with the right amount of sub-structures (and some scale at which transition 
from top-down to bottom up regime is visualized).}

\item{Moreover, interestingly enough, recent large high resolution $\Lambda$WDM N-body simulations allow to 
discriminate among thermal and non-thermal WDM (sterile neutrinos): Unlike conventional thermal relics, 
non-thermal WDM has a peculiar velocity distribution (a little skewed to low velocities) which translates 
into a characteristic linear matter power spectrum decreasing slowler across the cut-off free-streaming scale 
than the thermal WDM spectrum. As a consequence, the  radial distribution of the subhalos predicted by WDM 
sterile neutrinos remarkably reproduces the observed distribution of Milky Way satellites in the range above 
$\sim 40$ kpc, while the thermal WDM supresses subgalactic structures perhaps too much, by a factor $2-4$ 
than the observation. Both simulations were performed for a mass equal to 1 keV. 
Simulations for a mass larger than 1 keV  (in the range between 2 and 10 keV, say) should still improve these results.}

\item{The predicted $\Lambda$WDM galaxy distribution in the local universe (as performed by
CLUES simulations with a
a mass of $ m_{\rm WDM}=1 $ keV)  agrees well 
with the observed one in the ALFALFA survey. On the
contrary, $\Lambda$CDM predicts a steep rise in the velocity
function towards low velocities and thus forecasts much  more
sources than the ones observed by the ALFALFA survey (both in Virgo-direction as well as in 
anti-Virgo-direction). These results show again the $\Lambda$CDM problems, also shown in the spectrum of 
mini-voids. $\Lambda$WDM 
provides a natural solution to these problems.  WDM physics effectively acts as a truncation of the 
$\Lambda$CDM power spectrum. $\Lambda$WDM CLUES simulations with 
1 keV particles gives much better answer than $\Lambda$CDM when reproducing sizes of local 
minivoids. The velocity function of 1 keV WDM Local Volume-counterpart reproduces the observational 
velocity function remarkably well. 
Overall, keV WDM particles deserve dedicated experimental detection efforts and simulations.}

\item{The features observed in the cosmic-ray spectrum by Auger, Pamela, Fermi, HESS, 
CREAM and others can be all quantitatively well explained with the action of cosmic rays
 accelerated in the magnetic winds of very massive star explosions such as Wolf-Rayet stars, 
without any significant free parameter. All these observations of cosmic ray positrons and 
electrons and the like are due to normal astrophysical sources and processes, and do not require 
an hypothetical decay or annihilation of heavy CDM particles (wimps). 
The models of annihilation or decay of heavy CDM wimps
are highly tailored to explain these normal 
astrophysical processes and their ability to survive observations is more than reduced.}

\item{Theoretical analytic perturbative approachs for large scales in Eulerian and Lagrangian frameworks 
can be combined with phenomenological halo models to build unified schemes for describe all scales. The 
large $k$ Gaussian decay of Eulerian-space propagators (in agreement with simulations) is not a true loss of 
memory of the density field but a "sweepping" effect of the velocity field modes which coherently move density 
structures. Lagrangian-space propagators are not sensitive to such effect. Extending the Zeldovich  
approximation ("adhesion model") show that Eulerian propagators are in fact the velocity probability 
distribution, whereas Lagrangian propagators explicitely relate to the halo mass function and are most 
sensitive probes of the density field than their Eulerian conterparts. Extensions to the Vlasov-Poisson 
system deserve investigation.\\
Theoretical analytic modelling of the halo mass function on small scales gains insight with the use of 
the mathematical excursion set theory in a path integral formulation, which allows direct comparison with 
numerical simulations and observational results. Stochastic modeling of the halo collapse conditions can 
be easily implemented in this formalism.}

\item{{\it As an overall conclusion}, CDM represents the past and WDM represents the future in the DM research. 
CDM research is 20 years old. CDM simulations and their proposed baryonic solutions, 
and the CDM wimp candidates ($ \sim 100$ GeV)
are strongly pointed out by the galaxy observations as the {\it wrong} solution to DM. 
Theoretically, and placed in perspective after more than 20 years, the reason why CDM does not work appears simple and clear 
to understand and directly linked to the excessively heavy and slow CDM wimp, which determines an excessively small 
(for astrophysical structures) free streaming length, and unrealistic overabundance of structures at these scales.  
On the contrary, new keV WDM research, keV WDM simulations, and keV scale mass  
WDM particles are strongly favoured by galaxy observations and theoretical analysis, they naturally {\it work} 
and agree with the astrophysical observations at {\it all} scales, (galactic as well as cosmological scales). 
Theoretically, the reason why WDM works  so well
is clear and simple, directly linked to the keV scale mass and velocities of the WDM particles,
and free-streaming length.   The experimental search for serious WDM particle candidates (sterile neutrinos) 
appears urgent and important: it will be a fantastic discovery to detect dark matter in a beta decay. 
There is a formidable WDM work to perform ahead of us, these highlights point some of the directions where 
it is worthwhile to put the effort.}

\subsection{The present context and future in DM research.}

\item{Facts and status of DM research: Astrophysical observations
point to the existence of DM. Despite of that, proposals to
replace DM  by modifing  the laws of physics did appeared, however
notice that modifying gravity spoils the standard model of cosmology
and particle physics not providing an alternative.
After more than twenty active years the subject of DM is mature, (many people is involved in this problem, 
different groups perform N-body cosmological simulations and on the other hand direct experimental particle 
searches are performed by different groups, an important
number of conferences on DM and related subjects is held regularly). DM research  
appears mainly in three sets:
(a) Particle physics DM model building beyond the standard
model of particle physics, dedicated laboratory experiments,
annhilating DM, all concentrated on CDM and CDM wimps.
(b) Astrophysical DM: astronomical observations, astrophysical models.
(c) Numerical CDM simulations. : The results of (a) and (b)
do not agree and (b) and (c) do not agree neither at small scales.
None of the small scale predictions of CDM simulations 
have been observed: cusps and over abundance of substructures
differ by a huge factor with respect to those observed.
In addition, all direct {\it dedicated} searchs of CDM wimps from more than twenty years 
gave {\it null results}. {\it Something is going wrong in the CDM research and the right answer is: 
the nature of DM is not cold (GeV scale) but warm (keV scale)}.}

\item{Many researchers continue to work with heavy CDM candidates
(mass $ \gtrsim 1 $ GeV) despite the {\bf staggering} evidence that these
CDM particles do not reproduce the small scale astronomical observations
($ \lesssim 100 $ kpc). Why? [It is known now that the keV scale DM particles naturally produce the
observed small scale structure]. Such strategic
question is present in many discussions, everyday and off of the record (and on the record) talks in the field. 
The answer deals in large part with the inertia (material, intellectual, social, other, ...) that structured research 
and big-sized science in general do have, which involve huge number of people, huge budgets, longtime planned 
experiments, and the "power" (and the conservation of power) such situation could allow to some of 
the research lines following the trend; as long as budgets will allow to run wimp experimental searches and CDM simulations 
such research lines could not deeply change, although they would progressively decline.}

\item{Notice that in most of the DM litterature or conferences, wimps are still "granted" as "the" DM particle, 
and CDM as "the" DM; is only recently that the differences and clarifications are being clearly recognized 
and acknowledged. While wimps were a testable hypothesis at the beginning of the CDM research, 
one could ask oneself why they continue to be worked out and "searched" experimentally in spite of 
the strong astronomical and astrophysical  evidence against them. \\
Similar situations (although not as extremal as the CDM situation) happened in other branches of physics and cosmology:
Before the CMB anisotropy observations, the issue of structure formation was plugged with several alternative 
proposals which were afterwards ruled out. Also, string theory passed from being considered "the theory of 
everything" to "the theory of nothing" (as a physical theory), as no physical experimental evidence have been 
obtained and its cosmological implementation and predictions desagree with observations. (Despite all 
that, papers on such proposals continue -and probably will continue- to appear. But is clear that  big dedicated 
experiments are not planned or built to test such papers).  In science, what is today `popular' can be discarded 
afterwards; what is today `new' and minoritary can becomes `standard' and majoritarily accepted if verified 
experimentally. }
\end{itemize}

\bigskip

\begin{center}

 {\bf \em  `Examine the objects as they are and you will see their true nature;
look at them from your own ego and you will see only your feelings;
because nature is neutral, while your feelings are only prejudice and obscurity'}

\medskip

[Gerry Gilmore quoting Shao Yong, 1011-1077 in the 14th Paris Cosmology Colloquium Chalonge 2010
http://chalonge.obspm.fr/Programme\_Paris2010.html, arXiv:1009.3494].

\end{center}

\bigskip

\begin{center}

The Lectures of the Workshop can be found at:

\bigskip

{\bf http://www.chalonge.obspm.fr/Programme\_CIAS2011.html}

\bigskip

The photos of the Workshop can be found at:

\bigskip

{\bf http://www.chalonge.obspm.fr/albumCIAS2011/index.html}

\end{center}

\newpage

\section{List of Participants}

BEZRUKOV  Fedor,	Ludwig-Maximilians-Universitat, 	M\"unchen ,	Germany \\
\medskip

CHAUDHURY  Soumini,	Saha Institute of Nuclear Physics,	Kolkata,	India\\
\medskip

CONSELICE Christopher  J, School for Physics and Astronomy, Nottingham Univ Nottingham, 	United  Kingdom\\
\medskip

CORASANITI  Pier Stefano,	CNRS LUTH Observatoire de Paris, Meudon,  France\\
\medskip

DE VEGA Hector J.,	CNRS LPTHE UPMC, Paris, 	France\\
\medskip

DESTRI Claudio, INFN Dipt di Fisica G Occhialini, Univ. Milano-Bicocca, Milano, Italy\\
\medskip

DOMINGUEZ Mariano,	IATE, UNC-CONICET	Cordoba,	Argentina\\
\medskip

ETTORI Stefano, INAF, Osservatorio Astronomico di Bologna,	Bologna,	Italy\\
\medskip

GEWERING-PEINE, Alexander, Institute of Experimental Physics, Hamburg, Germany\\
\medskip

HESSMAN Frederic,	Georg-August-Universität G\"ottingen, G\"ottingen, 	Germany\\
\medskip

JIZBA Petr, FNSPE, Czech Technical University in Prague, Prague,	Czech Republic\\
\medskip

KAMADA Ayuki, Institute for the Physics and Mathematics of the Universe, Kashiwa	Japan\\
\medskip

KNUDDE Sylvain, LESIA Observatoire de Paris, Meudon,	France\\
\medskip

LETOURNEUR Nicole, CIAS Observatoire de Paris, Meudon,	France\\
\medskip

LOMBARDI  Marco,	Department of Physics, University of Milan,	Milan,	Italy\\
\medskip

MARKOVIC	Katarina,	Excellence Cluster Universe	Munich,	Germany\\
\medskip

MERLE Alexander,	Royal Institute of Technology (KTH), Stockholm,	Sweden\\
\medskip

MONI BIDIN Christian, Universidad de Concepci\'on, Depart. de Astromomia, Concepcion, Chile\\
\medskip

NOH, Hyerim	Korea Astronomy and Space Science Institute, Taejon, South Korea\\
\medskip

NUCCIOTTI  Angelo, Dip. di Fisica, Universit\`a di Milano-Bicocca, Milano,  Italia\\
\medskip

PADUROIU Sinziana, Observatoire de Gen\`eve,	Geneva, Switzerland\\
\medskip

PENARRUBIA, Jorge Institute of Astronomy, University of Cambridge,	Cambridge,UK\\
\medskip

PFENNIGER, Daniel,	University of Geneva, Geneva Observatory, Sauverny,	Switzerland\\
\medskip

PLANA Henri,	Lab. Astrofisica Teorica e Observacional - ILHEUS, Santa Cruz, Bahia, Brazil\\
\medskip

POMMIER  Mamta,	CRAL-Observatoire de Lyon	 Saint Genis,   France\\
\medskip

RAMON  MEDRANO  Marina,  Universidad  Complutense, Madrid, Spain\\
\medskip

RUCHAYSKIY  Oleg, CERN	Geneva,	Switzerland\\
\medskip

SALUCCI  Paolo,	SISSA,  Astrophysics Division,	Trieste, 	Italy\\
\medskip

SANCHEZ  Norma G., CNRS LERMA Observatoire de Paris,	Paris,	France\\
\medskip

SCHAYE  Joop, Leiden University,	Leiden, The Netherlands\\
\medskip

SMITH  Robert,	ITP, University of Zurich,	Zurich, 	Switzerland\\
\medskip

VALAGEAS Patrick, 	Institut de Physique Th\'eorique, CEA-Saclay, Gif-sur-Yvette, France\\
\medskip

ZANINI Alba,	INFN Sezione di Torino,	Turin,	Italy\\
\medskip

ZHOU  Shun,	MPI for Physics,	Munich,	Germany\\
\medskip

ZIAEEPOUR Houri,	Max Planck Institute für Extraterrestrische Physik,	Garching bei München, Germany\\
\medskip

ZIDANI  Djilali, CNRS LERMA Observatoire de Paris,	Paris,	France\\
\medskip

\begin{figure}[htbp]
\epsfig{file=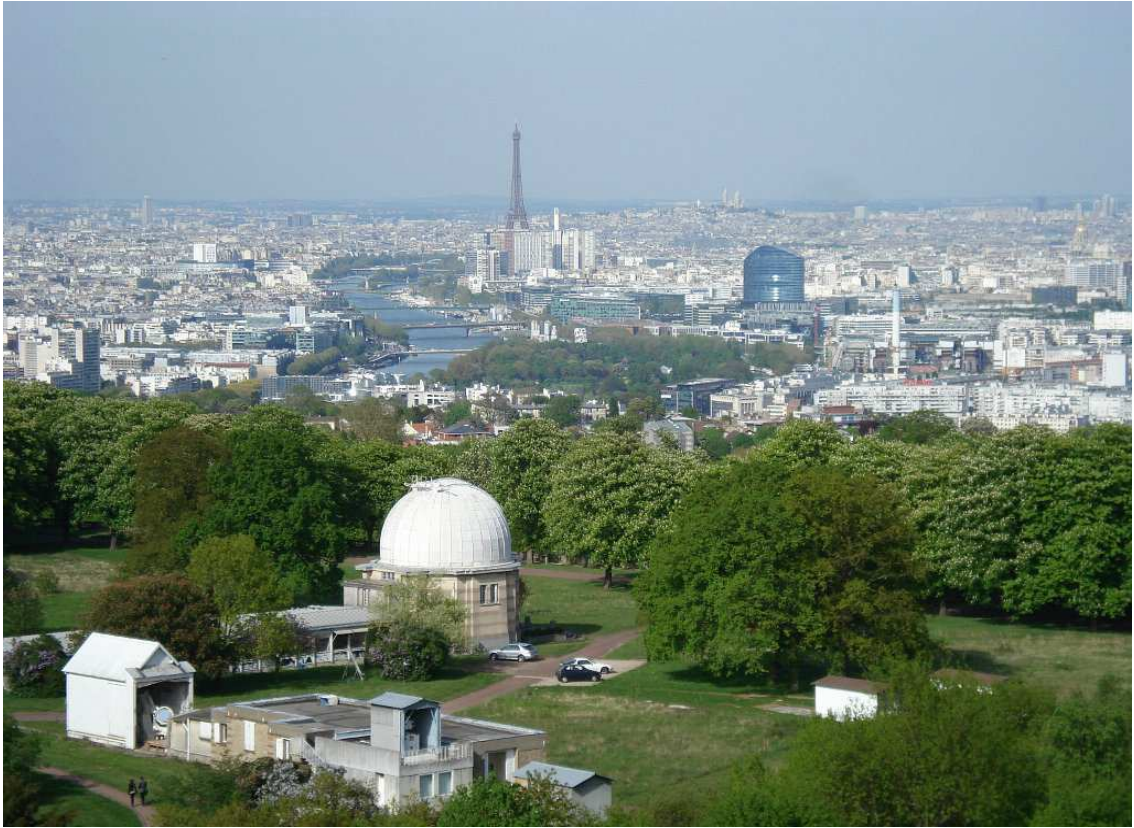,width=15cm,height=11cm}
\epsfig{file=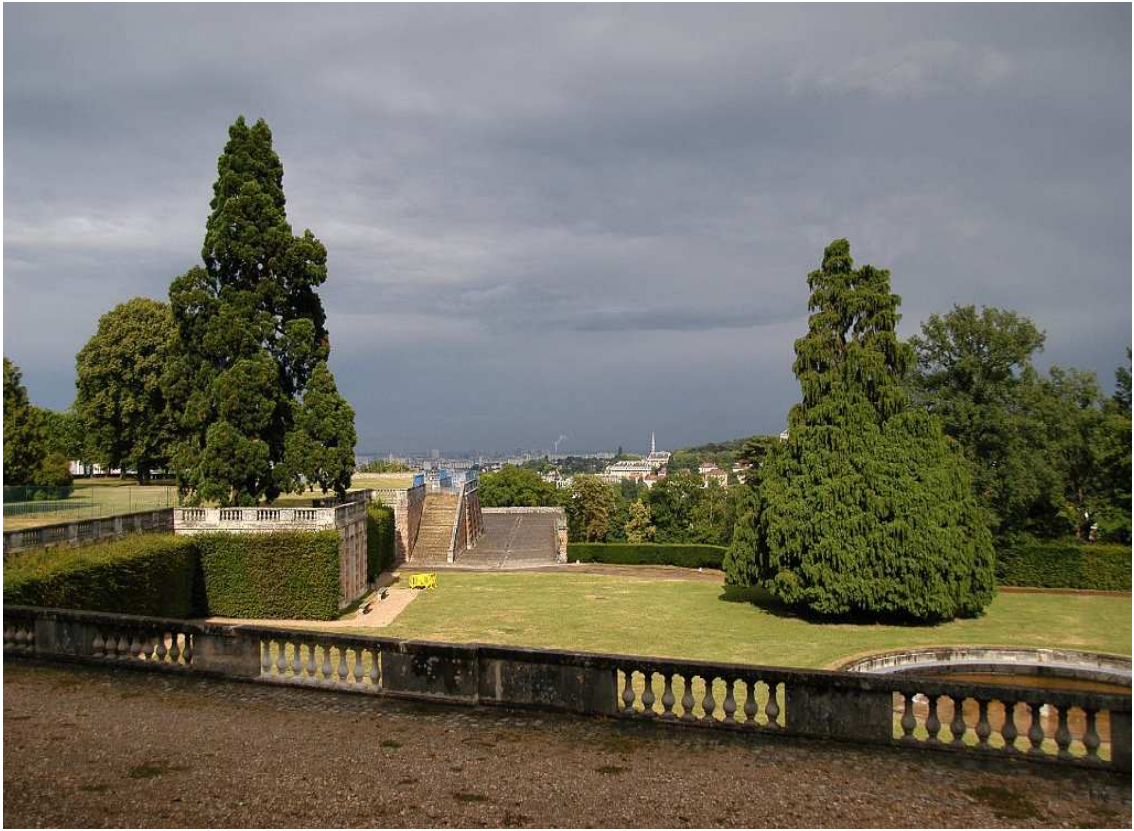,width=15cm,height=11cm}
\caption{Views from the Meudon Ch\^ateau}
\end{figure}

\end{document}